\documentclass{article}

\usepackage[margin=1.5in]{geometry}

\usepackage[utf8]{inputenc}
\usepackage{outlines}
\usepackage{amsmath,amssymb}
\usepackage{bm}
\usepackage{algorithm}
\usepackage{amssymb}
\usepackage{mathrsfs}
\usepackage{graphicx}
\usepackage{verbatim}
\usepackage{hyperref}
\usepackage{subfig}
\usepackage[normalem]{ulem}
\usepackage{authblk}

\newcommand{\R}{\mathbb{R}}

\usepackage{xcolor}

\usepackage{subfiles} 

\title{Deep Learning in Single-Cell Analysis}

\author[1]{Dylan Molho\thanks{Indicates equal contributions.}\thanks{molhodyl@msu.edu}}
\author[2]{Jiayuan Ding$^{*}$\thanks{dingjia5@msu.edu}}
\author[4]{Zhaoheng Li}
\author[2]{Hongzhi Wen}
\author[3]{Wenzhuo Tang}
\author[5]{Yixin Wang}
\author[1]{Julian Venegas}
\author[2]{Wei Jin}
\author[1]{Renming Liu}
\author[1,3]{Runze Su}
\author[8]{Patrick Danaher}
\author[9]{Robert Yang}
\author[6,7]{Yu Leo Lei}
\author[1,3]{Yuying Xie}
\author[2]{Jiliang Tang}

\affil[1]{Department of Computational Mathematics, Science and Engineering, Michigan State University, East Lansing, USA}
\affil[2]{Department of Computer Science and Engineering, Michigan State University, East Lansing, USA}
\affil[3]{Department of Statistics and Probability, Michigan State University, East Lansing, USA}
\affil[4]{Department of Biostatistics, University of Washington, Seattle, USA}
\affil[5]{Department of Bioengineering, Stanford University, Palo Alto, USA}
\affil[6]{Department of Periodontics and Oral Medicine, University of Michigan School of Dentistry, Ann Arbor, USA}
\affil[7]{University of Michigan Rogel Cancer Center, Ann Arbor, USA}
\affil[8]{NanoString Technologies, Seattle, USA}
\affil[9]{Johnson \& Johnson, Boston, USA}

\begin{document}
\maketitle

\begin{abstract}
Single-cell technologies are revolutionizing the entire field of biology.  The large volumes of data generated by single-cell technologies are high-dimensional, sparse, heterogeneous, and have complicated dependency structures, making analyses using conventional machine learning approaches challenging and impractical. 
In tackling these challenges, deep learning often demonstrates superior performance compared to traditional machine learning methods.  In this work, we give a comprehensive survey on deep learning in single-cell analysis.  We first introduce background on single-cell technologies and their development, as well as fundamental concepts of deep learning including the most popular deep architectures.
We present an overview of the single-cell analytic pipeline pursued in research applications while noting divergences 
due to data sources or specific applications. 
We then review seven popular tasks spanning through different stages of the single-cell analysis pipeline, including multimodal integration, imputation, clustering, spatial domain identification, cell-type deconvolution, cell segmentation, and cell-type annotation. Under each task, we describe the most recent developments in classical and deep learning methods and discuss their advantages and disadvantages. Deep learning tools and benchmark datasets are also summarized for each task. Finally, we discuss the future directions and the most recent challenges. This survey will serve as a reference for biologists and computer scientists, encouraging collaborations.

\end{abstract}



\section{Introduction}
As the basic building block of life,
cells assume dynamic and complex functional states to inform higher-order structures~\cite{mcmanus2015, patterson2003}.
Towards that end, the advance of single-cell sequencing and imaging technologies has revolutionized the investigation of the gene-expression behaviors of cells. The advent of single-cell sequencing technology occurred in the early 1990s for complementary DNA (cDNA)~\cite{Eberwine1992, Brady1990}. However, it was not until 2009, with the creation of the first single-cell RNA sequencing (scRNA-seq) method~\cite{Tang2009} that marked a true paradigm shift in the field. Since then, steady progress in the creation of new next-generation sequencing platforms has led to over one hundred currently existing techniques for single-cell sequencing~\cite{wang2015advances, liang2014single, wen2022recent}. These technologies measure a diverse collection of cell features including DNA sequences and epigenetic features, 
RNA expression, and profiles of surface proteins.  Recent technological advances have also enabled the augmentation of these features with additional data, e.g., multimodal sequencing platforms and spatial transcriptomic technology.

This paradigm shift comes from 
the quantity of available data 
using high throughput methods \cite{svensson2018exponential, kolodziejczyk2015technology}. 
For example, one bulk tissue RNAseq data \cite{li2021bulk} can only quantify the average gene expressions of a group of cells ignoring the cellular heterogeneity and hence can serve as one sample for downstream analyses. 
In contrast, single-cell sequencing technologies generate tens of thousands to millions of samples/cells in a given experiment. 
Deep learning methods, which have consistently shown cutting-edge performance in various big data applications \cite{pouyanfar2018survey, dong2021survey}, have fertile new ground for research that pushes the frontiers of biological science. 
Studies in single-cell data continue to expand exponentially and obtain new insights into immunology, oncology, developmental biology, pharmacology, 
and many other disciplines, 
just to name a few areas of applications \cite{chen2019revolutionizing, giladi2018single, valdes2018single}. 

Despite the success of single-cell data in numerous applications, difficulties arise due to the complexity of the data which requires advanced analysis pipelines with a number of steps. 
Single-cell data preprocessing includes many stages of data pruning, normalization, and often challenging machine learning tasks like batch effect correction, data imputation, or dimensionality reduction. 
Moreover, specialized types of single-cell data require further processing such as multimodal data integration and cell-type deconvolution for spatial transcriptomics. 
These steps are crucial to facilitate downstream tasks ranging from clustering and cell annotation, disease prediction, identifying gene coexpression networks, to the identification of developmental trajectories of cells transitioning between states  \cite{lahnemann2020}. 
For tasks with clear evaluation metrics, deep learning often achieves top performance against other classical machine learning techniques \cite{muzio2021}.
Deep learning can uniquely leverage its diverse architectures to capture networks of interdependencies between genes that alter other genes' expression levels \cite{bansal2007infer}, and cells that communicate with other cells through mechanisms like ligand-receptor pairs \cite{li2022decoding}.   
Due to the richness of deep learning architectures and the customization of hyper-parameters and loss functions, deep learning models can be more readily tailored to particular tasks in single-cell analysis compared to other machine learning methods.
Deep learning has already rapidly proliferated throughout the field, but due to the multidisciplinary nature of the work, many remain unaware of this burgeoning area of research. 
We write this survey as a bridge between two large research communities in single-cell biology and computer science. 
We provide background on deep learning to those in biology and less familiar with machine learning modeling, and also provide some history and summary of single-cell data to computer scientists who are looking for novel applications for their methods.  

In this survey, we review methods in the emerging use of deep learning for single-cell biology applications.
In Section (\ref{sc-tech}), we discuss the history and major technologies for single-cell sequencing. 
In Section (\ref{con_DL}), we give an overview of deep learning concepts and popular deep architectures.
Due to the categorization of the tasks involved in single-cell analysis, we group our review of deep learning methods by tasks.
We first give an overview of the pipeline in Section (\ref{pipeline}). 
Then, we describe the individual task objectives and highlight the alternative machine learning methods used for the task before detailing the deep learning methods.

\section{Single-Cell Technologies}\label{sc-tech}

\begin{figure*}[th]
    \begin{center}
         \centerline{
            \includegraphics[width=0.8\linewidth]{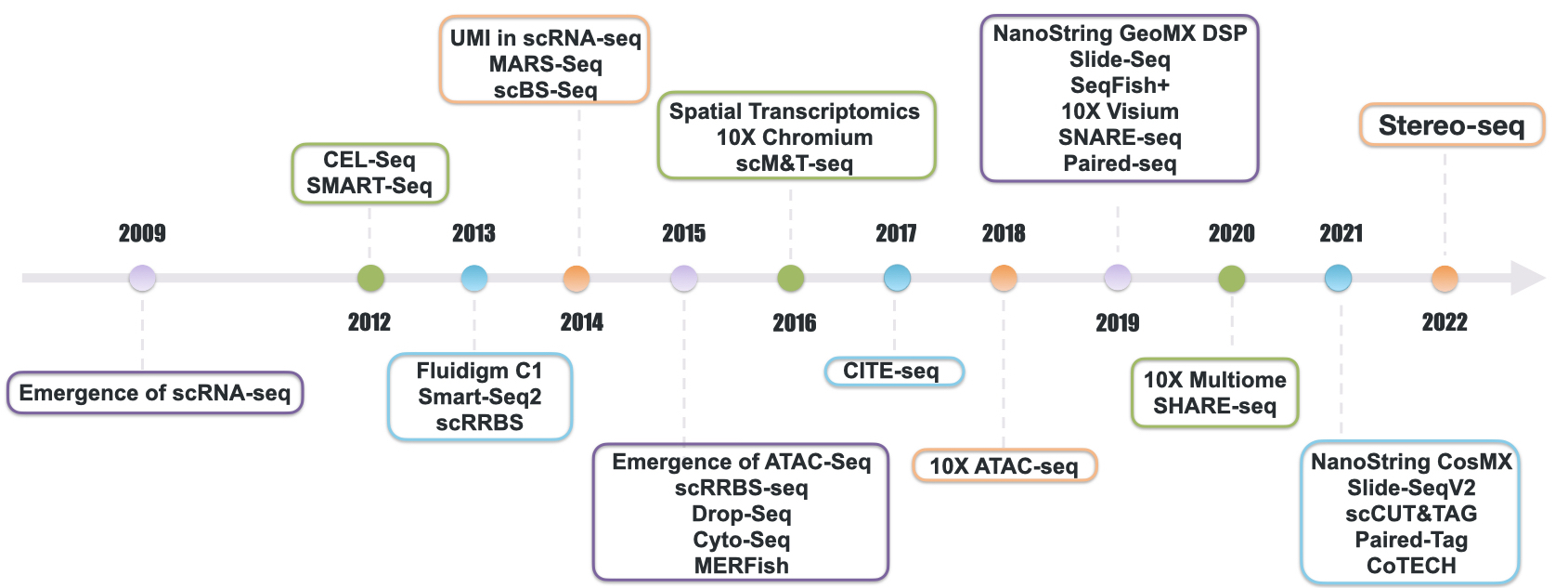} 
            }
        \vspace{-0.15in}
        \caption{Timeline of major developments in single-cell technologies}
        \label{fig:scTimeline}
    \end{center}
\end{figure*}

\begin{figure*}[th]
    \begin{center}
         \centerline{
            \includegraphics[width=0.8\linewidth]{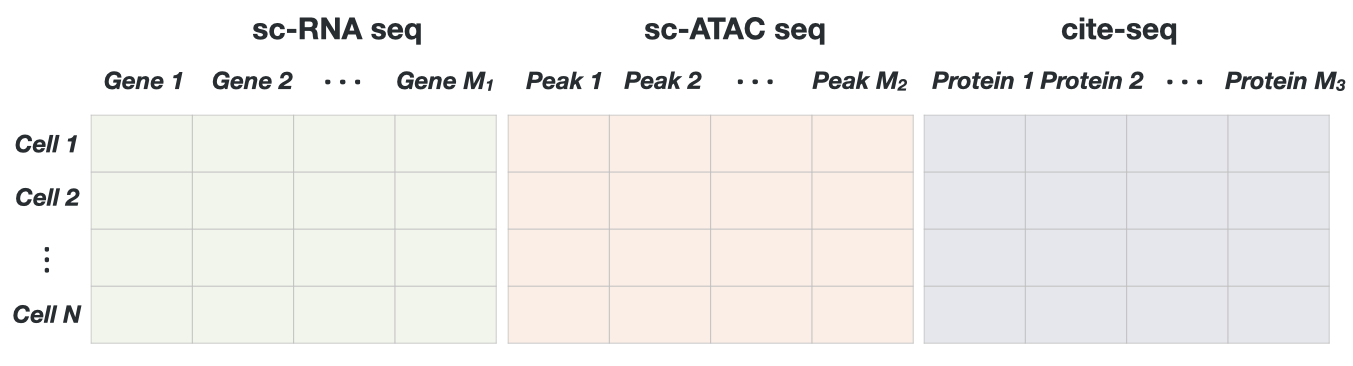} 
            }
        \vspace{-0.15in}
        \caption{An illustration of data matrices produced by single-cell technologies}
        \label{fig:scData}
    \end{center}
\end{figure*}

The goal of mapping genotypes to phenotypes presents a multitude of challenges to biologists performing transcriptome analysis \cite{Houle2010}.
The cells of an organism have nearly the same genotype, but the transcriptome is the result of gene regulatory networks in cells expressing only a subset of the total genes at any given time. 
With the advent of single-cell technologies, researchers have access to not just transcriptomic data at the cellular level, but also genomics and epigenomics data as shown in Figure \ref{fig:scTimeline}. 
Compared to bulk sampling technologies which measure the average transcriptome profiles of a group of cells, single-cell technologies provide a higher resolution of cell differences and can attribute biological behaviors to individual cells \cite{Goldman2019-cx,Kulkarni2019-qh,Stegle2015-hu,nguyenExperimentalConsiderationsSingleCell2018}.
We briefly discuss the history of single-cell technologies and the main technologies that are used in the applications we review. We summarize a timeline of their development in Figure~\ref{fig:scTimeline}.


\subsection{Single Modality Profiling}
Sequencing technology was first developed by James Eberwine et al. \cite{Eberwine1992} and Iscove et al.\cite{Brady1990}, by expanding the complementary DNAs (cDNAs). 
However, it wasn't until the creation of single-cell RNA sequencing (scRNA-seq) in 2009 \cite{Tang2009} that single-cell methods truly gained major traction. 
Since then, a few major branches in single-cell technologies have emerged, targeting different aspects of cells, such as RNA in 2009 \cite{Tang2009}), DNA methylation in 2013~\cite{Guo2013}, protein in 2015, DNA accessibility in 2015, and histone modifications in 2021 \cite{bartosovic2021single}. Single-cell data is often given in the form of a matrix, with features (e.g. genes, proteins, or DNA interval) corresponding to the columns and each cell as a row, as shown in Figure~\ref{fig:scData}. 

 Since its creation, scRNA-seq has had remarkable success in a number of different applications, such as cell developmental studies, classifying cell types, and gene regulation. 
For scRNA-seq, isolation of the cell is the first step for obtaining transcriptome information.  
Many technologies are differentiated according to their means of cell isolation before sequencing occurs. Earlier methods using serial dilution or robotic micromanipulation have low efficiency and throughput \cite{Brehm2004} when compared to more recent methods using microfluidic technologies \cite{Whitesides2006}. 
One promising microfluidic technique for single-cell isolation is using microdroplets \cite{Thorsen2001},  which creates the uniform dispersion of water droplets in a medium of oil, allowing the separation of cells into individual droplets. While commercial microfluidic platforms like Fluidigm C1, ICELL8, and Chromium can benefit from high throughput, they face the challenge of high cost and often the requirement of uniform cell size in the sample. 
Once a cell is separated and lysed, 
messenger RNAs in this cell are reverse transcripted into more stable cDNAs with a unique cell 'barcode'.  
The cDNAs are then amplified via Polymerase Chain Reaction (PCR) for better data capture before sequencing, which tends to introduce bias due to the uneven amplification efficiency. Therefore, besides the unique barcodes, the cDNA molecules in a cell are also given a Unique Molecular Identifier (UMI) to correct the amplification bias by collapsing the reads with the same UMI into one read. 
After debiasing, sequence reads are mapped to the genome and are grouped into genes for the creation of a count matrix \cite{wang2015advances}. 

Beyond recording RNA expression levels in a cell, technology may also capture information about the chromatin accessibility of a cell's chromosome.  
Eukaryotic genomes are hierarchically packaged into chromatin \cite{Kornberg1974}, and this packaging plays a central role in gene regulation \cite{Kornberg1992}. 
Buenrostro et al. created a means for sampling the epigenome at the single-cell level through the Assay for Transposase Accessible Chromatin using sequencing (ATAC-seq) \cite{buenrostro2013transposition} in 2013 . 
ATAC-seq allows the identification of accessible DNA, i.e. the nucleosome-free regions of the genome \cite{Hendrickson2018}.
DNA accessibility within the genome can be used to identify regulatory elements in different cell types which cause the  activation or repression of gene expression \cite{Thurman2012}. 
scATAC-seq produces a count matrix with a number of reads per open chromatin regions, which lead to very large matrices with hundreds of thousands of regions. 
Furthermore the data is known to be very sparse, where it is common to have the non-zero entries make up less than 3\% of the data \cite{li2021chromatin}. 

Gene expression can also be affected by a number of additional factors that are investigated under the umbrella of epigenetics, which studies mechanisms like DNA methylation and histone modification which do not change the DNA sequence, but can change gene activity and expression \cite{bird2007perceptions}. 
DNA methylation occurs when methyl groups are bonded to the DNA molecule, which can repress gene transcription, and is associated with a number of key biological processes \cite{moore2013dna}. 
In mammals, DNA methylation occurs most often in particular portions of the base pair sequence, namely CG (denoted CpG) portions where a cytosine is followed by a guanine \cite{singal1999dna}.
New technologies developed in the past decade for the profiling of DNA methylation use bisulfite sequencing (scBS-seq) \cite{farlik2016dna, smallwood2014single} or reduced representation busilfite sequencing (scRRBS-seq) \cite{farlik2015single, hou2016single} at a single-cell resolution. 
The output data for these are binary, indicating regions that are methylated by a 1, while 0 indicates no methylation.

\subsection{Multi-Modality Profiling}

In addition to the cell transcriptome and epigenome, cell proteome is another focus of single-cell technologies, which consist of the proteins that are encoded by the mRNA of the cell.
Comprehensive measurements of a cell’s proteome are integral to understanding how the genes respond to environmental changes, as well as for predicting cellular behavior since proteins are the functional units responsible for most of the cellular processes. 
While single-cell sequencing techniques for transcriptome measurements have widely proliferated, single-cell proteomics methods have made slower progress. 
Unlike most of the sequencing technologies which have a standard process, proteomic measurements are often bespoke and designed for specific applications \cite{vistain2021single}. However, some technologies developed have made significant strides in not only capturing protein information of cells but combining this with mRNA measurements. Specifically, Cellular Indexing of Transcriptomes and Epitopes by Sequencing(CITE-seq), a new technology introduced in 2017,  simultaneously sequences mRNA and measures the surface proteins on a cell \cite{stoeckius2017large}.
The method can sample over 1,000 genes and 80 proteins per cell, but like many other sequencing techniques, suffers from high noise. In addition, CITE-seq is incapable  to detect intracellular proteins \cite{baron2017new}.\\

The repertoire of multi-modal single-cell technologies bridges RNA expression not only to protein but also to DNA methylation, chromatin accessibility, and histone modifications. One of the first methods to simultaneously sequence RNA and chromatin accessibility is a droplet-based method named SNARE-seq. 
Published in 2019, it uses Tn5 transposase to capture accessible chromatin and creates shared barcodes between RNA and accessible regions~\cite{Chen2019}. 
The same year, paired-seq raised the throughput by two orders by combining a ligation-based combinatorial indexing strategy and an amplify-and-split library dedicating method \cite{Zhu2019}. 
SHARE-seq~\cite{Ma2020} further increased the throughput and resolution by adapting Paired-seq and SPLiT-seq~\cite{Rosenberg2018}, a scRNA-seq technology. In 2020, 10X also released 10X Multiome, a commercialized product for the joint profiling of RNA and chromatin accessibility. 
Beyond the co-profiling of RNA and chromatin accessibility, scM\&T-seq~\cite{Angermueller2016}, based on G\&T-seq~\cite{Macaulay2015}, allows for parallel analysis of single-cell RNA and DNA methylation. Another emerging field is the joint profiling of RNA and histone modifications. 
Example technologies in this category include Paired-Tag~\cite{Zhu2021} and CoTECH~\cite{Xiong2021}, both became available in 2021.

\subsection{Single-Cell Spatial Transcriptomics}
Single-cell technologies that capture transcriptomic, proteomic, or epigenetic information do so with great precision but with the loss of spatial information of the cells within the tissues. 
However, the cells' relative locations within tissue is critical to understanding normal development and disease pathology. 
With spatial transcriptomic technologies, researchers are able to measure transcriptomics and leverage the spatial information or relative locations of cells in a tissue for better performing downstream tasks \cite{crosetto2015spatially, moor2017spatial, wang2018multiplexed, marx2021method, asp2020spatially, waylen2020whole, teves2020mapping}.
For example, motivated by the fact that a pair of ligand and receptor with closer distance are easier to bind, HoloNet~\cite{li2022decoding} builds up a directed graph based on the expression of ligand–receptor gene lists and the physical distance between the sender cell and receiver cell to represent cell–cell communication events. 
However, the early generations of spatially resolved profiling technologies are not at the single-cell resolution but instead sampled in groups called `spots', which capture several cells.
It requires additional work to determine the cell type proportion in spots, a process called cell type deconvolution. 
Alternatively, many cell imaging platforms provide RNA spatial information at the cellular and subcellular level, but the individual cells must be identified through cell segmentation methods.   

Major technologies or platforms for spatial transcriptomics include multiplexed error-robust fluorescence in-situ hybridization (MERFISH), sequential fluorescence ISH (seqFISH+), Slide-Seq, Visium by 10x \cite{SpatialTrans}, GeoMx Digital Spatial Profiler (DSP) \cite{Merritt2020} by NanoString,and  CosMx Spatial Molecular Imager (SMI) by NanoString. 
MERFISH \cite{MOFFITT20161}, first introduced in 2015, is a  single-molecule-fluorescence-in-situ-hybridization (smFISH)-based technology that can be applied to fresh-frozen samples to provide subcellular resolution.
While traditionally the procedure of these smFISH-based technologies is complex, a number of commercialized platforms have emerged recently, such as Vizgen, Rebus Esper, Molecular Cartography, and Resolve Biosciences \cite{mosesMuseumSpatialTranscriptomics2022}, which allow more convenient sequencing of spatial transcriptomic at a lower cost.
As an alternative to MERFISH, seqFish+ \cite{Lubeck2014,Shah2016,eng2019transcriptome} employs 'pseudocolor' as a combination of colors to increase the amount of detectable transcripts \cite{Rao2021}. 


Beyond early in-situ hybridization methods, a number of sequence-based technologies have emerged. Closely related to scRNA-seq technologies, these sequencing-based methods barcode RNAs such that each read can be mapped to its corresponding spatial location through the associated barcodes. The rest of the sequencing read is mapped to the genome to identify the transcript of origin, collectively generating a gene-expression matrix. Stahl et al.~\cite{Sthl2016} first proposed this method, which has been adapted by commercial platforms such as 10X Visium. 
10x Visium fixes spatially barcoded oligos to each spot in a capture slide (area $6.5mm^2$), with the barcoding done through DNA extension and reverse transcription for formalin fixed paraffin embedded tissues (FFPE) and fresh frozen tissues respectively. 
In particular, the 10x Visium expression slide contains 4 capture slides, each with area  6.5 mm$^2$ where fresh frozen or FFPE tissues are placed. 
Each of the capture slides contain a grid of approximately 5000 barcoded spots that are $55\mu m$ in diameter with a center-to-center distance of $100 \mu m$ between any two adjacent spots. 
On average, there are $1-10$ cells in each of these spots, and $\sim 18,000$ unique genes in human ($\sim 20,000$ in mouse) can be quantified. 
Another major sequencing-based technology is Slide-Seq, which captures mRNA by placing barcoded beads on slides, which achieves a high resolution of 10 micron. Technological innovations further improved sequencing resolution in recent years. For instance, high-definition spatial transcriptomics (HDST) \cite{Rodriques2019} uses wells rather than slides, whereas built upon Slide-Seq, Slide-seqV2~\cite{Stickels2020} raised the resolution to near-cellular level while reaching RNA capture efficiency of roughly 50\% of scRNA-sequencing. Finally, spatio-temporal enhanced resolution omics sequencing (Stereo-seq)\cite{Chen2022} deposits barcoded DNA nanoballs in patterned arrays to achieve single-cell resolution while maintaining high sensitivity.


While 10x Visium and Slide-Seq do not profile at cellular resolution,  Nanostring's GeoMx DSP is capable of cellular resolution through user-drawn profiling regions. 
 Geomx DSP uses PC-linker to link barcodes via antibodies to proteins and RNA for identification.
The spatial regions of interest (ROI) on the tissue are flexible and can be user-defined, or with pre-defined layouts (such as a square grid). 
During imaging, the DSP barcodes from each ROI are UV-cleaved and collected for sequencing, and the spatial information is recorded. Due to the flexibility of the ROI definitions, the ROIs can be a range of sizes, from a single-cell or hundreds of cells. 
The RNA assay can quantify $>18,000$ target genes, and the protein assay can quantify $>96$ proteins.  

Though GeoMX can produce cellular-resolution sequencing, its scalability is limited. The most recent platform, CosMx Spatial Molecular Imager (SMI) \cite{lewis2022subcellular}, is able to profile consistently at single cell, and even subcellular resolution. 
CosMx SMI  follows much of the initial protocol as GeoMx DSP, with barcoding and ISH hybridization. However, the SMI instrument performs 16 cycles of automated cyclic readout, and in each cycle the set of barcodes (readouts) are UV-cleaved and removed. These cycles of hybridization and imaging yield spatial resolved profiling of RNA ($>980$ target genes) and protein ($>80$ validated proteins) at single-cell ($\sim 10 \mu m$) and subcellular ($\sim 1 \mu m$) resolution.


Multiplex imaging technologies have significantly advanced higher spatial resolution for single-cell profiling. Spatially resolved transcriptomic data, along with corresponding imaging data, enables single-cell or even subcellular analysis on both spatially morphological and pixel resolution information. 
Recently, antibody-based multiplexed imaging methods have dominated the multiplexing approaches, as they can capture cellular organization and tissue phenotypic heterogeneity at the protein level. They utilize various protein markers for cellular identification. 
Immunohistochemistry (IHC)\cite{coons1942demonstration}, first reported in 1942, is one of the most commonly used multiplexed imaging methods. It uses appropriately-labeled antibodies to bind specifically to their target antigens in situ (in the original site), which can be better captured by current light or fluorescence microscopy.
Due to the limited protein readouts, methods including multiplexed immunofluorescence (MxIF) \cite{Michael2013} and cyclic immunofluorescence (CyCIF) \cite{lin2015highly,lin2018highly} were proposed to add more new antibodies in multiple rounds of staining.
Another imaging platform, Co-Detection by IndeXing (CODEX)\cite{goltsev2018deep}, is designed for up to 40 proteins using cyclic detection of DNA-indexed antibody panels.
Imaging mass cytometry (IMC)\cite{giesen2014highly} is an evolutionary technology that leverages mass spectroscopy to obtain images from tissues with 40+ labels simultaneously. This vastly reduces data noise and enhances the multiplex capability. Multiplexed ion beam imaging (MIBI)\cite{keren2019mibi} is also performed by imaging tissues with secondary ion mass spectrometry based on metal-labeled antibodies.
These multiplexed imaging tools provide high-dimensionality imaging assays at the single-cell level and enable analyzing and understanding of the single-cell function and tissue structure.




\section{Concepts in Deep Learning}\label{con_DL}



Simultaneously with single-cell technologies, deep learning (DL)\cite{lecun2015deep} has redefined our capacity to analyze large-scale data by employing complex topologies for artificial neural networks. 
In this section, we present an overview of the fundamental concepts underlying a number of deep learning models that have been widely used for single-cell research.

\subsection{Feedforward Neural Networks}
\begin{figure*}[th]%
     \centering 
     \subfloat[Neuron]{\label{fig:neuron}{\includegraphics[width=0.25\linewidth]{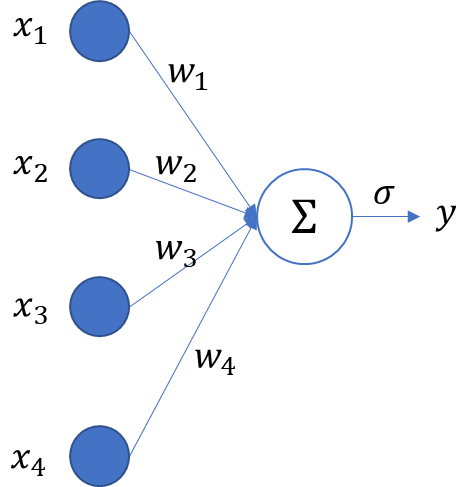} }}%
      \subfloat[MLP]{\label{fig:mlp}{\includegraphics[width=0.23\linewidth]{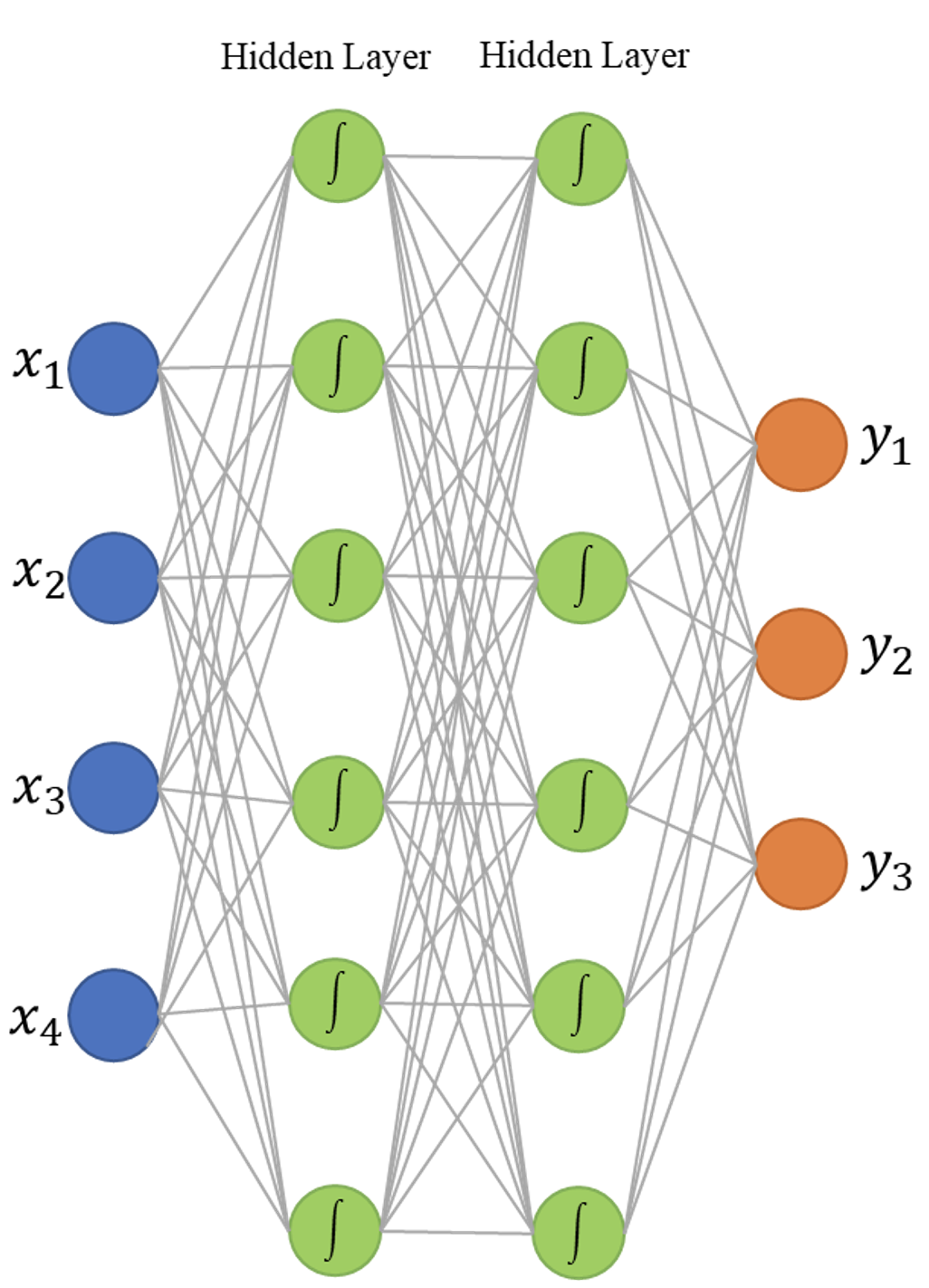} }}%
    \qquad
    \vskip -0.5em
\caption{An illustration of neuron and 
feedforward neural networks (also known as multi-layer perceptrons).}
\label{fig:neuron_mlp}
\vskip -1em
\end{figure*}

The Feedforward Neural Network is the simplest case of an artificial neural network (ANN). 
The term `feedforward networks' is used to distinguish the architecture from recurrent neural networks where the former has connections between the nodes that do not form a cycle, while the latter does. 
A feedforward neural network consists of multiple layers of neurons. Each neuron, as shown in Figure~\ref{fig:neuron}, is similar to
a perceptron~\cite{mcculloch43a}, which performs linear transformation over an input vector $\mathbf{x}$, and adds a non-linear activation function before output.  This is why it is often recognized as a `multi-layer perceptron (MLP)' . 

Formally, we denote the input of a neuron layer as a vector $\mathbf{x}$. Each dimension of $\mathbf{x}$ is fed to a neuron. A layer of neurons can thus be formulated as:
\begin{align}
 \mathbf{x}^{(i+1)} = \operatorname{\sigma}(\mathbf{W}^{(i)} \mathbf{x}^{(i)} + \mathbf{b}^{(i)} )  
\end{align}
\noindent where $\mathbf{x}^{(i)} \in \mathbb{R}^{d}$ is the input vector that feeds to the layer~$i$, $\mathbf{W}_{i} \in \mathbb{R}^{h \times d}$ and $\mathbf{b}_{i} \in \mathbb{R}^{h}$ are a weight matrix and a bias vector to be learned in the layer~$i$, and $\sigma$ is a non-linear activation function. $\mathbf{x}^{(i+1)} \in \mathbb{R}^{h}$ is the output vector of layer~$i$, which is also used as the input of the next layer~$i+1$.

Typically we stack multiple layers to form a feedforward neural network, as shown in Figure~\ref{fig:mlp}. 
The first layer is referred to as the input layer, the last as the output layer, and the layers in between as the hidden layers.
The output of each layer is fed to the next layer, thus the connections between neurons of adjacent layers are dense, which is also known as `fully-connected layers' or `dense layers'. Generally, a feedforward neural network as a whole is a non-linear transformation from the input space to the output space. For example, in Figure~\ref{fig:mlp}, the input of the whole network is vector $\mathbf{x}$ at the left side, and the output of the network is from the output layer at the right side, which is vector $\mathbf{y}$.


\subsection{Autoencoders}
\begin{figure*}[th]
    \begin{center}
         \centerline{
            \includegraphics[width=0.5\linewidth]{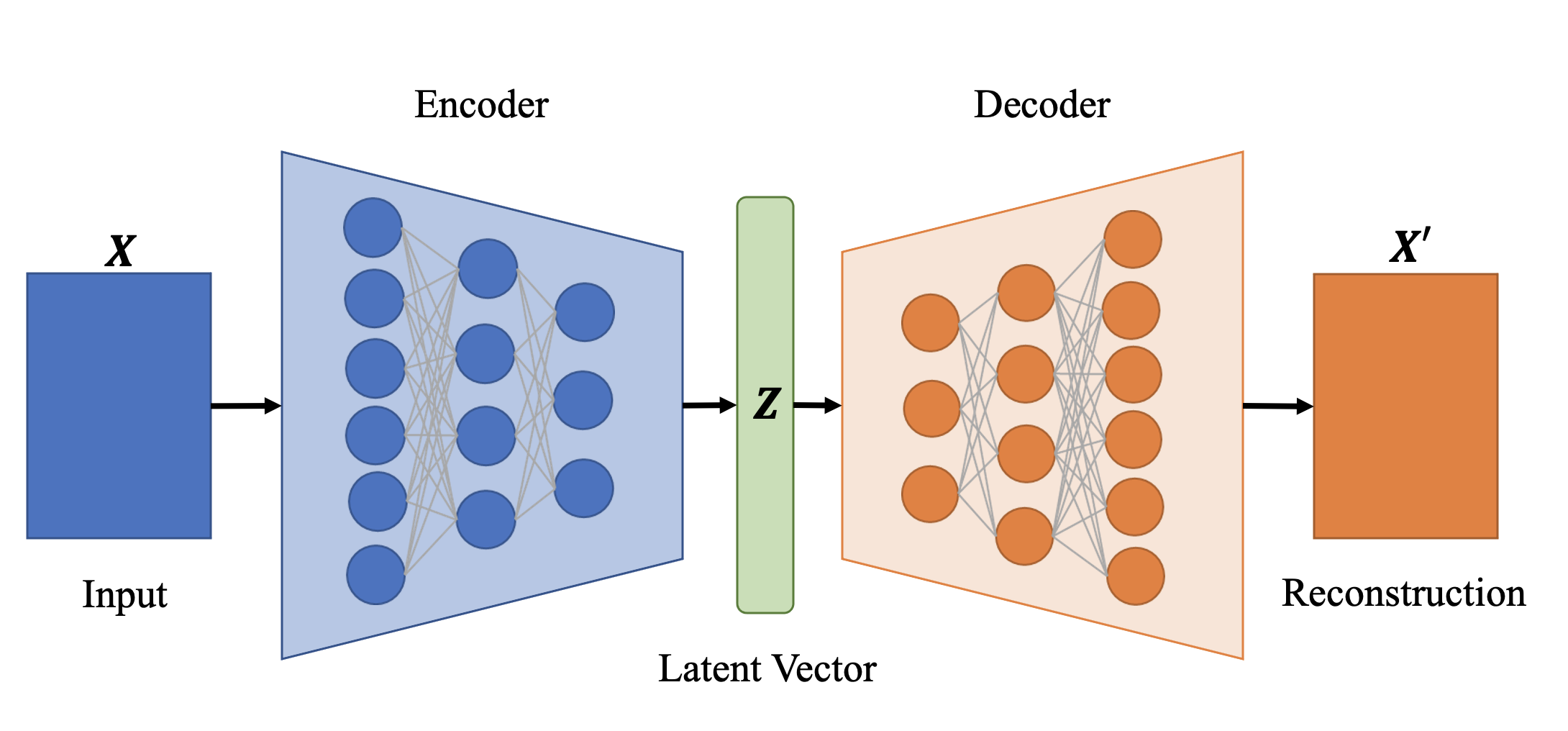} 
            }
        \vspace{-0.15in}
        \caption{An illustration of Autoencoders.}
        \label{fig:ae}
    \end{center}
\end{figure*}

Autoencoders (AEs) \cite{rumelhart1985learning} are neural networks to reduce dimensionality or reduce noise from different types of data via reconstructing the original input.
Conventional AEs consist of two parts: an encoder and a decoder as shown in Figure~\ref{fig:ae}. The encoder is to map the input data $\mathbf{X}$  to a latent space $\mathbf{Z}$, which is frequently significantly smaller than the original input space so that only the significant features of data variation are captured in the latent space.
The decoder will map the latent vector $\mathbf{Z}$ back to the original input dimension space and recover the original data as $\mathbf{X}^{'}$. We take an AE with only two layers as the simplest form as an example. The encoder with a single hidden layer maps the input data $\mathbf{X} \in \mathbb{R}^{N \times K}$ into the latent space representation $\mathbf{Z} \in \mathbb{R}^{N \times P}$ as defined in Eq.~\ref{encoder_eq}, where $N$ indicates the number of input samples, $K$ is the original feature dimension and $P$ is the compressed feature dimension.
\begin{align}
 \mathbf{Z} = \operatorname{\sigma}(\mathbf{X} \mathbf{W} + \mathbf{b} )  
\label{encoder_eq}
\end{align}
\noindent $\mathbf{W}$ is the weight matrix, $\mathbf{b}$ is the bias vector, and $\sigma$ is the activation function such as ReLU.
The decoder is responsible for mapping the latent representation $\mathbf{Z}$ back to the original input space defined as follows:
\begin{equation}
 \mathbf{X}^{'} = \operatorname{\sigma}(\mathbf{Z}\mathbf{W}^{'}  + \mathbf{b}^{'} ) 
\label{decoder_eq}
\end{equation}
where $\mathbf{X}^{'} \in \mathbb{R}^{N \times K}$ is the reconstructed data points.
The objective of typical AEs is to minimize the distance between the original input $\mathbf{X}$ and the recovered data points $\mathbf{X}^{'}$ defined as:

\begin{equation}
    \mathcal{L} =  \| \mathbf{X} - \mathbf{X}^{'} \|^{2} 
\label{AE_loss}
\end{equation}
In this simplest form, the parameters $\theta = \{\mathbf{W}, \mathbf{b}, \mathbf{W}^{'}, \mathbf{b}^{'}\}$ are optimized during training \cite{rumelhart1986learning}. 

Note that the model selection of the encoder and decoder of the AEs is very flexible and does not have to be consistent. For example, the encoder can consist of Graph Neural Networks \cite{scarselli2008graph} while the decoder can be multilayer perceptron \cite{popescu2009multilayer} in the same AE.
Furthermore, there are numerous architectural variants of AEs \cite{weng2017representation, jaiswal2018large, tran2018learning, zeune2020deep}. One popular variant is the variational AE (VAE) \cite{kingma2013auto, kingma2019introduction}.
VAEs account for the irregularity of the latent space by returning a distribution over the latent space as opposed to a single point.
To achieve this goal, a regularization term for the returning distribution is incorporated into the loss function to assure a more organized latent space.

\subsection{Generative Adversarial Networks}

Generative Adversarial Networks~\cite{goodfellow2014generative} (GANs) are a useful tool to generate realistic synthetic data. 
The general idea of GANs is to train two neural networks that compete against one another in a two-player game. 
The two networks in a GAN are referred to as the generator ($G$) and the discriminator ($D$). 
The training objective of $D$ is to discriminate between pseudo data and real data. 
On the other hand, the goal of $G$ is to generate pseudo samples that have the same distribution as the real data, thus can fool the discriminator into not being able to differentiate pseudo data from real data. 
Formally, the objective function can be written as:
\begin{equation}
\label{eq:minimaxgame-definition}
\min_G \max_D V(D, G) = \mathbb{E}_{\bm{x} \sim p_{\text{data}}(\bm{x})}[\log D(\bm{x})] + \mathbb{E}_{\bm{z} \sim p_{\bm{z}}(\bm{z})}[\log (1 - D(G(\bm{z})))].
\end{equation}
where $p_{\bm{z}}(\bm{z})$ is defined as a prior on input noise variables, $G$ is a generator that generates pseudo data given random vector $\mathbf{z}$, $D$ is a discriminator which is optimized to output one when given real data $\mathbf{x}$ and output zero when given pseudo data $G(\mathbf{z})$. Overall, the minmax objective function is optimizing $D$ and $G$ simultaneously, and finally, an optimal generator is produced.

\begin{figure*}[th]
    \begin{center}
         \centerline{
            \includegraphics[width=0.5\linewidth]{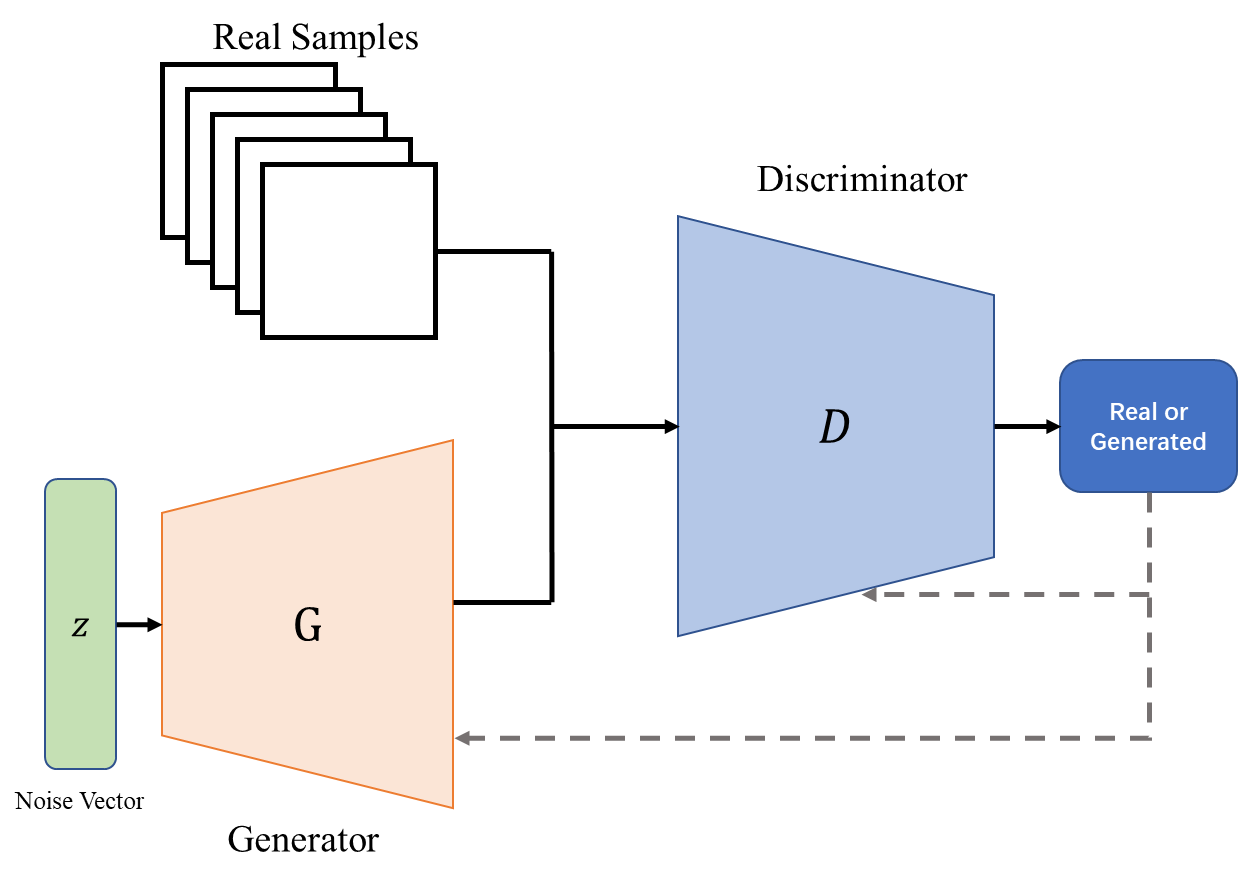} 
            }
        \vspace{-0.15in}
        \caption{An illustration of Generative Adversarial Networks.}
        \label{fig:gan}
    \end{center}
\end{figure*}

Typically, the parameters of $G$ and $D$ are iteratively updated during the training, where eventually both parameters would be optimized. During the process, the generator is gradually trained to generate realistic data which can cheat the discriminator to be classified as real samples, while the discriminator also progressively becomes stronger at distincting between real and pseudo data. Compared to other generative models, the adversarial training between $G$ and $D$ networks can get rid of the assumptions on the prior distribution, and thus can learn any distribution.

\subsection{Convolutional Neural Network}
A Convolutional Neural Network (CNN) \cite{lecun1995convolutional, o2015introduction} is a 
type of artificial neural network (ANN) typically used for analyzing image data. 
It aims to discover the ideal set of filters (weights) that can detect the 
required features for a given task (e.g. image classification).
A typical CNN architectural block consists of three distinct types of layers. 
These types consist of convolutional layers, pooling layers, and fully connected layers. 
By stacking these layers, a CNN architecture is produced. Next, we'll introduce them individually.

\begin{figure*}[t]
    \begin{center}
         \centerline{
            \includegraphics[width=0.5\linewidth]{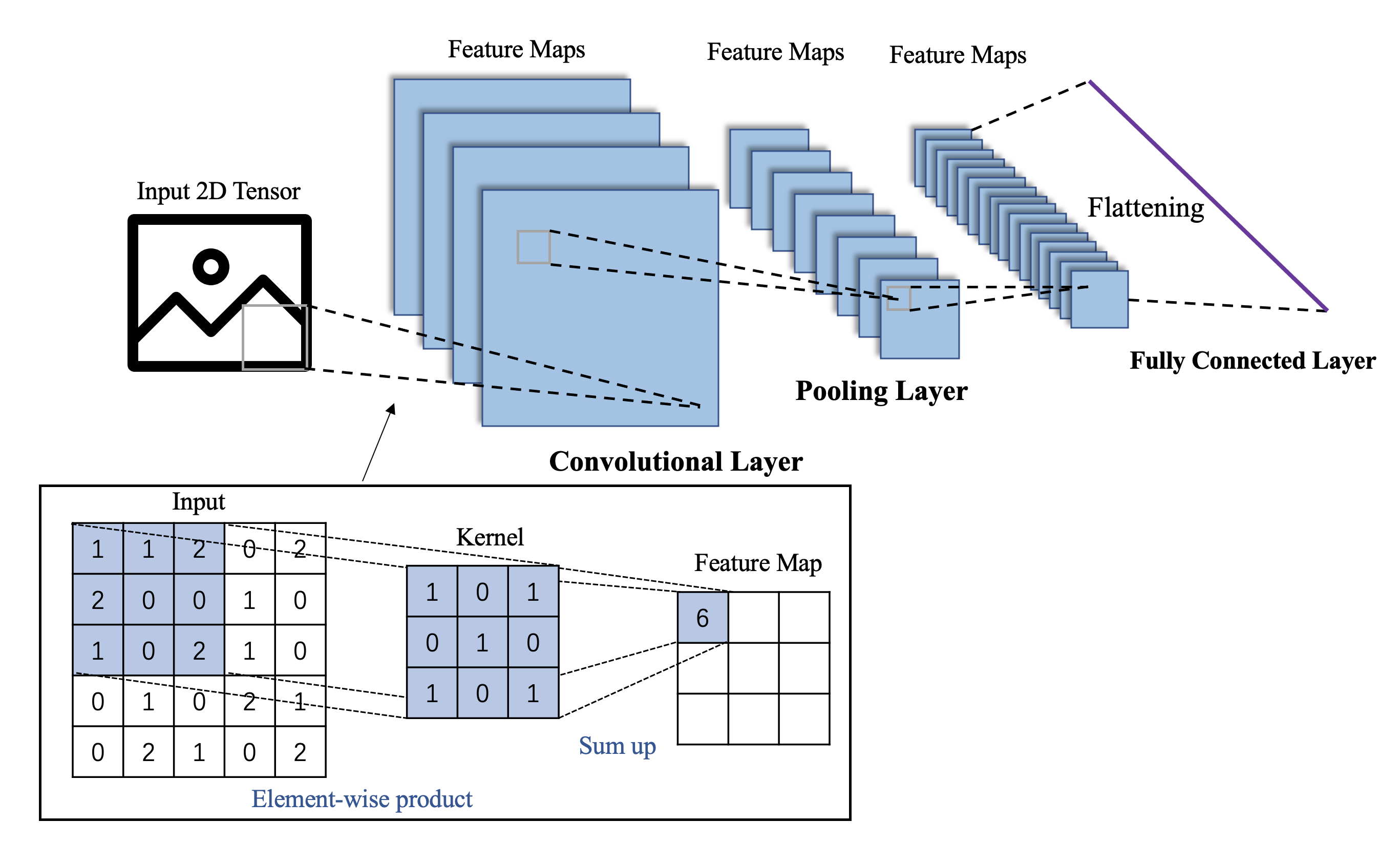} 
            }
        \vspace{0.07in}
        \caption{An illustration of Convolutional Neural Networks.}
        \label{fig:cnn}
    \end{center}
\end{figure*}

\begin{itemize}
    \item \textbf{Convolutional Layer:} The convolutional layer is the fundamental component of a CNN. 
    It requires three components: input data, a filter kernel, and a feature map. 
    If the input is an image, the feature filter kernel can be described as a two-dimensional array of weights,  which is applied to the image and shifted by a stride until the filter has swept across the entire image. 
    For each filter operation, a dot product is calculated between the input data and the filter and the result of the dot product is cached into an output. 
    The mapping from the original input or feature filter to the final output is called a feature map. As shown in Figure~\ref{fig:cnn}, a kernel size of 3*3 is defined to extract features from an image channel with a size of 5*5, and a feature map with a size of 3*3 is finally formed from this convolutional operation.
    
    Note that the first convolutional layer is typically responsible for capturing low-Level 
    characteristics such as edges, color, gradient direction, etc. 
    With additional layers, the architecture also responds to high-Level characteristics.
    
    \item \textbf{Pooling Layer:} 
    The pooling layer reduces the number of input parameters to achieve dimension reduction.
    In a manner comparable to the convolutional layer, the pooling procedure is performed by sweeping a weightless filter across the whole input.
    The pooling kernel aggregates the data in the receptive field to populate the output array.
    Max pooling and average pooling are the two main aggregation functions. 
    Max pooling chooses the highest value for transmission to the output, and average pooling computes the value that represents the receptive field's average intensity and then transmits that value.
    
    \item \textbf{Fully Connected Layer:} Each node in the output layer is directly connected to a node in the preceding layer in the fully-connected layer. When this layer occurs as the final 
    layer of the network, it performs classification or regression based on the features extracted through convolutional and pooling layers.
    
\end{itemize}
CNNs typically consist of a collection of stacked convolution layers, pooling layers and fully connected layers that generate the network's output. 

\subsection{Recurrent Neural Network}
\begin{figure*}[t]
    \begin{center}
         \centerline{
            \includegraphics[width=0.75\linewidth]{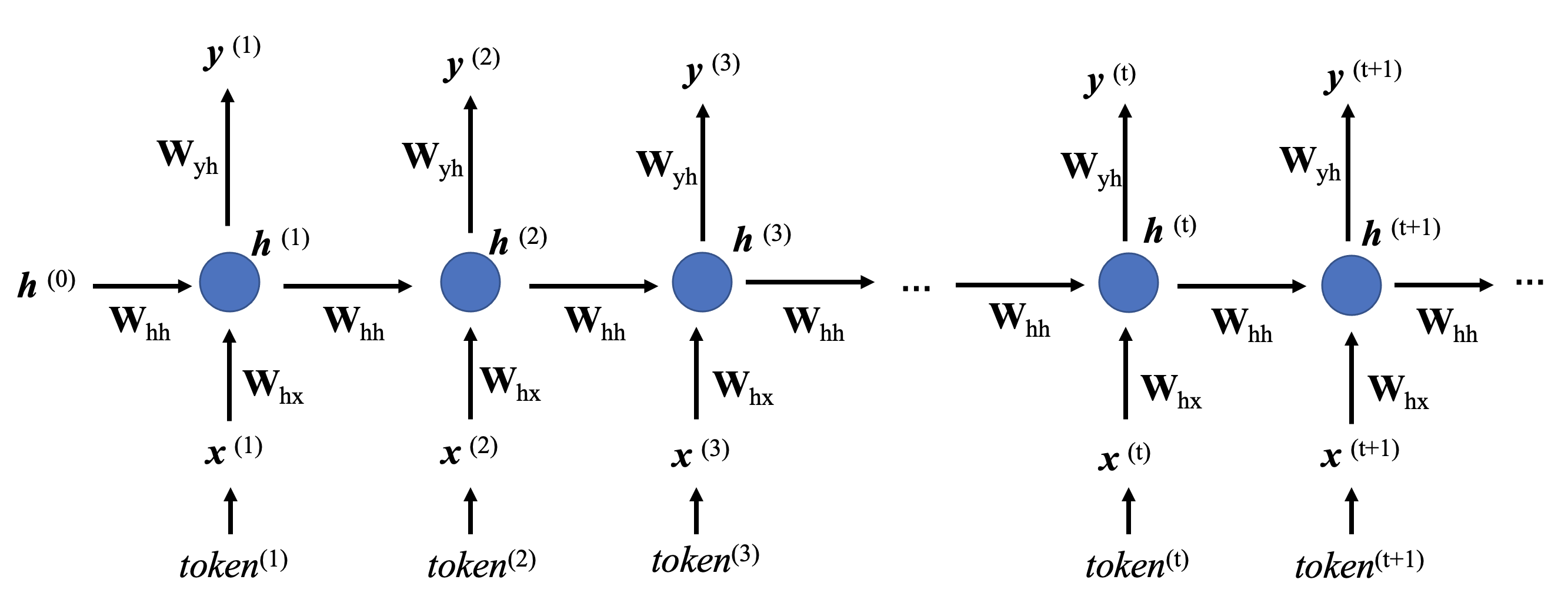} 
            }
        \vspace{0.07in}
        \caption{An illustration of Recurrent Neural Networks.}
        \label{fig:rnn}
    \end{center}
\end{figure*}

Recurrent Neural Network (RNN) \cite{medsker2001recurrent} is a type of artificial neural network which deals with sequential data or time series data. 
In traditional neural networks, the input and output are independent of each other, whereas in RNN, the output is dependent on previous elements in the sequence.
As shown in Figure~\ref{fig:rnn}, a sequence of tokens with length $n$ is an input of the model. Each token is first transformed into vector form, and then the input sequence can be denoted as $\left(\mathbf{x}^{(\mathbf{1})}, \mathbf{x}^{(\mathbf{2})}, \ldots, \mathbf{x}^{(\mathbf{t})}, \ldots, \mathbf{x}^{(\mathbf{n})}\right)$. 
The RNN model takes in a sequence of data and applies a chunk of neural networks to it one data point at a time.
Here, the chunk of neural networks in RNN is a single neural network layer. 
The inputs of the chunk of neural networks are the output from the preceding position as $\mathbf{h}^{(i-1)}$ and the current element in the sequence as $\mathbf{x}^{(i)}$. The outputs are the output layer as $\mathbf{y}^{(i)}$ and $\mathbf{h}^{(i)}$ to be transmitted to the next position.
The formulation for dealing with the $i-$th element in the sequence can be written as:
\begin{equation}
\mathbf{h}^{(i)}=\sigma_h\left(\mathbf{W}_{h h} \cdot \mathbf{h}^{(i-1)}+\mathbf{W}_{h x} \mathbf{x}^{(i)}+\mathbf{b}_h\right)
\label{rnn_1}
\end{equation}

\begin{equation}
\mathbf{y}^{(i)}=\sigma_y\left(\mathbf{W}_{y h} \mathbf{h}^{(i)}+\mathbf{b}_y\right)
\label{rnn_2}
\end{equation}
where $\mathbf{W}_{h h}, \mathbf{W}_{h x}$, and $\mathbf{W}_{y h}$ are learned to perform linear transformations on the output of previous position, current element in the sequence and the output of current element processing respectively; $\mathbf{b}_h$ and $\mathbf{b}_y$ are the bias term; and $\sigma_h()$ and $\sigma_y()$ are the activation functions. $\mathbf{h}^{(0)}$ is usually initialized as vector $\mathbf{0}$. We can see that when dealing with the current element, it not only considers the current element but also receives information transmitted from the preceding position in Eq.~\ref{rnn_1}. The neural network layer in the chain structure is identical for each element processing in the sequence. 

When the RNN processes long sequential data, two problems may be encountered during the training phase: vanishing gradient \cite{hochreiter1998vanishing} and exploding gradient \cite{pascanu2013difficulty}. Vanishing gradient problem refers to the issue of disappearing gradients. Gradients contain information used by the RNN, and when they become insignificantly small, parameter updates are rendered insignificantly. This makes it harder to learn long data sequences. If the slope continues to rise exponentially rather than decrease, this phenomenon is commonly referred to as an exploding gradient. This issue emerges when significant error gradients build, resulting in very large modifications to the weights of the neural network model during training. 
To address these problems, gated RNN models like Long Short Term Memory (LSTM) networks \cite{hochreiter1997long} and Gated Recurrent Unit (GRU) \cite{chung2014empirical} have been developed.

\subsubsection{Long Short Term Memory}
\begin{figure*}[t]
    \begin{center}
         \centerline{
            \includegraphics[width=0.65\linewidth]{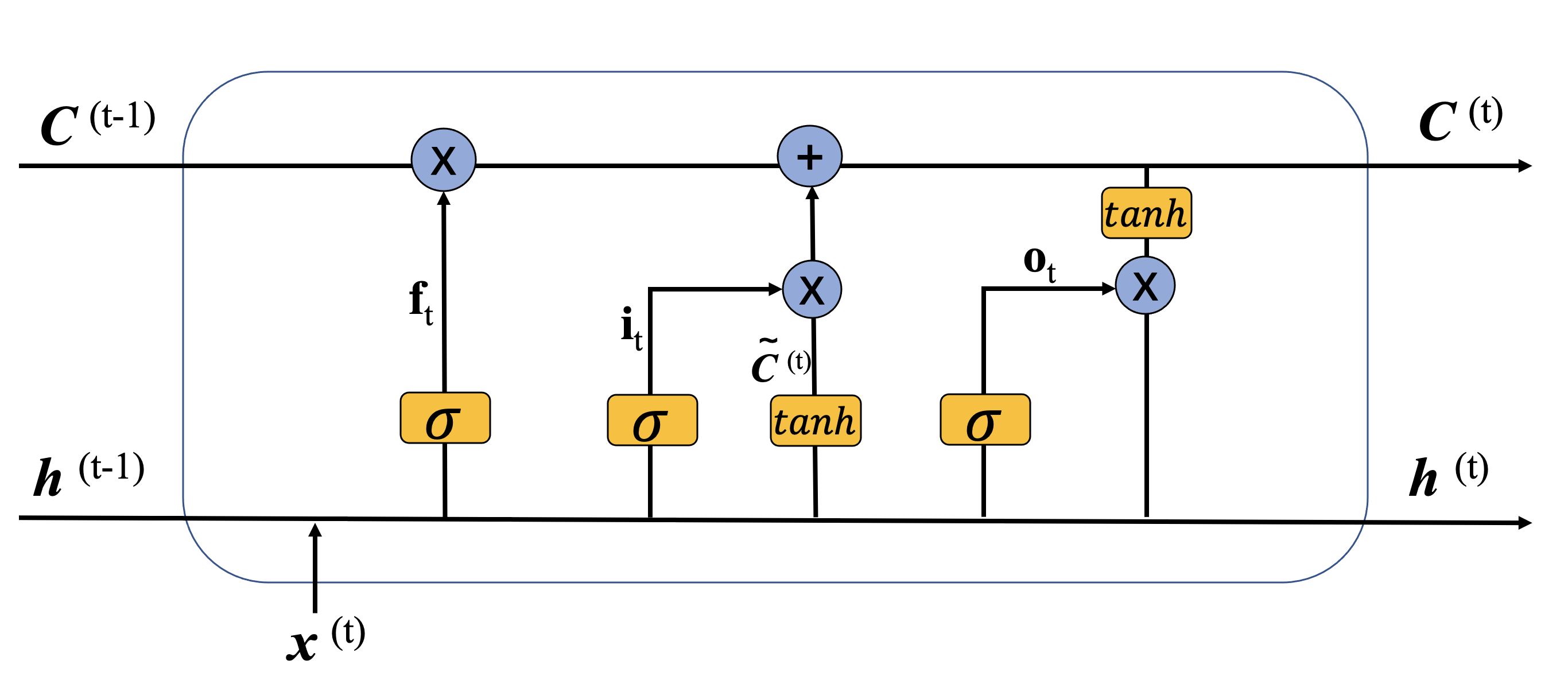} 
            }
        \vspace{0.07in}
        \caption{A block of LSTM.}
        \label{fig:lstm}
    \end{center}
\end{figure*}

Long Short Term Memory (LSTM) networks \cite{hochreiter1997long} is a special kind of RNN, capable of learning long sequence data. Similar to RNN, LSTM also uses a chain structure, with the same neural network blocks being applied to the sequence's individual elements. The block neural network in LSTM is much more complicated than that in RNN where the block neural network is just a simple single neural network layer. Instead, a collection of gating units are used to control the information flow in LSTM, which is the essential distinction. 
A bock of LSTM is shown in Figure~\ref{fig:lstm}. The information flowing across successive positions in a sequence consists of the cell state $\mathbf{C}^{(t-1)}$ and the hidden state $\mathbf{h}^{(t-1)}$ corresponding to the horizontal lines running through the top and bottom of the diagram respectively. 
The cell state functions as the information from prior states that is transmitted to the subsequent location,
and the hidden state contributes to the determination of how the information should be propagated. 
The LSTM is capable of removing or adding information to the cell state, under the careful control of structures known as gates. Gates are a way to allow information to pass through if desired. A sigmoid activation function and a pointwise multiplication procedure construct them. The sigmoid activation function outputs integers between 0 and 1 that describes the percentage of each component that should be allowed through. A value of zero indicates ``permit nothing to pass through," while a value of one indicates ``let everything pass through!". To protect and control the cell state, an LSTM has three of these gates, which are \textbf{forget gates}, \textbf{input gates} and \textbf{output gates} introduced below: 
\begin{itemize}
    \item \textbf{Forget Gate:} The initial phase of the LSTM is determining what information from the prior cell state will be discarded. The choice is determined by a forget gate. The forget gate generates a value between 0 and 1 for each element of the previous cell state $\mathbf{C}^{(t-1)}$ based on the prior hidden state $\mathbf{h}^{(t-1)}$ and the current input $\mathbf{X}^{(t)}$. The forget gate can be defined as:
    \begin{equation}
    \mathbf{f}_t=\sigma\left(\mathbf{W}_f \cdot \mathbf{x}^{(t)}+\mathbf{U}_f \cdot \mathbf{h}^{(t-1)}+\mathbf{b}_f\right)
    \end{equation}
    where $\mathbf{W}_f$ and $\mathbf{U}_f$ are the learned matrices, $\mathbf{b}_f$ is the bias term, and $\sigma()$ is the sigmoid activation function. 

    \item \textbf{Input Gate:} The following step is to determine what information will be stored in the cell state from new input $\mathbf{x}^{(t)}$. In a manner similar to the forget gate, the input gate is designed from the sigmoid activation function to make the decision. It is calculated as below:
    \begin{equation}
    \mathbf{i}_t=\sigma\left(\mathbf{W}_i \cdot \mathbf{x}^{(t)}+\mathbf{U}_i \cdot \mathbf{h}^{(t-1)}+\mathbf{b}_i\right)
    \end{equation}
    Next, a vector of potential new values as $\tilde{\mathbf{C}}^{(t)}$ is generated after a tanh layer, and it may be added to the cell state. The process of generating $\tilde{\mathbf{C}}^{(t)}$) is as:
    \begin{equation}
    \tilde{\mathbf{C}}^{(t)}=\tanh \left(\mathbf{W}_c \cdot \mathbf{x}^{(t)}+\mathbf{U}_c \cdot \mathbf{h}^{(t-1)}+\mathbf{b}_c\right)
    \end{equation}
    In the next step, we’ll combine these two to create an update $\mathbf{C}^{(t)}$ to the cell state as:
    \begin{equation}
    \mathbf{C}^{(t)}=\mathbf{f}_t \odot \mathbf{C}^{(t-1)}+\mathbf{i}_t \odot \tilde{\mathbf{C}}^{(t)}
    \end{equation}
    where $\odot$ denotes element-wise multiplication.
    
    \item \textbf{Output Gate:} 
    As the last step, a hidden state $\mathbf{h}^{(t)}$ is generated, and it can flow to the position that comes after it and serve as the position's output.
    The hidden state is derived from the updated cell state $\mathbf{C}^{(t)}$, and an output gate chooses which portions of the cell state are to be maintained. 
    The output gate is defined as:
    \begin{equation}
    \mathbf{o}_t=\sigma\left(\mathbf{W}_o \cdot \mathbf{x}^{(t)}+\mathbf{U}_o \cdot \mathbf{h}^{(t-1)}+\mathbf{b}_o\right)
    \end{equation}
    The new hidden state $\mathbf{h}^{(t)}$ is then defined as:
    \begin{equation}
    \mathbf{h}^{(t)}=\mathbf{o}_t \odot \tanh \left(\mathbf{C}^{(t)}\right)
    \end{equation}

\end{itemize}

\subsubsection{Gated Recurrent Unit}
\begin{figure*}[t]
    \begin{center}
         \centerline{
            \includegraphics[width=0.65\linewidth]{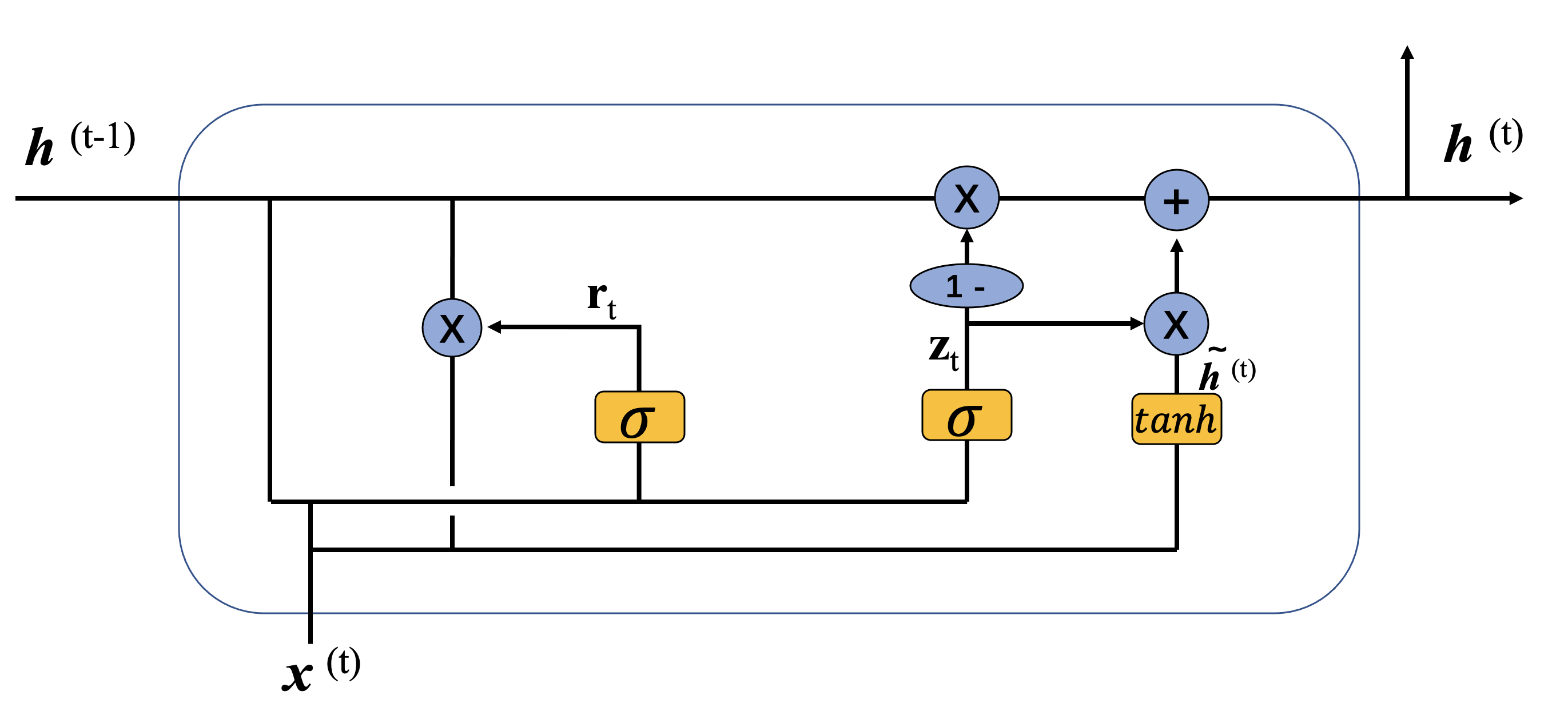} 
            }
        \vspace{0.07in}
        \caption{A block of GRU.}
        \label{fig:gru}
    \end{center}
\end{figure*}

Gated Recurrent Unit (GRU) \cite{chung2014empirical} is a variant of the LSTM, but has fewer parameters than LSTM. 
In GRU, \textbf{update gate} and \textbf{reset gate} are employed to alleviate the problem of vanishing gradient, which is typical in traditional RNNs. In their most basic form, they are two vectors that control the information that is passed on to the output.
A bock of GRU is shown in Figure~\ref{fig:gru}, those two gates are described below:
\begin{itemize}
    \item \textbf{Update Gate:} 
    Similar to forget and input gates in LSTM, the update gate in GRU helps the model estimate how much information from the past (the previous position or time step) must be transmitted to the future. This is extremely potent 
    since the model can choose to reuse all it's ever learned and avoid the vanishing gradient problem.
    The update gate is formulated as:
    \begin{equation}
    \mathbf{z}_t=\sigma\left(\mathbf{W}_z \cdot \mathbf{x}^{(t)}+\mathbf{U}_z \cdot \mathbf{h}^{(t-1)}+\mathbf{b}_z\right)
    \end{equation}
    where $\sigma$ is employed to squash the output between 0 and 1 to determine which information to discard from $\mathbf{h}^{(t-1)}$ and which information to add from $\mathbf{x}^{(t)}$.
    
    \item \textbf{Reset Gate:} The reset gate is another gate that determines how much historical data to forget. The reset gate is formulated as:
    \begin{equation}
    \mathbf{r}_t=\sigma\left(\mathbf{W}_r \cdot \mathbf{x}^{(t)}+\mathbf{U}_r \cdot \mathbf{h}^{(t-1)}+\mathbf{b}_r\right)
    \end{equation}
    Next, the reset gate would be employed to store the relevant information from the past in a new memory content. The calculation is as follows:
    \begin{equation}
    \tilde{\mathbf{h}}^{(t)}=\tanh \left(\mathbf{W} \cdot \mathbf{x}^{(t)}+\mathbf{U} \cdot\left(\mathbf{r}_t \odot \mathbf{h}^{(t-1)}\right)+\mathbf{b}\right)
    \end{equation}
    
    As the final stage, the network must calculate the final output  $\mathbf{h}^{(t)}$ — a vector that contains information about the current time step and transmits it to the network. To accomplish this, the update gate is required. It selects which information to acquire from the present time step $\tilde{\mathbf{h}}^{(t)}$ and which information to collect from the preceding steps $\mathbf{h}^{(t-1)}$.
    \begin{equation}
    \mathbf{h}^{(t)}=\left(\mathbf{1}-\mathbf{z}_t\right) \odot \tilde{\mathbf{h}}^{(t)}+\mathbf{z}_t \odot \mathbf{h}^{(t-1)}
    \end{equation}
    
\end{itemize}

\subsection{Graph Neural Networks}
Graph Neural Networks (GNNs) \cite{scarselli2008graph} are a class of neural networks that operate on graph-structured data.
They often employ a message-passing mechanism in which the representation of a node is derived from the representations of its neighbors via a recursive computation. 
So, eventually, the node representation can encode high-order structural information via several aggregation layers.
Currently, GNNs have demonstrated promising performance for graph-structured data \cite{henaff2015deep, ma2021deep, zhang2020deep}.


\begin{figure*}[h]
    \begin{center}
         \centerline{
            \includegraphics[width=0.55\linewidth]{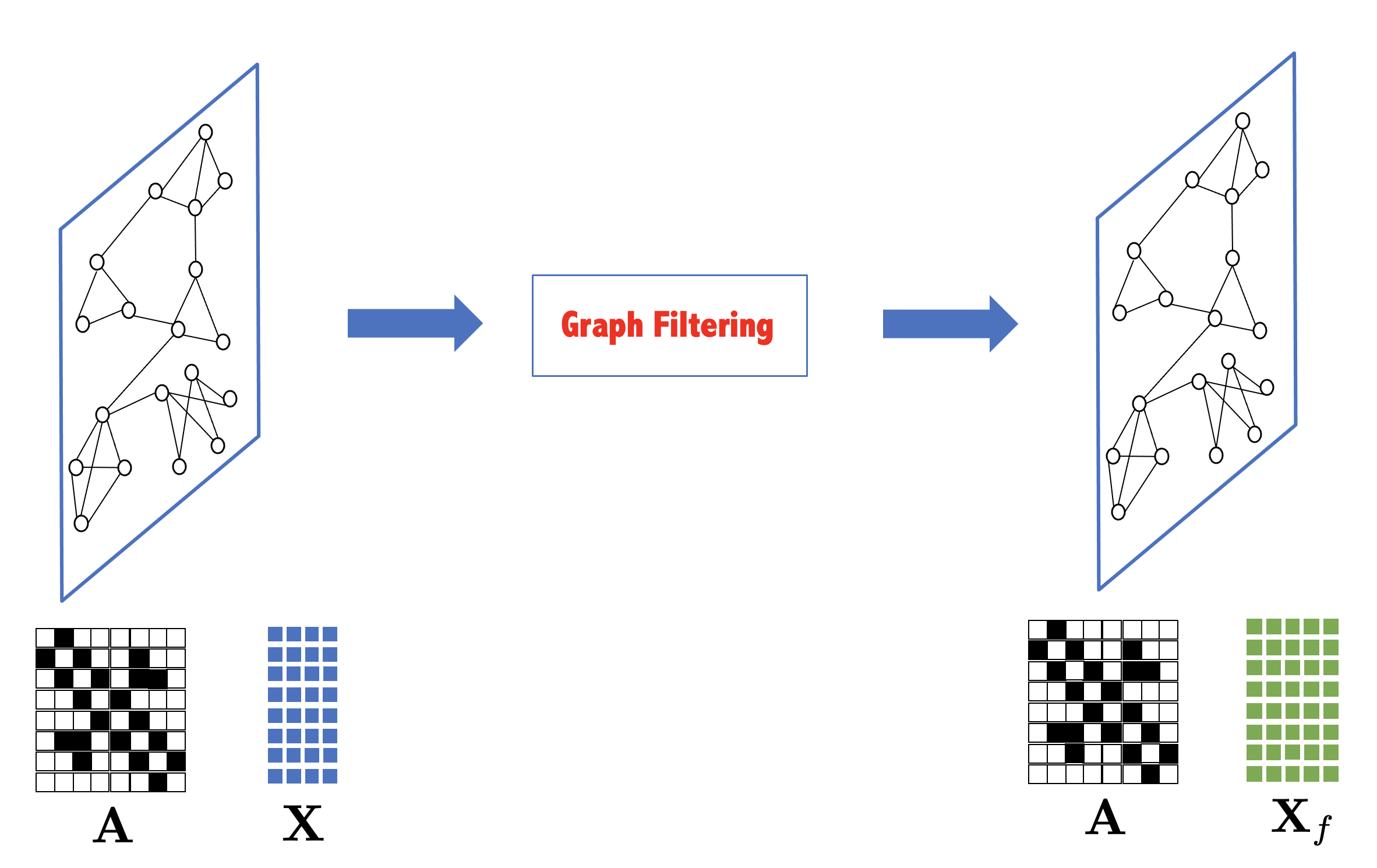} 
            }
        \caption{An illustration of Graph Neural Networks.}
        \label{fig:gnn}
    \end{center}
\end{figure*}

Let $\mathcal{G}=(\mathcal{V},\mathcal{E})$ be a graph, where $\mathcal{V}$ is the set of $N$ nodes $\{v_1, v_2, ..., v_N\}$ and $\mathcal{E}$ is the set of edges. 
The node feature matrix can be denoted as $\mathbf{X} \in \mathbb{R}^{N \times K}$, where $K$ is the number of node features.
The graph structure can be described by adjacency matrix $\mathbf{A} \in \mathbb{R}^{N \times N}$, where $\mathbf{A}_{i,j}$ denotes the relationship between nodes ${v_i}$ and ${v_j}$. Thus a graph can also be denoted as $\mathcal{G}$ = ($\mathbf{A}$, $\mathbf{X}$).

As shown in Figure~\ref{fig:gnn} from \cite{ma2021deep}, graph filtering including feature transformation and feature aggregation would be performed on node features $\mathbf{X}$ and graph structure $\mathbf{A}$. After graph filtering, graph structure $\mathbf{A}$ would be kept the same while original node features $\mathbf{X}$ would be updated as $\mathbf{X}_\textbf{f}$. Here, we take graph convolution network (GCN) \cite{kipf2016semi} as an example to elaborate more details on graph filtering operation.
In Eq.~\ref{gcn_eq}, $\mathbf{Z}$ is the new graph embedding for each node learned from a two-layer GCN.

\begin{equation}
 \mathbf{Z} = \operatorname{\sigma}(\mathbf{\widetilde{A}}  \operatorname{\sigma}(\mathbf{\widetilde{A}} \mathbf{X} \mathbf{W}_\textbf{1} )  \mathbf{W}_\textbf{2} ) 
\label{gcn_eq}
\end{equation}
where $\mathbf{\widetilde{A}} = \mathbf{D}^{-1/2} \mathbf{A}\mathbf{D}^{-1/2}$ is the normalized adjacency matrix, and $\mathbf{D}$ is the degree matrix.
$\sigma$ is the activation function such as ReLU.
$\mathbf{W}_\textbf{1}$ and $\mathbf{W}_\textbf{2}$ are the learned weight matrices in the first and second GCN layer training.

\subsection{A Summary of Popular Deep Learning Frameworks}
We summarize popular frameworks for deep learning in Table~\ref{tool_dl}.

\begin{table}[h]\normalsize
\centering
\caption{A summary of existing popular frameworks for deep learning.}
\vskip 1em
\resizebox{15cm}{!}{%
\begin{tabular}{llllll}
\hline
\textbf{Tool}       & \textbf{Category}   & \textbf{Lannguage}  & \textbf{Availability} \\ \hline
\begin{tabular}[c]{@{}l@{}}Pytorch\cite{paszke2019pytorch}\\ \end{tabular}    & Deep Learning      & \begin{tabular}[c]{@{}l@{}}C++, Python\end{tabular}                                                                              & \begin{tabular}[c]{@{}l@{}}https://github.com/pytorch/pytorch\end{tabular} \\ \hline

\begin{tabular}[c]{@{}l@{}}Tensorflow\cite{abadi2016tensorflow}\\ \end{tabular}    & Deep Learning      & \begin{tabular}[c]{@{}l@{}}C++, Python\end{tabular}                                                                              & \begin{tabular}[c]{@{}l@{}}https://github.com/tensorflow/tensorflow\end{tabular} \\ \hline

\begin{tabular}[c]{@{}l@{}}Keras\cite{chollet2018keras}\\ \end{tabular}    & Deep Learning      & \begin{tabular}[c]{@{}l@{}}Python\end{tabular}                                                                              & \begin{tabular}[c]{@{}l@{}}https://github.com/keras-team/keras\end{tabular} \\ \hline

\begin{tabular}[c]{@{}l@{}}MXNet \cite{chen2015mxnet}\\ \end{tabular}    & Deep Learning      & \begin{tabular}[c]{@{}l@{}}C++, Python\end{tabular}                                                                              & \begin{tabular}[c]{@{}l@{}}https://github.com/apache/incubator-mxnet\end{tabular} \\ \hline

\begin{tabular}[c]{@{}l@{}}DGL\cite{wang2019deep}\\ \end{tabular}    & Graph Learning     & \begin{tabular}[c]{@{}l@{}}Python\end{tabular}                                                                              & \begin{tabular}[c]{@{}l@{}}https://github.com/dmlc/dgl\end{tabular} \\ \hline

\begin{tabular}[c]{@{}l@{}}PyG\cite{fey2019fast}\\ \end{tabular}    & Graph Learning     & \begin{tabular}[c]{@{}l@{}}Python\end{tabular}                                                                              & \begin{tabular}[c]{@{}l@{}}https://github.com/pyg-team/pytorch\_geometric\end{tabular} \\ \hline

\begin{tabular}[c]{@{}l@{}}networkx\cite{hagberg2008exploring}\\ \end{tabular}    & Graph Learning     & \begin{tabular}[c]{@{}l@{}}Python\end{tabular}                                                                              & \begin{tabular}[c]{@{}l@{}}https://github.com/networkx/networkx\end{tabular} \\ \hline

\begin{tabular}[c]{@{}l@{}}PyTorch-VAE
\cite{Subramanian2020}\\ \end{tabular}    & Variational AutoEncoders     & \begin{tabular}[c]{@{}l@{}}Python\end{tabular}                                                                              & \begin{tabular}[c]{@{}l@{}}https://github.com/AntixK/PyTorch-VAE\end{tabular} \\ \hline


\begin{tabular}[c]{@{}l@{}}OpenCV
\cite{culjak2012brief}\\ \end{tabular}    & Image Processing     & \begin{tabular}[c]{@{}l@{}}C++, Python\end{tabular}                                                                              & \begin{tabular}[c]{@{}l@{}}https://github.com/opencv/opencv\end{tabular} \\ \hline

\end{tabular}%
}
\label{tool_dl}
\end{table}

\section{Single-Cell Analysis Pipeline}\label{pipeline}

Single-cell data is used in a wide range of applications and requires a number of different stages of data processing to prepare the data and develop biological analysis. 
The pipeline consists of tasks that can be grouped by overall goal; we illustrate a typical single-cell analysis pipeline  in Figure~\ref{pipeline}.  
The first collection of tasks consists of applying pre-processing functions that improve the data quality and normalize the information.
Then a number of transformations of the data are performed to generate a better representation that eliminates possible instrumentation noise from the biologically salient signal. 
Given these commonly performed tasks, the user then chooses select analysis, based on the application, to capture high-level biological features, e.g. discovering the types of cells in the population. 
The final stage takes the output data for the chosen application that pursues scientific insights in fields like oncology, immunology, etc.  

\begin{figure*}[t]
    \begin{center}
         \centerline{
            \includegraphics[width=0.6\linewidth]{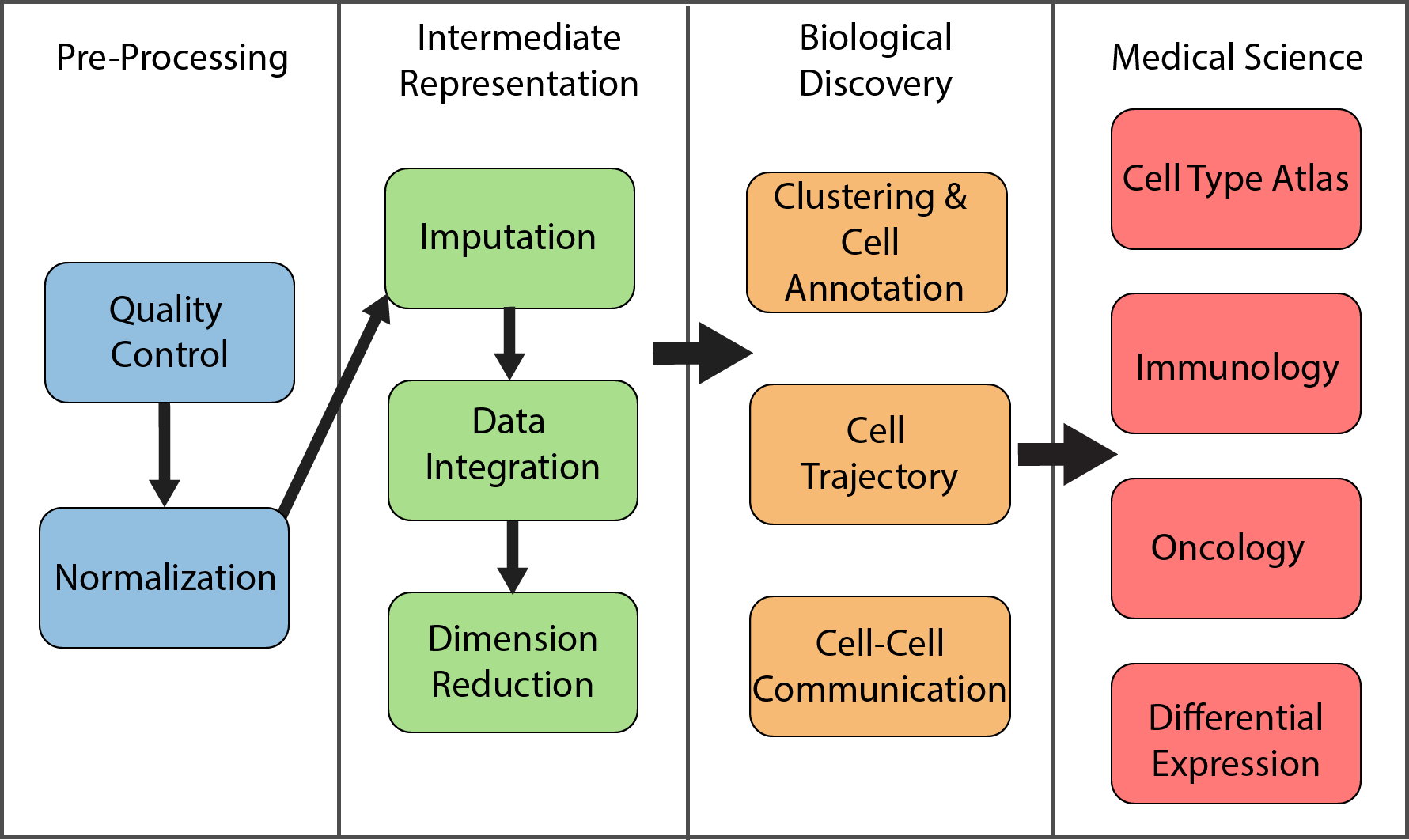} 
            }
        \caption{An overview of the single-cell data pipeline.}
        \label{fig:pipeline}
    \end{center}
\end{figure*}

The data type of many single-cell technologies consists of sparse matrices with rows and columns indicating cells and features (usually counts of genes expressed), respectively. 
For technologies that mark cells via barcodes, errors leading to empty or doubly marked cells require quality control to remove them and other low information data from the dataset\cite{kolodziejczyk2015technology}.
Cells with a very low percentage of non-zero features may be removed, as well as cells with more individually expressing features but a small total count of features\cite{islam2011characterization, natarajan2019single}.
Further data normalization is also necessary for mitigating instrumentation variation between cell samples\cite{hou2016single}. 
When samples are measured in different batches, large biases can be introduced between the groups, leading to a large body of techniques dedicated to removing these `batch effects'\cite{haghverdi2018batch}.
A careful balance needs to be struck between applying enough normalization to remove technical artifacts, but not so much that natural biological variation is also lost. 
While a standard collection of transforms have been established in the field, recent research suggests that accepted normalization techniques should be carefully chosen based on the dataset; 
different depth normalization and variance stabilization methods can have large effects on downstream interpretation, see \cite{booeshaghi2022depth, cole2019performance}.

After initial pre-processing is completed, the data is further transformed and simplified to better capture biological information and remove instrumentation biases. 
Sparsity in the data is near ubiquitous in data produced from single-cell technologies.  
The main method for addressing this gap is via imputation, a principled means of replacing zeros in data with non-zero values. 
When data is sampled in batches, large biases can be introduced that need to be addressed via batch effect removal.
Similarly, multi-modal technologies can produce multiple types of data from the same cell, which requires a means to combine the data types into a single representation. 
Lastly, in order to reduce the number of parameters, and to better capture information and remove noise, dimension reduction techniques are employed to obtain a low-dimensional representation of the input data.
All these processes aim to obtain a better representation of data for chosen downstream tasks, and many modern methods perform multiple of these data transformations simultaneously.
Machine learning becomes especially important in these transformations, and deep learning in particular has shown remarkable success. In the following sections, we will survey two representative data transformation tasks, i.e.,  
multi-modal data integration in Section (\ref{multimodal}), and imputation in Section (\ref{imputation}).   

Given the intermediate representation obtained by the above transformations, the pipeline begins to branch depending on the type of data and the end goals for scientific inquiry.  
Generally, these methods involve some form of data annotations to inform the biological properties of the cells. 
The most common form of this is clustering and cell annotation, which involve partitioning cells by their cell types or gene expression behavior and other biological markers.  
Cell annotation (Section \ref{clustering}) specifically uses known cell markers and other prior knowledge to classify the sampled cells \cite{shao2020sccatch}, whereas single-cell RNA-seq clustering (Section (\ref{annotation}) is unsupervised and only uses the given data to group cells.
The type of clustering that is performed can depend on the type of technology used to sample the cells.
For image-based single-cell profiling, the clustering of the pixels of the image into distinct cells is known as the cell segmentation problem. 
As well the newly developed spatial transcriptomic single-cell sequencing presents unique challenges that require alterations to the clustering methods applied to scRNA-seq data.
The spatial domain task partitions the sample into geometric clusters dependent on the samples' spatial positions and according to their gene expression behavior.
Furthermore, the spatial samples mix multiple cells together in spots, which requires the identification of the cell types in each spot; this task is known as cell deconvolution. 
Since all of these tasks are well designed for the application of deep learning, we review them in Section (\ref{domain}) for the spatial domain problem, Section (\ref{segmentation}) for cell segmentation, and Section (\ref{deconvolution}) for cell deconvolution.

Beyond clustering and annotation, if cells are sampled multiple times throughout stages of cell growth and differentiation, the gene expression between similar cells indicates cell developmental trajectories.
Connecting cells over time as being part of a cell trajectory is a central part of understanding stem cell behavior\cite{tritschler2019concepts}, and cancer\cite{fan2020single}. 
Cell-cell communication (CCC) and interactions are additional biological signals that can be captured that indicates information about tissues, cell processes and intercellular relationships \cite{almet2021landscape}.
These relations given between cells determined in trajectory analysis and CCC give a means to understand how `close' cells are to one another. 
In contrast to spatial transcriptomics, scRNA-seq does not impart spatial information, so the above computations can be used in place of underlying geometric information. 

The end goal of any single-cell analysis is to solve problems and address hypotheses developed in highly important fields of biology.
Already there have been major accomplishments made in oncology, immunology, differential expression analysis, and the creation of cell-type atlases.
In oncology, single-cell sequencing has allowed for the identification of rare cells in tumors and how they contribute to tumor progression \cite{chung2017single}. 
As well, single-cell analysis has the potential to uncover causes of metastasis and therapy resistance in cancer cells, which could lead to novel treatments and therapies \cite{navin2015first}. 
Single-cell data has also been a boon for immunology, which studies the biology of immunity and can provide insights into diseases and diagnostics. 
Previous bulk cell measurements were only able to take average responses from populations of heterogeneous cells, which obfuscates the behavior of important rare immune cells \cite{chattopadhyay2014single}. 
Differential expression analysis is fundamental to identifying genes causally linked to phenotype states and changes, and the analysis of gene expression behavior on the level of individual cells allows for novel discoveries \cite{soneson2018bias, van2020trajectory}. 
Single-cell data has led to the creation of massive cell atlases that serve as a foundational tool for studying complex tissues and cancers \cite{wagner2019single, travaglini2020molecular}.
These applications, and many others, will continue to see novel research gains as single-cell analysis techniques continue to advance.


\section{Multimodal Integration}\label{multimodal}

The scope and resolution of biology research are being revolutionized by contemporary single-cell omics technology. For instance, scRNA-seq~\cite{hwang2018single} has recently developed into a useful technique for studying biological responses and accelerating innovation in medicine. In the meantime, techniques for profiling cellular properties from different omics are continuously being adapted to work at the single-cell level (e.g. scDNA-seq~\cite{mallory2020methods}, scATAC-seq~\cite{labib2020single}, REAP-seq~\cite{peterson2017multiplexed}, etc). More recently, the quick development of single-cell technology has made it possible to measure a variety of omics in a cell concurrently, including gene expression, protein abundance, and chromatin accessibility. For instance, cellular indexing of transcriptomes and epitopes by sequencing (CITE-seq)~\cite{stoeckius2017simultaneous} allows for the joint profiling of mRNA expression and surface protein abundance, while sci-CAR~\cite{cao2018joint}, Paired-seq~\cite{zhu2019ultra}, and SNARE-seq~\cite{chen2019high} allow for the simultaneous quantification of mRNA expression and chromatin accessibility.
There are many important applications and achievements of multimodal single-cell analysis. For example, the joint analysis of single-cell multi-omics data can discover novel cell populations~\cite{hao2021integrated} or regulatory networks \cite{ma2020chromatin, duren2018integrative, zeng2019, jansen2019building}.  Since most of the methods in the downstream analysis are also commonly used in the unimodal analysis, for simplicity, we concentrate on the core issue with multimodal single-cell analysis., i.e., multimodal integration. 

Multimodal data integration is not a single, clearly defined task because of the variety of modalities and biological problems~\cite{rautenstrauch2021intricacies}. Existing methods can be roughly classified into two categories. One is the integrated analysis of several unimodal datasets, and the other is the analysis of multimodal datasets. Both can offer fresh insights into cellular states, thus benefiting downstream analysis. Next, we summarize the most common settings of existing methods. Formally, the objective of data alignment among unimodal datasets is to find a function that can project multiple sources into a shared latent space. Given $k$ datasets $\mathcal{D}_1$, $\mathcal{D}_2$, ..., $\mathcal{D}_k$, the goal is to find an embedding $\mathbf{H} \in \mathcal{R}^{n \times d}$ for all the data points, where $n=\sum^k_{i=1}{|\mathcal{D}_i|}$, and $d$ is a predefined number of dimensions. Some alternative objectives include: (a) finding popularity-to-popularity correspondence which can be summarized as $p$ clusters $\mathcal{C} = \{c_1, c_2, ..., c_p\}$ that are shared among datasets, so that $\mathcal{D}_1 \cup \mathcal{D}_2 \cup ... \cup \mathcal{D}_k 	\mapsto \mathcal{C}$; and (b) finding cell-to-cell correspondence which can be described as a pair-wise similarity matrix $\mathbf{S} \in \mathcal{R}^{n \times n}$. Note that the alternative objectives can be easily achieved given the embeddings obtained from the first objective. Therefore many methods start with the first objective, but there are also methods that directly get to the latter two goals. Meanwhile, the objective of data integration for multimodal datasets is to obtain a joint embedding, which can reduce the dimensions and preserve essential information from multi-modalities to improve cell state identification. Formally, given $k$ modalities $M_1 \in \mathcal{R}^{n \times d_1}$, $M_2 \in \mathcal{R}^{n \times d_2}$, ..., $M_k \in \mathcal{R}^{n \times d_k}$, where $d_i$ is the feature dimensions of the $i$-th modality, and $n$ is the cell number. In this setting, the correspondence between cells among different modalities is already known. Therefore for each cell, we have features from all the modalities. The goal is to find an embedding matrix $\mathbf{H} \in \mathcal{R}^{n \times d}$ for all cells so that each cell would be represented by a $d$-dimensional vector in the downstream analysis.

One challenging problem for multimodal data integration is that the resulting embedding $\mathbf{H}$ is hard to be evaluated due to the lack of ground truth. One common evaluation method is to calculate Normalized Mutual Information (NMI) between embedding-based clustering results and predefined cell type labels. However, in most cases those cell-type labels are acquired from clustering algorithms as well, resulting in a double-dipping issue. Another way is to evaluate the outcome of multimodal data integration in a specific downstream analysis. However, it is not conducive to a unified benchmark for comparison. One potential way to benchmark multimodal integration, as suggested in a recent work~\cite{luecken2021sandbox}, is to leverage multi-omics aligned data, where two omics are simultaneously measured in each cell (e.g. CITE-seq\cite{stoeckius2017simultaneous}), to provide ground truth for multimodal integration. More specifically, to comprehensively evaluate the power of various integration methods, three key tasks are defined, i.e. modality prediction, modality matching, and joint embedding.  
The modality prediction task is to predict one omic from another omic. The modality matching task is to align observations in different modalities that are actually from the same cells, while the ground truth label is given. The joint embedding task is to comprehensively evaluate the embedding $\mathbf{H}$ by various metrics based on biological states of cells and batch effects removal.

Since the joint measurement technology is not yet widespread enough, most existing works focus on identifying cell correspondences across modalities, in order to support the joint analysis with multiple unimodal sources. We categorize this type of research as ``data alignment among unimodal datasets" in the following subsection. Because multimodal assays are becoming more and more popular, we categorize methods that focus on leveraging aligned data as a separate category and discuss them in Section~\ref{sec_pair}.

\subsection{Data Alignment among Unimodal Datasets}

The general idea of data alignment is to identify cell-to-cell correspondences or similarities. More concretely, given two or more datasets, the integration methods sought to find a unified interface for data points from all the datasets, so that downstream analysis can get rid of the discrepancy between different modalities. For instance, projecting all modalities into a shared latent space is one of the most popular methodologies in this task.
Due to the uncertainty of ground truth, there are various underlying key assumptions for multimodal integration, which can be broadly divided into two categories. First, feature correspondence exists across modalities. For example, features from different modalities are controlled by the same gene or there exist certain correlations between features. Typical approaches include feature correlation-based methods~\cite{jansen2019building, lin2022scjoint, zhang2022scdart} and non-negative matrix factorization (NMF)-based methods~\cite{jin2020scai, duren2018integrative, liu2020, kriebel2022uinmf}. Second, an underlying manifold is shared among different modalities, or in other words, some shared low-dimensional latent factors exist in different modalities. Manifold alignment~\cite{stanley2020, cao2020unsupervised, jain2021} and coupled clustering~\cite{zeng2019, zeng2021} are well-received methods based on this assumption. 

\subsubsection{Traditional Methods}
The study of multimodal integration started with some classic algorithms, such as statistical modeling. MATCHER~\cite{welch2017}, one of the earliest works in this field, uses a Gaussian process latent variable model (GPLVM) to model high-dimensional single-cell data as a function of latent variables (e.g. pseudo time). scACE~\cite{lin2020model} develops a statistical model-based joint clustering method that is specifically designed for single-cell genomic data. scAMACE~\cite{wangwu2021scamace} extends scACE~\cite{lin2020model} to broader data types while performing joint clustering in a similar statistical way. MOFA+~\cite{argelaguet2020mofa+} uses a GPU-accelerated efficient variational inference algorithm to infer a small number of latent factors as integrated representation.

The most popular group of methods is matrix factorization. scAI~\cite{jin2020scai} introduces an adapted matrix factorization that is designed specifically for the integration of epigenetic and transcriptomic data. CoupledNMF~\cite{duren2018integrative} relies on an alignment between gene expressions and accessible chromatin regions, obtained from the paired expression and chromatin accessibility (PECA) model~\cite{duren2017modeling}. LIGER~\cite{liu2020} introduces integrative non-negative matrix factorization (iNMF~\cite{yang2016non}) which jointly decomposes data matrices from different modalities to get a cell representation in a shared latent space. UINMF~\cite{kriebel2022uinmf} follows iNMF and further separates the metagene matrix into shared and unshared components, thus enabling unshared features to inform the factorization. BindSC~\cite{dou2020} aligns both cells and features with a bi-order canonical correlation analysis (CCA) algorithm, where rows and columns from two data matrices are iteratively aligned by CCA. Seurat v3~\cite{stuart2019} applies canonical correlation analysis (CCA) and mutual nearest neighbors (MNNs) to identify ‘anchors’ (i.e. correspondences of cells across datasets). SingleCellFusion~\cite{luo2022single} uses CCA and a restricted MNN to impute features in one modality from another modality. Hence, it requires cell correspondences between modalities. 

Another important group of methods is manifold alignment. Harmonic alignment~\cite{stanley2020} projects data points to a shared embedding using principle components analysis (PCA) and removes multi-dataset-specific effects iteratively. UnionCom~\cite{cao2020unsupervised} first generates a distance matrix within the same dataset. Then it matches the distance matrices across datasets via matrix optimization. MMD-MA~\cite{singh2020} first introduces Maximum Mean Discrepancy (MMD) loss to achieve similar distributions in the latent space across different datasets. Then it adds two extra losses. One to preserve structure between the input space and the latent space, and the other to avoid collapse to a
trivial solution. MultiMAP~\cite{jain2021} recovers geodesic distances on a shared latent manifold that involves all datasets and then acquires low-dimensional embedding based on a neighboring graph on the manifold. 

Some of the remaining methods also explore coupled clustering which skips the step of projection to a low-dimensional latent space. DC3~\cite{zeng2019} adds a coupling term to the cost function to improve joint clustering, where the coupling term is derived from empirical correlations between modalities. coupleCoC+~\cite{zeng2021} is a co-clustering framework that requires feature correspondences but also takes advantage of unlinked features. Gradient-boosted regression~\cite{lake2018} implements a gradient-boosting classification tree model to predict clustering labels on one modality from another modality, correspondingly determining the correspondence between cells from the resulting decision tree.

\subsubsection {Deep Learning Methods}

Compared with traditional methods, deep learning methods are receiving increasing attention, and many of them are capable of jointly handling unimodal and multimodal datasets. Here, we introduce three representative deep learning methods that focus on the alignment of unimodal datasets.
%

SOMatic~\cite{jansen2019building} links scATAC-seq regions with scRNA-seq genes using self-organizing maps (SOMs). SOMs are a type of artificial neural networks which are referred to as Kohonen networks~\cite{kohonen1982self}. They are unsupervised methods to generate low-dimensional data representation. SOMatic starts with separate training and clustering SOMs for each modality and then links the clusters from both modalities via a linking algorithm from GREAT~\cite{mclean2010great}.

SCIM~\cite{stark2020scim} matches cells from multiple datasets based on low-dimensional latent representations, which are obtained from an autoencoder framework with an adversarial objective. 
During training autoencoders for each modality, the latent spaces are aligned to have a comparable structure using a neural network discriminator and an adversarial loss.
The autoencoders and the discriminator are simultaneously trained, resulting in a joint embedding space.

GLUE~\cite{cao2022multi} initializes omics-specific cell embeddings with variational autoencoders. Then, in order to link the latent spaces for different modalities, GLUE makes use of a prior knowledge-based guidance graph to learn feature embeddings. They replaced the decoders in previous omics-specific autoencoders with an inner product between feature embeddings and cell embeddings. Furthermore, a discriminator is implemented to align the latent space of different omics via adversarial learning. 

 
    

\subsection{Data Integration for Multimodal Dataset} \label{sec_pair}
Increasingly sophisticated co-assay techniques, such as CITE-seq and SNARE-seq, bring us unprecedented multimodal data at the single-cell level. It provides new insights into the interaction between different modalities, as well as a comprehensive understanding of the cellular system.
In light of this, a new group of methods dedicates to making good use of paired data to improve cell state identification. In paired data, we observe features of more than one modality for every single cell. Therefore, the cell-to-cell correspondence is given, and the integration approaches can focus on learning high-quality data representations that encode the fundamental cellular states and their collective roles in tissues. 
In this category, we review representative methods that are able to take advantage of paired data. Some of them are completely supervised, while most of them are semi-supervised and utilize paired data to improve unpaired integration.  Most methods are deep learning methods, except Pamona~\cite{cao2022}. Pamona is a non-deep learning approach. It uses a partial Gromov-Wasserstein distance-based manifold alignment framework to project multi-modalities to a shared latent space, preserving both common and modality-specific structures. 


Among the deep learning methods, the vast majority of them are based on auto-encoders, except for one GNN model~\cite{wen2022graph}. We roughly divide all deep learning models into two classes according to whether cell correspondence is indispensable.
The first class of methods is flexible enough to leverage both aligned and unaligned data. 
Cross-modal Autoencoders~\cite{yang2021} uses autoencoders to map vastly different modalities (including images) to a shared latent space. Specifically, a discriminator and adversarial loss are added to force the distributions of different modalities to be matched in the latent space. To make use of prior knowledge, an additional loss term can further be added to align specific markers or anchoring cells. 
MAGAN~\cite{amodio2018magan} is one of the earliest works in this area featuring a very natural design. It consists of two generative adversarial networks (GANs) that learn mutual mappings between two domains. In addition to the traditional adversarial loss from generators and discriminators, a reconstruction loss is introduced. Namely, if the first generator can generate modality 2 from modality 1, then the second generator should generate modality 1 from modality 2,  where the back translation should be the same as the original features. To exploit the paired data, MAGAN can further involve a supervised loss for the generator.
Cobolt \cite{gong2021cobolt}, uses autoencoders as well, to project two modalities into shared latent space. It fuses modalities and prior by taking the posterior mean of multimodal distributions as a summary of the shared representation passing to the decoder. Therefore, the framework is flexible  to integrate both paired and unpaired data, and more than two modalities simultaneously.
scDART~\cite{zhang2022scdart} starts with a gene activity module, where a neural network transforms ATAC-seq data into scRNA-seq data and is regularized by a pre-defined gene-activity matrix (GAM). Next, a projection module projects data into low-dimensional latent space, where trajectory structure  is preserved by adding a cell pair-wise distance loss. Besides, a Maximum Mean Discrepancy (MMD) loss is also added to project multiple datasets into shared space. With respect to paired multimodal data, an additional anchor loss can be added to utilize the alignment information. The advantage of scDART is the preservation of cell trajectories in continuous cell populations.
scJoint~\cite{lin2022scjoint} takes the gene activity score matrix and gene expression matrix as input. They first project data into latent space through neural-network-based dimension reduction, where three losses are jointly enabled to train an ideal encoder. Neural-networks-based dimension reduction (NNDR) loss is used to force features in the low-dimensional embedding space to be orthogonal.  A cosine similarity loss is used to maximize the similarity between the best-aligned multimodal data pair, which helps utilize alignment information. A cross-entropy loss with cell-type annotations further enhances the representation of scRNA-seq datasets. After the projection, it transfers cell-type labels through KNN in embedding space and further improves the modalities mixing with transferred cell-type labels in a metric training loss.
SMILE~\cite{xu2022smile} uses contrastive learning to maximize the mutual information between original feature space and latent space, in order to learn discriminative representations for cells, which is clustering-friendly. Besides, its contrastive learning method utilizes paired data to improve integration.

The other class of models accepts only aligned data for training, although trained models can be used on unaligned data.
BABEL~\cite{wu2021babel} trains two neural-network-based encoders and two decoders on the paired data to translate data from one modality to the other and to reconstruct itself, thus eventually obtaining shared embedding. Both BABEL and Cross-modal Autoencoders consider autoencoders as the projecting function towards latent space, yet they use discriminative loss and translation loss to align embedding from multimodalities.
scMM~\cite{minoura2021mixture} leverages a mixture-of-experts multimodal variational autoencoder to explore the latent dimensions associated with multimodal regulatory programs. It models the raw count data of each modality via different probability distributions in an end-to-end way.
scMVAE~\cite{zuo2021deep} adapts variational autoencoders to a Product of Expert (PoE) inference network to estimate the joint posterior from all omics, combined with a Gaussian Mixture Model (GMM) prior. Two interesting points are that the model uses library factors to normalize input expressions and re-scales the output, and instead of direct reconstruction, the decoder predicts the parameters of a probabilistic distribution to fit the input expressions, which is the training objective.
DCCA~\cite{zuo2021dcca} combines autoencoders for each modality with attention transfer that enables autoencoders to learn through mutual supervision across modalities based on the similarity of embeddings. Namely, instead of having a shared latent space, they use the scRNA-seq data to train a teacher network, guiding a student network that works with scATAC-seq data.
scMoGNN~\cite{wen2022graph} constructs a cell-feature graph based on paired data to capture both cell similarity and gene similarity, incorporating feature relations as well. A graph encoder is applied to reduce the dimensionality. The proposed method works on large-scale paired data and achieves good performance on comprehensive benchmarking tasks.

\subsection{Tools and Datasets}
We summarize representative tools or methods for multimodal integration in Table~\ref{integration_tool_table} and useful benchmarks in Table~\ref{integration_dataset_table}.

\begin{table}[ht]\normalsize
\caption{A summary of multimodal integration tools.}
\vskip 1em
\resizebox{\textwidth}{!}{%
\begin{tabular}{lllll}
\hline
\textbf{Tool} & \textbf{Algorithm}                                         & \textbf{Description}                                                                                                                                                                                                                                         & \textbf{Language} & \textbf{Availability}                                                                                                                    \\ \hline
MAGAN     & GAN                                                  &  \begin{tabular}[c]{@{}l@{}} A GAN model with a reconstruction loss\\ to enhance alignment.  \end{tabular}   & Python            & \begin{tabular}[c]{@{}l@{}} \href{https://github.com/KrishnaswamyLab/MAGAN}{{\textcolor{blue}{MAGAN}}}~\cite{amodio2018magan} \end{tabular}           \\ \hline
SOMatic     & SOM                                                  & \begin{tabular}[c]{@{}l@{}} A SOM model that links scATAC-seq regions\\ with scRNA-seq genes\end{tabular}                                                                                                                                  & C++            &  \href{https://github.com/csjansen/SOMatic}{{\textcolor{blue}{SOMatic}}}~\cite{jansen2019building}          \\ \hline
SCIM      & Classical                                                  & \begin{tabular}[c]{@{}l@{}} Autoencoders aligned with a adversarial discriminator \end{tabular}                                   & Python            & \href{https://github.com/ratschlab/scim}{{\textcolor{blue}{SCIM}}}~\cite{stark2020scim}\\ \hline
CMAE    & AutoEncoder                                                  & Autoencoders with adversarial loss, alignment loss and prior      & Python        & \href{https://github.com/uhlerlab/cross-modal-autoencoders}{{\textcolor{blue}{CMAE}}}~\cite{yang2021}; \href{https://github.com/OmicsML/dance}{{\textcolor{blue}{DANCE}}}~\cite{link2dance}                                                                                \\ \hline
BABEL    & AutoEncoder                                                  & \begin{tabular}[c]{@{}l@{}} Two encoders and two decoders for translation and reconstruction \end{tabular}       & Python      & \href{https://github.com/wukevin/babel}{{\textcolor{blue}{BABEL}}};~\cite{wu2021babel}  \href{https://github.com/OmicsML/dance}{{\textcolor{blue}{DANCE}}}~\cite{link2dance}                                        \\ \hline
Cobolt       & AutoEncoder                                                  & \begin{tabular}[c]{@{}l@{}} Modality fusion with posterior mean of multimodal distributions \end{tabular}                                                     & Python  &               \href{https://github.com/epurdom/cobolt\_manuscript}{{\textcolor{blue}{Cobolt}}}~\cite{gong2021cobolt}
                    \\ \hline
scMM    & \begin{tabular}[c]{@{}l@{}} AutoEncoder\end{tabular} & \begin{tabular}[c]{@{}l@{}} A mixture-of-experts multimodal variational autoencoder \end{tabular}                                                                   & Python            &  \href{https://github.com/kodaim1115/scMM}{{\textcolor{blue}{scMM}}}~\cite{minoura2021mixture}; \href{https://github.com/OmicsML/dance}{{\textcolor{blue}{DANCE}}}~\cite{link2dance}             \\ \hline
scMVAE    & \begin{tabular}[c]{@{}l@{}} AutoEncoder \end{tabular} &  A PoE inference network with variational autoencoders    & Python            &  \href{https://github.com/cmzuo11/scMVAE}{{\textcolor{blue}{scMVAE}}}~\cite{zuo2021deep};  \href{https://github.com/OmicsML/dance}{{\textcolor{blue}{DANCE}}}~\cite{link2dance}           \\ \hline
DCCA     & \begin{tabular}[c]{@{}l@{}} AutoEncoder\end{tabular}  & \begin{tabular}[c]{@{}l@{}} Autoencoders with cross-modal knowledge transfer \end{tabular}              & Python & \href{https://github.com/cmzuo11/DCCA}{{\textcolor{blue}{DCCA}}}~\cite{zuo2021dcca} ; \href{https://github.com/OmicsML/dance}{{\textcolor{blue}{DANCE}}}~\cite{link2dance}              \\ \hline
scJoint      & MLP                   & \begin{tabular}[c]{@{}l@{}}MLP-based dimension reduction with regularization losses \end{tabular} & Python            & \href{https://github.com/SydneyBioX/scJoint}{{\textcolor{blue}{scJoint}}}~\cite{lin2022scjoint}                        \\ \hline
SMILE     & MLP                  & \begin{tabular}[c]{@{}l@{}} MLP with contrastive learning \end{tabular} & Python            & \href{https://github.com/rpmccordlab/SMILE}{{\textcolor{blue}{SMILE}}}~\cite{xu2022smile}              \\ \hline
scMoGNN      & GNN                                                        & \begin{tabular}[c]{@{}l@{}} A gnn model with cell-feature heterogeneous graph \end{tabular} & Python            & \href{https://github.com/OmicsML/dance}{{\textcolor{blue}{DANCE}}}~\cite{link2dance}                       \\ \hline
GLUE & GNN, Autoecndoers & \begin{tabular}[c]{@{}l@{}} An autoencoder framework with prior knowledge-based guidance graph \end{tabular} & Python            & \href{https://github.com/gao-lab/GLUE}{{\textcolor{blue}{GLUE}}}~\cite{cao2022multi}                       \\ \hline
scDART        & MLP                                                       & \begin{tabular}[c]{@{}l@{}} MLP-based modality translator regularized\\ by gene-activity matrix \end{tabular} & Python            & \href{https://github.com/PeterZZQ/scDART\_test}{{\textcolor{blue}{scDAT}}}~\cite{zhang2022scdart}                       \\ \hline

\end{tabular}%
}

\label{integration_tool_table}
\end{table}



\begin{table}[ht]\normalsize
\caption{A summary of multimodal integration datasets.}
\vskip 1em
\resizebox{\textwidth}{!}{%
\begin{tabular}{llllll}
\hline
\textbf{Dataset}                                                                    & \textbf{Species} & \textbf{Tissue}                                                            & \textbf{\begin{tabular}[c]{@{}l@{}}Dataset\\ Size\end{tabular}}                                                              & \textbf{Protocol}       & \textbf{Availability}                                                                                        \\ \hline
sci-CAR &   Human, Mouse   & \begin{tabular}[c]{@{}l@{}} Human lung, human kidney,\\ mouse kidney\end{tabular} & \begin{tabular}[c]{@{}l@{}}4,825 aligned human cells, \\11,296 aligned mouse cells \end{tabular} & RNA-seq and ATAC-seq              & \href{https://www.ncbi.nlm.nih.gov/geo/query/acc.cgi?acc=GSE117089}{{\textcolor{blue}{sci-CAR}}}~\cite{cao2018joint}  \\ \hline
PBMC CITE-seq &     Human    & PBMC  & 33,455 aligned cells & RNA-seq, ADT and HTO           & \href{https://www.ncbi.nlm.nih.gov/geo/query/acc.cgi?acc=GSE128639}{{\textcolor{blue}{PBMC CITE-seq}}}~\cite{stuart2019} 
 \\ \hline
10x Genomics CITE-seq &  Human & PBMC   &  5,527 aligned cells & RNA-seq and ADT              & \href{https://support.10xgenomics.com/single-cell-gene-expression/datasets/3.1.0/5k\_pbmc\_protein\_v3\_nextgem}{{\textcolor{blue}{10x Genomics CITE-seq}}}~\cite{10x_Genomics_CITE-seq}  \\ \hline
10x Multiome Chromium X &   Human &  PBMC   &  11,984 aligned cells & RNA-seq and ATAC-seq & \href{https://support.10xgenomics.com/single-cell-multiome-atac-gex/datasets/2.0.0/10k\_PBMC\_Multiome\_nextgem\_Chromium\_X}{{\textcolor{blue}{Multiome Chromium X}}}~\cite{Multiome_Chromium_X}  \\ \hline
10x Multiome unsorted &   Human    & PBMC  & 12,012 & RNA-seq and ATAC-seq              & \href{https://www.10xgenomics.com/resources/datasets/pbmc-from-a-healthy-donor-no-cell-sorting-10-k-1-standard-2-0-0}{{\textcolor{blue}{Multiome unsorted}}}~\cite{Multiome_unsorted}  \\ \hline
SNARE-seq &   Human, Mouse   & \begin{tabular}[c]{@{}l@{}} Human cell line mixture, \\mouse brain \end{tabular} & \begin{tabular}[c]{@{}l@{}}1,047 aligned human cells,\\ 15,390 aligned mouse cells \end{tabular} & RNA-seq and ATAC-seq              & \href{https://www.ncbi.nlm.nih.gov/geo/query/acc.cgi?acc=GSE126074}{{\textcolor{blue}{SNARE-seq}}}~\cite{chen2019high}  \\ \hline
SHARE-seq &  Mouse & Mouse skin  & 34,774 mouse skin cells & RNA-seq and ATAC-seq              & \href{https://www.ncbi.nlm.nih.gov/geo/query/acc.cgi?acc=GSE140203}{{\textcolor{blue}{SHARE-seq}}}~\cite{ma2020chromatin}  \\ \hline
scNMT-seq &     Mouse    &  \begin{tabular}[c]{@{}l@{}} Embryonic stem cells, \\embryoid body cells \end{tabular}  & Less than 100 & RNA-seq, BSseq and ATAC-seq              & \href{https://www.ncbi.nlm.nih.gov/geo/query/acc.cgi?acc=GSE109262}{{\textcolor{blue}{scNMT-seq}}}~\cite{clark2018scnmt} \\ \hline
Multimodal PBMC &   Human, Mouse   & Mainly PBMC  & \begin{tabular}[c]{@{}l@{}} 4644 cells of CITE-seq,\\ 4502 cells of ASAP-seq\\ (after quality control) \end{tabular} & RNA-seq, ADT and ATAC-seq  &    \href{https://www.ncbi.nlm.nih.gov/geo/query/acc.cgi?acc=GSE156478}{{\textcolor{blue}{Multimodal PBMC}}}~\cite{mimitou2021scalable} \\ \hline
\end{tabular}%
}
\label{integration_dataset_table}
\end{table}

\section{Imputation}\label{imputation}

The increased resolution of single-cell RNA sequencing methods comes at a cost in the form of the increased sparsity of the data.  The sequencing technology may fail to capture a number of the expressed genes of an individual cell due to low RNA capture rate, as well as the stochasticity of mRNA expression.
This artificially large proportion of false zero counts is defined as ‘dropout’\cite{svensson2017, kharchenko2014}.  The zero-inflated data can have sparsity rates from 40\% to anywhere as high as 85\% to 90\%\cite{qiu2020embracing}.
This requires several preprocessing steps like gene selection and dimension reduction to improve the data signal and make downstream tasks more viable.
Single-cell DNA methylation sequencing technologies like scBS-seq and scRRBS-seq have limited amounts of DNA starting material per cell, so methylation sequencing can be limited by the degree of CpG coverage. 
This leads to missing data across the genome in each cell, requiring imputation for further analysis.
A direct way to address dropout in both cases is to perform imputation, a method to correct false zeros by estimating realistic values for those gene-cell pairs. 
For scRNA-seq data, imputation creates artificial count values for genes that fail to express, while for DNA methylation the imputation will only provide the binary one or zero. 
However, imputation methods must be careful to distinguish the zeros in the data that represent the true
absence of gene expression or methylation in specific cells from the zero counts due to dropout.
Otherwise, information will be lost during analysis from removing the truly expressed zeros.

Imputation methods can be trained and evaluated in a couple of different ways. 
Training an imputation method frequently occurs as an unsupervised learning problem for a single dataset. 
If simulated datasets are used, the underlying true counts are known, so reconstruction error (usually mean squared error) can be used for evaluation. On empirical datasets, there is no knowledge of the true counts, so other methods have to be used. 
One choice is to induce artificial dropout on the nonzero values so that reconstruction error can be computed on those values as an evaluation metric. 
Alternatively, some metrics may be used to determine how well the imputation process preserves or enhances certain biological features of the data. 
For example, if cell types or sampling times of the cells are known in a dataset, then some clustering metrics may be used to evaluate the success of the imputation, including adjusted rand index (ARI), cluster accuracy, and silhouette score. 
Similarly, single-cell data sets may have genes that are known to differentially express (DEGs) if bulk RNA-sequencing is done alongside scRNA sequencing\cite{soneson2018bias}.  Such differential expression can be evaluated by how well DEGs are detected after imputation, using AUC or F-scores. 

\subsection{Traditional Methods}
A large number of methods exist for the task of scRNA-seq data imputation, mainly focusing on generative probability models or matrix factorization. One of the early methods is MAGIC, which uses Markov diffusion networks to replace zeros in gene expression count data.
MAGIC\cite{van2017magic} calculates distances between cell expression vectors, converts the distances using the Gaussian kernel, and then uses them as a weighted adjacency matrix to perform Markov diffusion.
This method is shown to have shortcomings in that it replaces all zeros despite some being true zeros and replaces the nonzero values with new potentially biased values. 
Many other methods also use cell-cell graphs to capture information of similar cells to determine how genes with dropout behave in similar cells that have nonzero expression.
netSmooth\cite{ronen2018netsmooth} imputes using a diffusion process similar to MAGIC. DrImpute  \cite{gong2018drimpute} takes clusters of cells and simply takes averages over the clusters to find imputation values.

Another popular imputation technique uses matrix factorization techniques. VIPER \cite{chen2018viper} develops a nonnegative sparse regression model to learn the underlying count data. 
scRMD \cite{chen2020scrmd} leverages regularized robust matrix optimization, netNMF-sc \cite{elyanow2020netnmf} adopts nonnegative matrix factorization with network regularizations, and McImpute \cite{mongia2019mcimpute} builds low-rank matrix factorization. CMF-Impute \cite{xu2020cmf}  uses matrix optimization via singular value decompositions (SVD), and ALRA \cite{linderman2022zero} applies SVD and thresholding.   

SAVER \cite{huang2018saver} is a generative method that attempts to model the probability distribution of the gene expression count data. The negative binomial distribution is used to model the data but is treated equivalently as a Poisson distribution with a Gamma prior on the Poisson parameter. 
Using the data, an empirical Bayes framework is used to estimate the Gamma prior via Poisson Lasso regression. Then the posterior distribution is used to output the expression as the posterior mean with uncertainity quantification given by the distribution.
bayNorm \cite{tang2020baynorm} similarly utilizes a  Bayesian model to generate imputed values. 
scImpute \cite{li2018accurate} combines a few techniques by creating neighbor relations between cells leading to clusters, before using a probabilistic model to determine which zeros are dropout events, and then using non-negative least squares regression to impute the values.
CIDR \cite{Lin2017} likewise employs a generative model to give a probability of dropout for zero values, which is determined via non-linear least-squares regression, before using the probabilities to calculate a weighted mean for the imputation value.  

While the preponderance of imputation methods is applied to scRNA-seq data, a couple of methods have been developed for single-cell DNA methylation data. Melissa \cite{kapourani2019melissa} imputes values based on Bayesian estimation, while CaMelia \cite{tang2021camelia} makes use of the machine learning technique of boosting. 

\subsection{Deep Learning Methods}

A natural deep learning architecture for the task of imputation is the autoencoder, due to its prevalence in data denoising and missing data applications \cite{pereira2020reviewing, gondara2017multiple, gondara2018mida, beaulieu2017missing, boquet2019missing}.
One of the early autoencoder methods for scRNA imputation is AutoImpute~\cite{talwar2018autoimpute}. It is unlike many denoising autoencoders since it employs an overcomplete AE model rather than the usual undercomplete one. The AutoImpute model is a simple two-layer AE, one layer for decoder and encoder each. 
The hidden layer of the encoder output is designed to be larger than the input dimension size dependent on the training dataset. Only the encoder has an activation function and the decoder is entirely linear. 
The objective function is commonly used $L_2$ reconstruction loss penalty function, with the addition of $L_2$-norm regularization terms on the weights of the encoder and decoder. 

The deep count autoencoder (DCA) \cite{eraslan2019single}, is another autoencoder-based imputation method. In contrast to the overcomplete model of AutoImpute, DCA has the standard undercomplete architecture. 
DCA consists of four layers, two for the decoder and two for the encoder, and takes a form similar to a standard VAE. 
Instead of modeling a distribution and attempting reconstruction of the input, DCA directly predicts the parameters of a chosen probability distribution of the input data. 
By choosing a probability distribution like the zero-inflated negative binomial (ZINB) as the generating data distribution, the DCA outputs the three parameters of the distribution, and employs a log-likelihood loss function to train the model. 
This triple output requires three separate layers transforming the output of the decoder.
Other potential probability distributions include the negative binomial (NB), which may better represent the count data of scRNA-seq even with dropout, as well as the Poisson distribution. 
Additional regularization can be placed on the loss function in the case of ZINB, penalizing the $L_2$ weight of the zero-inflation parameter. 

scScope is a combination autoencoder and recurrent neural network that performs imputation along with batch effect removal \cite{deng2018massive}. 
Like some of the above methods, scScope operates in an iterative manner, taking the output of the model and running it through the model sequentially until some stopping condition is met. However, a recurrent element combines original input with sequential output. 
The input has an initial batch effect correction for input data $x\in \R^d$ using learnable parameter $B\in \R^{d\times K}$ of the form 
\[
x_c = \text{ReLU}(x - Bu)
\]
where $K$ is the number of batches, and $u$ is a one-hot vector indicating the batch of $x$. 
The batch-corrected vector is then input to the AE which has a single-layer encoder and a single layer 
decoder. 
The final layer performs a fully connected transformation with ReLU as the activation function, and then makes all entries that are not originally zero equal to zero, giving a vector of imputed values on all the potential drop-out entries. 
This vector is added to $x_c$, and then put through the AE model again. 
The objective function is the standard $L_2$-error reconstruction loss on all the non-zero entries of $x_c$. 
SAVER-X, an adaptation of the SAVER model, combines an autoencoder with a Bayesian hierarchical model, while also using transfer learning.
The autoencoder is pretrained on a reference dataset to leverage the information in some potentially larger reference datasets before being further trained on the target dataset.

scGNN \cite{wang2021scgnn} utilizes a graph autoencoder trained in an iterative loop with other standard autoencoders.  
The initial training of scGNN feeds the data through three models that give their output to the next model for input, and this continues until convergence.
The first model is a standard undercomplete autoencoder with two layers for the encoder and decoder, and is trained using reconstruction loss with an additional penalty term that captures regulatory signals in the gene expression data. 
A KNN cell graph is created based on the embedding learned by the feature autoencoder, and pruned of outliers using the Isolation Forest algorithm. 
Then a graph autoencoder takes the output graph and the reconstructed output of the first autoencoder and runs both through a graph autoencoder which has two layers using graph convolutions, and a decoder which attempts to reconstruct the graph from the embedding using a sigmoid activation function on the gram matrix of the embedding.   
The loss function of the graph autoencoder is cross-entropy between the original graph and the reconstructed graph.
After output from graph embedding is given, a $k$-means clustering method is used to cluster the cells embedding, with the number of clusters determined by Louvain algorithm. 
Then the final model has one autoencoder trained for each cell cluster, using the same architecture and loss function as the first autoencoder except with no regularization added.
After the iterative process stops, a final autoencoder is trained for imputation.
It uses the same architecture as before, but also adopts three additional regularizers in its training objective function derived from the cell graph created in the iterative process.

Another method that uses a graphical convolution network  with an autoencoder is GraphSCI\cite{rao2021imputing}.
This method constructs a graph from the data with genes taken as the nodes and the edges between them given by the Pearson correlation coefficient (PCC) of the (log-normalized) expression data, so that if the correlation of two genes is $> .3$ or $<-.3$, an edge is established between them. 
Given this graph, the autoencoder and graph autoencoder make use of the expression data and graph data to reconstruct its input. The graph autoencoder applies a two-layer GCN to the gene expression matrix using the adjacency matrix of the graph, and used the encoding as the means and variances to Gaussian distributions. 
The distributions are used to generate a reconstruction of the adjacency matrix, where each edge of the graph follows a Gaussian distribution parameterized by a pair of mean and variance parameters given in the encoding. 
A reparameterization method forgoes directly sampling from these distributions, and instead computes the reconstruction of the adjacency matrix as the sigmoid function applied to the inner product of latent vectors, one computed for each gene. 
The autoencoder is designed similarly, taking the gene expression matrix as input, and applying two layers to yield an encoding that gives the parameters of a ZINB distribution. 
The first layer takes the adjacency matrix into account by taking the element-wise product with the weight matrix. 
The reparameterization computes the reconstruction of the expression matrix as a collection of latent vectors times a diagonal size matrix that is given by the size factors of the input cells.
The training of the model is iterative, using the reconstruction of the gene expression matrix and the adjacency matrix in each iteration. The loss function uses the variational lower bound on both autoencoders, using the KL divergence on the encoded variables for the ZINB and Gaussian distributions.

For the application of DNA methylation imputation, DeepCpGm \cite{angermueller2017deepcpg} is a method that takes both DNA sequences and methylation states as inputs, and outputs imputed methylation states. 
The DNA sequence is input into a CNN via a one-hot encoding of base-pairs in the DNA sequence, so the four base pairs would be encoded by vectors $(1,0,0,0)$, $(0,1,0,0)$, $(0,0,1,0)$ and $(0,0,0,1)$. 
Each CpG site a sequence of 1001 base pairs centered on the site would be input into the DNA CNN module. The CNN consists of two pairs of a convolution layer followed by a pooling layer, before going into a fully connected layer. 
The methylation state sequence over the BpG sites is first transformed before being input into a bidirectional gated recurrent network (GRU).
The states are given by $T$ vectors representing the methylation states around a target CpG site for $T$ different cells. 
A fully connected layer transforms the vectors into a representation that seeks to capture interactions between mehtylation states and distances within cells, before the sequence of vectors is fed into the GRU, which seeks to model dependencies between cells.
The GRU is bidirectional to analyze the dependencies between cells independent of the order of the cells given in the data. 
The forward and backward GRUs produce $T$ hidden state vectors, with the last hidden states of the GRUs concatenated and given as output. 
Finally, the output of the CNN and GRU modules are concatenated and transformed by a two fully connected layer NN, with the final layer output using the sigmoid activation function to produce a dimension $T$ vector giving the predictions of the methylation rate in [0,1] of a chosen CpG site for $T$ cells.
The training uses a negative log-likelihood loss function with a $L_2$ weight-regularized penalty function.





\subsection{Tools and Datasets}

We summarize all tools or baseline methods for imputation in Table~\ref{imputation_tool_table} and useful benchmarks in Table~\ref{imputation_dataset_table}.

\begin{table}[ht]\normalsize 
    \caption{A summary of popular imputation tools}
    \vskip 1em
    \resizebox{\textwidth}{!}{
        \begin{tabular}{lllll}
            \hline
            \textbf{Tools} & \textbf{Algorithm} & \textbf{Description} & \textbf{Language} & \textbf{Availability} \\ \hline

    MAGIC       & Classical &    \begin{tabular}[c]{@{}l@{}}  MAGIC uses a diffusion model to smooth values \\ and impute zeros.   \end{tabular} &   R  & \begin{tabular}[c]{@{}l@{}} \href{https://github.com/KrishnaswamyLab/MAGIC}{{\textcolor{blue}{MAGIC}}}~\cite{van2017magic} \end{tabular}           \\ \hline

    DrImpute     & Classical &    \begin{tabular}[c]{@{}l@{}}  DrImpute clusters the data and averages the values \\ of the clusters.   \end{tabular} &     R  & \begin{tabular}[c]{@{}l@{}} \href{https://github.com/gongx030/DrImpute}{{\textcolor{blue}{DrImpute}}}~\cite{gong2018drimpute} \end{tabular}           \\ \hline
    
    scImpute        & Classical &   \begin{tabular}[c]{@{}l@{}} scImpute uses a statistical model to identify dropouts, \\ and non-negative least squares regression to impute.  \end{tabular} & R & \begin{tabular}[c]{@{}l@{}} \href{https://github.com/Vivianstats/scImpute}{{\textcolor{blue}{scImpute}}}~\cite{li2018accurate} \end{tabular} \\ \hline

    CIDR        & Classical &   \begin{tabular}[c]{@{}l@{}} CIDR uses a statistical model to identify dropouts, \\ and weighted means in clusters are used for imputation values.  \end{tabular} &     R    & \begin{tabular}[c]{@{}l@{}} \href{https://github.com/VCCRI/CIDR}{{\textcolor{blue}{CIDR}}}~\cite{lin2017cidr} \end{tabular}           \\ \hline
    
    ALRA        & Classical &    \begin{tabular}[c]{@{}l@{}} ALRA uses Singular Value Decomposition  \\ to get low-rank representation of data, \\ then attempts to restore biological zeros using thresholding.  \end{tabular}   &     R    & \begin{tabular}[c]{@{}l@{}} \href{https://github.com/KlugerLab/ALRA}{{\textcolor{blue}{ALRA}}}~\cite{linderman2022zero} \end{tabular}           \\ \hline
    
    CMF-Impute        & Classical & \begin{tabular}[c]{@{}l@{}}  CMF-Impute performs a matrix optimization to find a regularized SVD \\ for the scRNA-seq expression data.  \end{tabular} &  Matlab  & \begin{tabular}[c]{@{}l@{}} \href{https://github.com/xujunlin123/CMFImpute}{{\textcolor{blue}{CMFImpute}}}~\cite{xu2020cmf} \end{tabular} \\ \hline

    scRMD       & Classical & \begin{tabular}[c]{@{}l@{}} scRMD finds a robust regularized matrix decomposition using \\ nuclear and $\ell_1$-norm regularization.   \end{tabular} &   R  & \begin{tabular}[c]{@{}l@{}} \href{https://github.com/XiDsLab/scRMD}A{{\textcolor{blue}{scRMD}}}~\cite{chen2020scrmd} \end{tabular}           \\ \hline

    McImpute      & Classical &  \begin{tabular}[c]{@{}l@{}} McImpute performs a  nuclear norm minimization \\ optimization for regularized matrix completion     \end{tabular}     &     Matlab  & \begin{tabular}[c]{@{}l@{}}  \href{https://github.com/aanchalMongia/McImpute\_scRNAseq}{{\textcolor{blue}{McImpute}}}~\cite{mongia2019mcimpute} \end{tabular}           \\ \hline

    netSmooth       & Classical &    \begin{tabular}[c]{@{}l@{}}  netSmooth  performs network-diffusion on the \\ expression data with priors for the covariance structure. \end{tabular} &     R  & \begin{tabular}[c]{@{}l@{}} \href{https://github.com/BIMSBbioinfo/netSmooth}{{\textcolor{blue}{netSmooth}}}~\cite{ronen2018netsmooth} \end{tabular}           \\ \hline

    SAVER      & Classical &    \begin{tabular}[c]{@{}l@{}}  SAVER is a probability model that \\ generates counts using negative binomial distribution.  \end{tabular} &     R  & \begin{tabular}[c]{@{}l@{}} \href{https://github.com/mohuangx/SAVER}{{\textcolor{blue}{SAVER}}}~\cite{huang2018saver} \end{tabular}           \\ \hline

    LATE/TRANSLATE      & AutoEncoder & \begin{tabular}[c]{@{}l@{}} LATE is a undercomplete autoencoder \\ trained on reconstructing the nonzero inputs, \\ while TRANSLATE combines this autoencoder with transfer learning by \\ using a reference data set to initialize parameters. \end{tabular} & Python       & \begin{tabular}[c]{@{}l@{}} \href{https://github.com/audreyqyfu/LATE}{{\textcolor{blue}{LATE}}}~\cite{badsha2020imputation} \end{tabular}          \\ \hline

    AutoImpute     & AutoEncoder & \begin{tabular}[c]{@{}l@{}} AutoImpute uses an overcomplete autoencoder trained on  \\ non-zero entries for the imputation of dropout locations.  \end{tabular} & Python & \begin{tabular}[c]{@{}l@{}} \href{https://github.com/divyanshu-talwar/AutoImpute}{{\textcolor{blue}{AutoImpute}}}~\cite{talwar2018autoimpute} \end{tabular} \\ \hline

    scScope  & AutoEncoder & \begin{tabular}[c]{@{}l@{}} scScope uses an iterative autoencoder \\ that cycles output into input \\ while applying batch effect correction. \end{tabular} & Python & \begin{tabular}[c]{@{}l@{}} \href{https://github.com/AltschulerWu-Lab/scScope}{{\textcolor{blue}{scScope}}}~\cite{deng2019scalable} \end{tabular} \\ \hline

    DCA   & AutoEncoder & \begin{tabular}[c]{@{}l@{}} DCA is an undercomplete autoencoder that predicts \\ parameters of chosen distributions like ZINB \\ to generate the imputed data.  \end{tabular} & Python & \begin{tabular}[c]{@{}l@{}} \href{https://github.com/theislab/dca}{{\textcolor{blue}{DCA}}}~\cite{eraslan2019single} \end{tabular} \\ \hline

    scGNN   & Graph AE & \begin{tabular}[c]{@{}l@{}} scGNN uses three autoencoders in a cycle - \\ using a graph autoencoder, and plain autoencoder, \\ and a cluster-specific collection of autoencoders.\\ The convergence of a cell graph is \\ used in the regularization of a final autoencoder that \\ performs the imputation. \end{tabular} & Python & \begin{tabular}[c]{@{}l@{}} \href{https://github.com/juexinwang/scGNN}{{\textcolor{blue}{scGNN}}}~\cite{wang2021scgnn} ; \href{https://github.com/OmicsML/dance}{{\textcolor{blue}{DANCE}}}~\cite{link2dance} \end{tabular} \\ \hline

    GraphSCI   & Graph AE & \begin{tabular}[c]{@{}l@{}} GraphSCI uses two autoencoders, one being a\\ graph autoencoder on  a cell graph, the other \\ reconstructs the input using the graph as additional input.  \end{tabular} & Python & \begin{tabular}[c]{@{}l@{}} \href{https://github.com/biomed-AI/GraphSCI}{{\textcolor{blue}{GraphSCI}}}~\cite{rao2021imputing} ; \href{https://github.com/OmicsML/dance}{{\textcolor{blue}{DANCE}}}~\cite{link2dance} \end{tabular} \\ \hline

    scIGANs   & GAN & \begin{tabular}[c]{@{}l@{}} scIGANs uses a GAN to model the generation \\ of scRNA-seq data using the generated expression data for the imputation. \end{tabular} & Python & \begin{tabular}[c]{@{}l@{}} \href{https://github.com/xuyungang/scIGANs}{{\textcolor{blue}{scIGANs}}}~\cite{xu2020scigans}  \end{tabular} \\ \hline

\end{tabular}
}
\label{imputation_tool_table}
\end{table}

\begin{table}[ht]\normalsize
    \caption{A summary of imputation datasets}
    \vskip 1em
    \resizebox{\textwidth}{!}{
        \begin{tabular}{llllll}
            \hline
            \textbf{Dataset} & \textbf{Species} & \textbf{Tissue} & \textbf{\begin{tabular}[c]{@{}l@{}}Dataset\\ Dimensions\end{tabular}} & \textbf{Protocol} & \textbf{Availability} \\ \hline
            
            Tabula Muris & Mouse & 20 tissues & $53,760$ cells & \begin{tabular}[c]{@{}l@{}}SMART-Seq2\\ 10x Genomics\end{tabular} & \href{https://tabula-muris.ds.czbiohub.org/}{{\textcolor{blue}{Tabula Muris}}}~\cite{tabula2018single} \\ \hline

            Tabula Sapiens  & Human & 24 tissues & $483,152 $ cells & \begin{tabular}[c]{@{}l@{}}Illumina NovaSeq 6000 \end{tabular} & 
            \href{https://www.ncbi.nlm.nih.gov/geo/query/acc.cgi?acc=GSE201333}{{\textcolor{blue}{Tabula Sapiens}}}~\cite{Tabula2018} \\ \hline

            10X PBMC 4K  & Human & PBMC & \begin{tabular}[c]{@{}l@{}}$4,271$ cells\\ $16,653$ genes \end{tabular} & 10x Genomics & 
            \begin{tabular}[c]{@{}l@{}} \href{https://support.10xgenomics.com/single-cell-gene-expression/datasets/2.1.0/pbmc4k}{{\textcolor{blue}{10X PMBC 4K}}}~\cite{Zheng2017} \end{tabular} \\ \hline
    
            Human Cell Atlas  & Human &  & 30.6 million cells & \begin{tabular}[c]{@{}l@{}} HiSeq X Ten \end{tabular} &\begin{tabular}[c]{@{}l@{}} 
            \href{https://www.ncbi.nlm.nih.gov/geo/query/acc.cgi?acc=GSE159929}{{\textcolor{blue}{Human Cell Atlas}}}~\cite{regev2017science} \end{tabular} \\ \hline
   
        \end{tabular}
    }
\label{imputation_dataset_table}
\end{table}

\section{Clustering}\label{clustering}

One critical phase of single-cell analysis is to characterize cell types within a given tissue. Cell-type groups are typically made up of cells with similar transcriptome files. Knowing specific cellular types will help us to reveal the diversity between different groups and predict gene expression of incoming cells~\cite{Gan2022}. Single-cell sequencing technology offers gene expression at the single-cell level. It allows researchers to elucidate the transcriptome heterogeneity between individual cells, which unveils the underlying subgroup structure~\cite{Yu2022}. As an unsupervised learning method, clustering has been proven to be the most effective methodology for identifying cell subgroups utilizing transcriptomic data. Classical clustering methods like K-means, Leiden and Louvain are widely applied in single-cell analysis, while the latter two graph-based methods become the dominant choice when utilized as an additional clustering module. ScDeepCluster~\cite{Tian2019} summarized some popular benchmark datasets like 10X PBMC 4K~\cite{Zheng2017}, Mouse Bladder Cells~\cite{Han2018} and Worm Neuron Cells~\cite{Cao2017}. Based on those benchmarks, there are popular metrics to evaluate clustering performance: the adjusted Rand index (ARI), the accuracy (ACC) and the normalized mutual information (NMI). A high metric score indicates that the model successfully identifies the underlying subgroup structure and satisfyingly recovers the cell-type group assignment.

\subsection{Traditional Methods}

Hierarchical clustering~\cite{Johnson1967} has been widely adopted by early works. BISCUIT~\cite{Prabhakaran2016} incorporates parameters denoting technical variation into a hierarchical Dirichlet process mixture model. SINCERA~\cite{Guo2015} presents a generally applicable analytic pipeline for processing single-cell data with hierarchical clustering. CIDR~\cite{Lin2017} performs hierarchical clustering on the first few principal coordinates of the dissimilarity matrix between the imputed gene expression. Corr~\cite{Jiang2018} calculates cell-pair differentiability correlation as the input of the hierarchical clustering module. Corr integrates variance analysis with hierarchical clustering by calculating the proportion of between-group variance to decide where to cut the hierarchical dendrogram.

Other classical clustering methods like K-means~\cite{MacQueen1967} and shared nearest neighbor (SNN) clustering~\cite{Jarvis1973} have been applied to single-cell data to identify cell types. RaceID~\cite{Chlis2017} utilizes the distance matrix as the input of a K-means algorithm for rare cell type identification. Similar to K-means, K-branches clustering algorithm~\cite{Chlis2017} locally fits half-lines as the representatives of branches in the differentiation trajectory of cells. Based on the SNN clustering, SNN-clique~\cite{Xu2015} identifies clusters in the SNN graph by finding maximal quasi-cliques. 

There are also other clustering methods for single-cell analysis. SOUP~\cite{Zhu2019} identifies group structure with a non-negative membership matrix representing the membership of cells to clusters. SOUP recovers the membership matrix from the top eigenvectors of the similarity matrix, where the number of eigenvectors is determined by the number of clusters. Based on non-negative matrix factorization (NMF), scRNA~\cite{Mieth2019} factorizes the source dataset into a gene-independent target data matrix and a data size independent dictionary, derives a new expression matrix from those two parts and performs clustering on the reconstructed data.

\subsection{Deep Learning Methods}
In single-cell analysis, traditional clustering methods may lead to suboptimal results, since single-cell data usually contains a large number of zero elements and its high heterogeneity even makes things harder. To overcome this challenge, deep learning methods have been adopted. DEC~\cite{Xie2016} proposes a new KL (Kullback-Leibler) divergence loss between the Student’s $t$-distribution of embedding space and the auxiliary target distribution. The proposed KL divergence loss makes it possible for models to achieve self-training under an unsupervised setting. DCA~\cite{Eraslan2019} introduced a deep count autoencoder (AE) network for denoising and imputation. The noise model is based on zero-inflated negative binomial (ZINB) distribution, and the mean, dropout and dispersion of data are estimated by different activation functions respectively. The ZINB model-based denoising autoencoder and the KL divergence loss then became the backbone of many popular clustering methods.

ScDeepCluster~\cite{Tian2019} adds a clustering layer to DCA. ScDeepCluster performs clustering based on soft assignment of denoised embedding. The model is optimized by minimizing the KL divergence loss of the embedded space and the ZINB loss of the denoising module. Based on scDeepCluster, scDCC~\cite{Tian2021} integrates prior information into the modeling process by merging pairwise constraints into the training process. A pairwise constraint can have two types: must-link (ML) and cannot-link (CL). As a trade-off, the ML constraint forces the paired instances to have similar soft labels, while the loss of the CL constraint encourages different soft labels. In another aspect, scziDesk~\cite{Chen2020} adds a weighted soft K-means clustering algorithm with inflation operation to the DCA structure. The model is optimized by minimizing the ZINB loss, KL divergence loss and weighted K-means clustering loss. ScVI~\cite{Lopez2018} uses variational inference to approximate the posterior distribution, where the variational lower bound is optimized by stochastic optimization. The conditional distributions are specified by ZINB model-based variational autoencoder (VAE), and the cell-type clustering is based on the K-means algorithm on the latent space. With a focus on the batch effect, DESC~\cite{Li2020} initializes parameters obtained from a ZINB model-based autoencoder and iteratively optimizes the KL divergence loss. This iterative procedure moves each cell to its nearest cluster centroid, and simultaneously removes batch effect over iterations. To conduct clustering and time-trajectory inference, scDHA~\cite{Tran2021} consists of a non-negative kernel autoencoder and a stacked Bayesian autoencoder. The kernel autoencoder shrinks the non-negative coefficients of the less important features toward zero, and the stacked Bayesian autoencoder is a VAE with two stacked decoders obtained by two different transformations.

Graph neural networks (GNN) have been applied to single-cell clustering. GraphSCC~\cite{Zeng2020} accounts structural relations between cells by constructing a cell-cell k-nearest neighbor (KNN) graph, processes it through a graph convolutional network (GCN), and optimizes the network by a dual self-supervised module together with a denoising autoencoder network. The target distribution is calculated from the latent space of the autoencoder, and two KL divergence losses are calculated using Student’s $t$-distribution of embedded spaces of an autoencoder and GCN respectively. Inspired by weighted cell-gene graph introduced by scdeepsort~\cite{shao2021scdeepsort}, graph-sc~\cite{Ciortan2022} models the expression data as a gene-to-cell graph by a graph autoencoder network and then clusters embeddings with K-means or Leiden algorithm. The encoder iterates only over the cell nodes, and the decoder is trained to reconstruct the normalized adjacency matrix. Focusing on cell-cell KNN graph, scTAG~\cite{Yu2022} optimizes a ZINB-based topology adaptive graph convolutional autoencoder with clustering KL divergence loss, ZINB reconstruction loss and graph reconstruction loss. The topology adaptive graph convolutional encoder processes the node features with a polynomial convolution kernel and produces latent embeddings for soft assignment clustering. With a concentration in cell-cell KNN graph, scDSC~\cite{Gan2022} consists of a GNN module and a ZINB-based autoencoder, and adapts a multi-module mutual supervision strategy to achieve end-to-end training. The mutual supervision consists of GNN KL divergence loss, binary cross entropy loss between target distribution and soft label, ZINB reconstruction loss and expression matrix reconstruction loss.


\subsection{Tools and Datasets}
We summarize some popular clustering analysis tools in Table~\ref{clustering_tools_table} and available datasets in Table~\ref{clustering_datasets_table}.

\begin{table}[ht]\normalsize
    \caption{A summary of clustering analysis tools}
    \vskip 1em
    \resizebox{\textwidth}{!}{
        \begin{tabular}{lllll}
            
            \hline
            \textbf{Tools} & \textbf{Algorithm} & \textbf{Description} & \textbf{Language} & \textbf{Availability} \\ \hline
            SC3 & Classical & Consensus clustering of single-cell RNA-seq data. & R &  \href{https://github.com/hemberg-lab/sc3}{{\textcolor{blue}{SC3}}}~\cite{Kiselev2017} \\ \hline
            
            Seurat & Classical & Integrated analysis of multimodal single-cell data. & R & \href{https://github.com/satijalab/seurat}{{\textcolor{blue}{Seurat}}}~\cite{hao2021integrated}  \\ \hline
            
            scedar & Classical & \begin{tabular}[c]{@{}l@{}} Scalable Python package for single-cell RNA-seq data. \end{tabular} & Python & \href{https://pypi.org/project/scedar}{{\textcolor{blue}{scedar}}}~\cite{Zhang2020}\\ \hline
            
            SIMLR & Classical & Multi-kernel learning tool for large-scale genomic analysis & \begin{tabular}[c]{@{}l@{}} R \\ Matlab \end{tabular} & \href{https://github.com/BatzoglouLabSU/SIMLR}{{\textcolor{blue}{SIMLR}}}~\cite{Wang2018} \\ \hline
            
            SPRING & Classical & Interface for visualizing high dimensional expression. & \begin{tabular}[c]{@{}l@{}} Python \\ Matlab \end{tabular} & \href{https://github.com/AllonKleinLab/SPRING}{{\textcolor{blue}{SPRING}}}~\cite{Weinreb2018}\\ \hline
            
            ASAP & Classical & \begin{tabular}[c]{@{}l@{}} Platform for analysis and visualization \\ of single-cell RNA-seq data. \end{tabular} & \begin{tabular}[c]{@{}l@{}} R \\ Python \end{tabular} &  \href{https://github.com/DeplanckeLab/ASAP}{{\textcolor{blue}{ASAP}}}~\cite{Gardeux2017} \\ \hline
            
            RaceID & Classical &  \begin{tabular}[c]{@{}l@{}} Branching point detection in single-cell data \\ by K-branches clustering. \end{tabular} & R & \href{https://github.com/theislab/kbranches}{{\textcolor{blue}{RaceID}}}~\cite{Chlis2017}  \\ \hline
            
            CIDR & Classical & \begin{tabular}[c]{@{}l@{}} Fast clustering through imputation \\ for single-cell RNA-seq data. \end{tabular} & R & \href{https://github.com/VCCRI/CIDR}{{\textcolor{blue}{CIDR}}}~\cite{Lin2017}\\ \hline
            
            SOUP & Classical & Semisoft clustering of single-cell data. & R &  \href{https://github.com/lingxuez/SOUPR}{{\textcolor{blue}{SOUP}}}~\cite{Zhu2019} \\ \hline
            
            scRNA & Classical & \begin{tabular}[c]{@{}l@{}} Clustering by transfer learning on single-cell \\ RNA-seq data. \end{tabular} & Python &  \href{https://github.com/nicococo/scRNA}{{\textcolor{blue}{scRNA}}}~\cite{Mieth2019} \\ \hline
            
            scziDesk & AutoEncoder & \begin{tabular}[c]{@{}l@{}} Deep soft K-means clustering with self-training for \\ single-cell RNA sequence data. \end{tabular} & Python &  \href{https://github.com/xuebaliang/scziDesk}{{\textcolor{blue}{scziDesk}}}~\cite{Chen2020} \\ \hline
            
            scVI & AutoEncoder & Deep generative modeling for single-cell transcriptomics. & Python & \href{https://github.com/scverse/scvi-tools}{{\textcolor{blue}{scVI}}}~\cite{Lopez2018} \\ \hline
            
            DESC & AutoEncoder & \begin{tabular}[c]{@{}l@{}} Clustering and batch effect removal by autoencoder. \end{tabular} & Python & \href{https://eleozzr.github.io/desc/}{{\textcolor{blue}{DESC}}}~\cite{Li2020} \\ \hline
            
            scDHA & AutoEncoder & \begin{tabular}[c]{@{}l@{}} Hierarchical autoencoder for single-cell data analysis. \end{tabular} & Python & \href{https://github.com/duct317/scDHA}{{\textcolor{blue}{scDHA}}}~\cite{Tran2021} \\ \hline
            
            scDeepCluster & AutoEncoder & \begin{tabular}[c]{@{}l@{}} Soft clustering by autoencoder on single-cell \\ RNA-seq data. \end{tabular} & Python & \begin{tabular}[c]{@{}l@{}} \href{https://github.com/ttgump/scDeepCluster}{{\textcolor{blue}{scDeepCluster}}}~\cite{Tian2019} \\ \href{https://github.com/OmicsML/dance}{{\textcolor{blue}{DANCE}}}~\cite{ link2dance} \end{tabular} \\ \hline
            
            scDCC & AutoEncoder & \begin{tabular}[c]{@{}l@{}} Autoencoder-based deep embedding for constrained \\ clustering analysis of single cell RNA-seq data. \end{tabular} & Python & \begin{tabular}[c]{@{}l@{}} \href{https://github.com/ttgump/scDCC}{{\textcolor{blue}{scDCC}}}~\cite{Tian2021} \\ \href{https://github.com/OmicsML/dance}{{\textcolor{blue}{DANCE}}}~\cite{ link2dance} \end{tabular} \\ \hline
            
            GraphSCC & \begin{tabular}[c]{@{}l@{}} GNN \\ AutoEncoder \end{tabular} & \begin{tabular}[c]{@{}l@{}} Graph convolutional network for clustering \\ single-cell RNA-seq data. \end{tabular} & Python & \href{https://github.com/biomed-AI/GraphSCC}{{\textcolor{blue}{GraphSCC}}}~\cite{Zeng2020} \\ \hline
            
            scGNN & \begin{tabular}[c]{@{}l@{}} GNN \\ AutoEncoder \end{tabular} & \begin{tabular}[c]{@{}l@{}} Graph neural network framework for \\ imputation and clustering. \end{tabular} & Python & \begin{tabular}[c]{@{}l@{}} \href{https://github.com/juexinwang/scGNN}{{\textcolor{blue}{scGNN}}}~\cite{wang2021scgnn} \\  \href{https://github.com/OmicsML/dance}{{\textcolor{blue}{DANCE}}}~\cite{ link2dance} \end{tabular} \\ \hline
            
            graph-sc & \begin{tabular}[c]{@{}l@{}} GNN \\ AutoEncoder \end{tabular} & GNN-based embedding for clustering scRNA-seq data. & Python & \begin{tabular}[c]{@{}l@{}}  \href{https://github.com/ciortanmadalina/graph-sc}{{\textcolor{blue}{graph-sc}}}~\cite{Ciortan2022} \\ \href{https://github.com/OmicsML/dance}{{\textcolor{blue}{DANCE}}}~\cite{ link2dance} \end{tabular} \\ \hline
            
            scTAG & \begin{tabular}[c]{@{}l@{}} GNN \\ AutoEncoder \end{tabular} & \begin{tabular}[c]{@{}l@{}} ZINB-based graph embedding autoencoder for \\ single-cell RNA-seq interpretations. \end{tabular} & Python & \begin{tabular}[c]{@{}l@{}} \href{https://github.com/Philyzh8/scTAG}{{\textcolor{blue}{scTAG}}}~\cite{Yu2022} \\ \href{https://github.com/OmicsML/dance}{{\textcolor{blue}{DANCE}}}~\cite{ link2dance} \end{tabular} \\ \hline
            
            scDSC & \begin{tabular}[c]{@{}l@{}} GNN \\ AutoEncoder \end{tabular} & \begin{tabular}[c]{@{}l@{}} Deep structural clustering for single-cell RNA-seq data \\ jointly through autoencoder and graph neural network. \end{tabular} & Python & \begin{tabular}[c]{@{}l@{}} \href{https://github.com/DHUDBlab/scDSC}{{\textcolor{blue}{scDSC}}}~\cite{Gan2022} \\  \href{https://github.com/OmicsML/dance}{{\textcolor{blue}{DANCE}}}~\cite{ link2dance} \end{tabular} \\ \hline
        \end{tabular}
    }
    \label{clustering_tools_table}
\end{table}

\begin{table}[ht]\normalsize
    \caption{A summary of clustering datasets}
    \centering
    \vskip 1em
    \resizebox{12cm}{!}{
        \begin{tabular}{llllll}
            \hline
            \textbf{Dataset} & \textbf{Species} & \textbf{Tissue} & \textbf{\begin{tabular}[c]{@{}l@{}}Dataset\\ Dimensions\end{tabular}} & \textbf{Protocol} & \textbf{Availability} \\ \hline
            
            Tabula Muris & Mouse & 20 tissues & $53,760$ cells & \begin{tabular}[c]{@{}l@{}}SMART-Seq2\\ 10x Genomics\end{tabular} & \href{https://tabula-muris.ds.czbiohub.org/}{{\textcolor{blue}{Tabula Muris}}}~\cite{Tabula2018} \\ \hline
            
            10X PBMC 4K & Human & PBMC & \begin{tabular}[c]{@{}l@{}}$4,271$ cells\\ $16,653$ genes \end{tabular} & 10x Genomics & \href{https://support.10xgenomics.com/single-cell-gene-expression/datasets/2.1.0/pbmc4k}{{\textcolor{blue}{10X PBMC 4K}}}~\cite{Zheng2017} \\ \hline
            
            Mouse Bladder Cells & Mouse & Bladder & \begin{tabular}[c]{@{}l@{}}$2,746$ cells\\ $20,670$ genes \end{tabular} & Microwell-seq & \href{https://figshare.com/s/865e694ad06d5857db4b}{{\textcolor{blue}{Mouse Bladder Cells}}}~\cite{Han2018} \\ \hline
            
            Worm Neuron Cells & Worm & Nerve & \begin{tabular}[c]{@{}l@{}}$4,186$ cells\\ $13,488$ genes \end{tabular} & sci-RNA-seq &  \href{http://atlas.gs.washington.edu/worm-rna/docs/}{{\textcolor{blue}{Worm Neuron Cells}}}~\cite{Cao2017} \\ \hline
            
            Human Kidney Cells & Human & Kidney & \begin{tabular}[c]{@{}l@{}}$5,685$ cells\\ $33,658$ genes \end{tabular} & RNA-seq & \href{https://github.com/xuebaliang/scziDesk/tree/master/dataset/Young}{{\textcolor{blue}{Human Kidney Cells}}}~\cite{Chen2020} \\ \hline
            
            \begin{tabular}[c]{@{}l@{}}Mouse Embryonic\\ Stem Cells \end{tabular} & Mouse & Embryo & \begin{tabular}[c]{@{}l@{}}$2,717$ cells\\ $24,175$ genes \end{tabular} & Droplet Barcoding & \href{https://www.ncbi.nlm.nih.gov/geo/query/acc.cgi?acc=GSE65525}{{\textcolor{blue}{Mouse ES Cells}}}~\cite{Klein2015} \\ \hline
            
            Adam's GSE94333 & Mouse & Kidney & \begin{tabular}[c]{@{}l@{}}$3,660$ cells\\ $23,797$ genes \end{tabular} & Drop-seq & \href{https://www.ncbi.nlm.nih.gov/geo/query/acc.cgi?acc=GSE94333}{{\textcolor{blue}{Adam}}}~\cite{Adam2017} \\ \hline
            
            Muraro's GSE85241 & Human & Pancreas & \begin{tabular}[c]{@{}l@{}}$2,122$ cells\\ $19,046$ genes \end{tabular} & CEL-seq2 & \href{https://www.ncbi.nlm.nih.gov/geo/query/acc.cgi?acc=GSE85241}{{\textcolor{blue}{Muraro}}}~\cite{Muraro2016} \\ \hline
            
            Romanov's GSE74672 & Mouse & Hypothalamus & \begin{tabular}[c]{@{}l@{}}$2,881$ cells\\ $21,143$ genes \end{tabular} & RNA-seq & \href{https://www.ncbi.nlm.nih.gov/geo/query/acc.cgi?acc=GSE74672}{{\textcolor{blue}{Romanov}}}~\cite{Romanov2017} \\ \hline
        \end{tabular}
    }
    \label{clustering_datasets_table}
\end{table}

\section{Spatial Domain}\label{domain}

Spatial domain aims to partition spatial data into a series of meaningful clusters. Each identified cluster from this analysis is considered as a spatial domain.
Spots in the same spatial domain are similar with each other and coherent in gene expression and histology, and are dissimilar to those in different spatial domains\cite{hu2021spagcn}. 
Hidden spatial domains to be identified cannot be directly observed from gene expression or other data. A lot of methods study how to infer latent variables from those gene expression along with other types of data to characterize the hidden spatial domains.
The identification of spatial domain plays a crucial role in medical analysis.
Histological staining of cancer tissue slides can serve as a stimulating example \cite{kather2019predicting, majeed2019quantitative}. Due to the varying affinities of staining agents, cancer areas and normal tissues can be discriminated visually. 
This enables pathologists to grade and stage individual slides of cancer tissue depending on the size and location of malignant spots.
The inputs of the task consist of three components: 1) gene expression at spot or cell level, 2) spot coordinates to indicate spatial information of each spot, and 3) histology image. The goal of the task is to cluster or group spots based on the input features of each spot. In order to evaluate the effectivfeness of various spatial domain techniques, ARI\cite{yeung2001details} is used to determine how well an algorithm's projected clustering labels match the actual labels.

\subsection{Traditional Methods}
One of broadly used traditional methods is based on the Louvain clustering method \cite{blondelvd2008fast}. It comes from the community detection family to extract community structure from large networks.
The normal scRNA-seq data clustering methods\cite{blondelvd2008fast, kanungo2002efficient}, on the other hand, tend to focus on gene expression levels, whereas the spatial domain requires taking into account both spatial information and morphological image aspects.
stLearn \cite{pham2020stlearn} first develops a distance measure utilizing morphological similarity and neighborhood smoothing for gene expression normalisation. The normalized data is then utilized to identify clusters that represent the transcriptional patterns of particular cell types and phenotypes. 
If cells are spatially separated, clusters are further sub-clustering.
Apart from these, BayesSpace \cite{zhao2021spatial} and a hidden Markov random field method (HMRF) \cite{zhu2018identification} together assume that latent variables can characterize hidden spatial domains from gene expression data, and the latent variables can't be seen in the wild, but inferred from gene expression data.
BayesSpace is a Bayesian statistical method and assigns greater weights to geographical areas that are physically closer for resolution enhancement of spatial transcriptomics and for better clustering analysis.
The computationally intensive Markov chain Monte Carlo (MCMC) component in BayesSpace, however, makes it impractical for use with high throughput spatial transcriptomics data.
An hidden Markov random fields (HMRF) based method \cite{zhu2018identification} utilizes hidden Markov random fields (HMRF) to identify spatial domains. 
They first build a neighborhood graph to show how the captured locations are related spatially. The label of the immediate neighbor nodes affects the cell states. 
The HMRF model dictates that there is genetic and spatial cohesion within the clusters.
The graph would be divided into several components via HMRF model, and each component corresponds to a spatial domain.
What's more, spatial clustering can also be performed by the creation of a single cell ''neighborhood" matrix with columns providing neighborhood characteristics, such as the percentage of each cell type or the average gene expression within a radius \cite{schurch2020coordinated, singhal2022banksy}. The matrix can then be directly clustered.

\subsection{Deep Learning Methods}
Graph neural networks (GNNs) based methods recently played a more and more important role in the task of the spatial domain. Even though several GNN models \cite{hu2021spagcn, dong2022deciphering, li2022cell, fu2021unsupervised} have been proposed on this task, their ways to construct graphs are different. 
CCST \cite{li2022cell} constructs a graph only based on spatial information, and each node in the graph represents one spot. Gene expression would be considered as the initial node representation for further feature aggregation. Clustering methods would be finally performed based on newly learned spot representation after feature aggregation.
SpaGCN \cite{hu2021spagcn} first constructs an undirected weighted graph of spots from the spatial transcriptomics data. The edge weight between two nodes would be determined by the distance between them. Then GCN \cite{kipf2016semi} is utilized to aggregate spot gene expression from neighborhoods and update spot gene expression. Finally unsupervised clustering method would be applied to new learned spot gene expression for clustering iteratively.
Different from the aggregation method used in SpaGCN, STAGATE \cite{dong2022deciphering} is a graph attention-based auto-encoder framework to learn low-dimensional latent embeddings with both spatial information and gene expressions.
With the help of The attention mechanism, which is located in the middle layer of the encoder and decoder, it would be easier to learn the edge weights of the networks and employ them to update the spot representation via collective aggregation of information from neighbors.
SEDR\cite{fu2021unsupervised}is a spatial embedding representation of both gene expression and spatial data. SEDR first uses a deep autoencoder to create a low-dimensional latent representation of gene expression, and then it uses a variational graph autoencoder to integrate this representation with the accompanying spatial information.




\subsection{Tools and Datasets}
We summarize all tools or baseline methods for the spatial domain task in Table~\ref{spatial_domain_tool_table} and useful benchmarks in Table~\ref{spatial_domain_dataset_table}.

\begin{table}[h]\normalsize
\caption{A summary of spatial domain tools.}
\vskip 1em
\resizebox{\textwidth}{!}{%
\begin{tabular}{lllll}
\hline
\textbf{Tool} & \textbf{Algorithm}                                         & \textbf{Description}                                                                                                                                                                                                                                         & \textbf{Language} & \textbf{Availability}                                                                                                                    \\ \hline
Louvain      & Classical                                                  & \begin{tabular}[c]{@{}l@{}}
Iterative modularity optimization for \\network community detection
\end{tabular}                                                                                                                     & Python            & \begin{tabular}[c]{@{}l@{}} \href{https://github.com/taynaud/python-louvain}{{\textcolor{blue}{louvain}}}~\cite{blondelvd2008fast}; \href{https://github.com/OmicsML/dance}{{\textcolor{blue}{DANCE}}}~\cite{link2dance}\end{tabular}           \\ \hline
K-means     & Classical                                                  & \begin{tabular}[c]{@{}l@{}}Iterative observation assignment to the cluster\\ with the nearest mean\end{tabular}                                                                                                                                  & Python            & \begin{tabular}[c]{@{}l@{}} \href{https://scikit-learn.org/stable/modules/generated/sklearn.cluster.KMeans.html}{{\textcolor{blue}{K-means}}}~\cite{kanungo2002efficient}\end{tabular}          \\ \hline
stLearn    & Classical                                                  & \begin{tabular}[c]{@{}l@{}}
Integration of spatial location, morphology, and\\ gene expression to determine cell-cell interactions\\, cell types and spatial trajectories
\end{tabular}                                   & Python            & \begin{tabular}[c]{@{}l@{}} \href{https://github.com/BiomedicalMachineLearning/stLearn}{{\textcolor{blue}{stLearn}}}~\cite{pham2020stlearn}; \href{https://github.com/OmicsML/dance}{{\textcolor{blue}{DANCE}}}~\cite{link2dance}\end{tabular} \\ \hline
BayesSpace  & Classical                                                  & \begin{tabular}[c]{@{}l@{}}A 
Low-dimensional representation clustering with \\the gene expression and spatial information
\end{tabular}                                           & R                 & \href{https://github.com/edward130603/BayesSpace}{{\textcolor{blue}{BayesSpace}}}~\cite{zhao2021spatial}                                                                                            \\ \hline
SpatialDE   & Classical                                                  & \begin{tabular}[c]{@{}l@{}}Gene identification using spatial coordinates\end{tabular}                                                                                   & Python            &  \href{https://github.com/Teichlab/SpatialDE}{{\textcolor{blue}{SpatialDE}}}~\cite{svensson2018spatialde}                                                                                           \\ \hline
Giotto    & Classical                                                  & \begin{tabular}[c]{@{}l@{}}
Giotto Analyzer and Viewer to process, analyze,\\ and visualize single-cell spatial expression data
\end{tabular}                                                     & R                 & \href{https://github.com/RubD/Giotto}{{\textcolor{blue}{Giotto}}}~\cite{dries2021giotto}                                                                                                       \\ \hline
SpaGCN     & \begin{tabular}[c]{@{}l@{}}GNN \\ AutoEncoder\end{tabular} & \begin{tabular}[c]{@{}l@{}}
A GCN-based method via integrating gene \\expression and histology to find spatial domains \\and variable genes
\end{tabular}                                                                   & Python            & \begin{tabular}[c]{@{}l@{}} \href{https://github.com/jianhuupenn/SpaGCN}{{\textcolor{blue}{SpaGCN}}}~\cite{hu2021spagcn}; \href{https://github.com/OmicsML/dance}{{\textcolor{blue}{DANCE}}}~\cite{link2dance}\end{tabular}               \\ \hline
STAGATE     & \begin{tabular}[c]{@{}l@{}}GNN \\ AutoEncoder\end{tabular} & \begin{tabular}[c]{@{}l@{}}
A graph attention auto-encoder to learn low-dimensional\\ latent embeddings from gene expression and spatial\\ information
\end{tabular}                                                                         & Python            & \begin{tabular}[c]{@{}l@{}} \href{https://github.com/zhanglabtools/STAGATE}{{\textcolor{blue}{STAGATE}}}~\cite{dong2022deciphering}; \href{https://github.com/OmicsML/dance}{{\textcolor{blue}{DANCE}}}~\cite{link2dance}\end{tabular}            \\ \hline
SEDR     & \begin{tabular}[c]{@{}l@{}}GNN\\ AutoEncoder\end{tabular}  & \begin{tabular}[c]{@{}l@{}}
An autoencoder for learning gene representations \\while variational graph autoencoder for learning \\geographical information
\end{tabular}              & Python            & \href{https://github.com/JinmiaoChenLab/SEDR}{{\textcolor{blue}{SEDR}}}~\cite{fu2021unsupervised}                                                                                           \\ \hline
CCST     & GNN                                                        & \begin{tabular}[c]{@{}l@{}}
A GCN-based unsupervised cell clustering method \\to improve ab initio cell grouping and find novel \\sub cell types
\end{tabular} & Python            & \href{https://github.com/xiaoyeye/CCST}{{\textcolor{blue}{CCST}}}~\cite{li2022cell}                                                                                                 \\ \hline
\end{tabular}%
}
\label{spatial_domain_tool_table}
\end{table}

\begin{table}[h]\normalsize
\centering
\caption{A summary of spatial domain datasets.}
\vskip 1em
\resizebox{12cm}{!}{%
\resizebox{\textwidth}{!}{%
\begin{tabular}{llllll}
\hline
\textbf{Dataset}                                                                    & \textbf{Species} & \textbf{Tissue}                                                            & \textbf{\begin{tabular}[c]{@{}l@{}}Dataset\\ Dimensions\end{tabular}}                                                              & \textbf{Protocol}       & \textbf{Availability}                                                                                        \\ \hline
\begin{tabular}[c]{@{}l@{}}Mouse Posterior Brain \\ 10x Visium Data\end{tabular}    & Mouse            & Posterior brain                                                            & \begin{tabular}[c]{@{}l@{}}3,353 spots\\ 31,053 genes\end{tabular}                                                                 & 10X Visium              & \begin{tabular}[c]{@{}l@{}}\href{https://support.10xgenomics.com/spatial-gene-expression/datasets/1.0.0/V1_Mouse_Brain_Sagittal_Posterior}{{\textcolor{blue}{MPB10xV}}}~\cite{link2MPB10xV}\end{tabular} \\ \hline
\begin{tabular}[c]{@{}l@{}}LIBD Human Dorsolateral\\ Prefrontal Cortex\end{tabular} & Human            & \begin{tabular}[c]{@{}l@{}}Dorsolateral\\ prefrontal cortex\end{tabular}   & \begin{tabular}[c]{@{}l@{}}Slice 151673:\\ 3,639 spots\\ 33,538 genes\\ \\ Slice 151507:\\ 4,226 spots\\ 33,538 genes\end{tabular} & 10X Visium              & \href{http://research.libd.org/spatialLIBD/}{{\textcolor{blue}{spatialLIBD}}}~\cite{link2spatialLIBD}                                                                        \\ \hline
\begin{tabular}[c]{@{}l@{}}Human Primary Pancreatic\\ Cancer ST Data\end{tabular}   & Human            & \begin{tabular}[c]{@{}l@{}}Primary pancreatic\\ cancer tissue\end{tabular} & \begin{tabular}[c]{@{}l@{}}224 spots\\ 16,448 genes\end{tabular}                                                                   & Spatial Transcriptomics & \href{https://www.ncbi.nlm.nih.gov/geo/query/acc.cgi?acc=GSE111672}{{\textcolor{blue}{HPPCST}}}~\cite{chen2019high}                                                  \\ \hline
\end{tabular}%
}
}
\label{spatial_domain_dataset_table}
\end{table}

\section{Cell-type Deconvolution}\label{deconvolution}

Cell-type deconvolution is the task of estimating cell-type proportions in mixed-cell RNA sequencing data, such as bulk RNA-seq or spatial transcriptomics data. In particular, a primary goal of spatial transcriptomics is to determine cellular structure across tissues. While single-cell RNA sequencing can help determine cell types within a tissue, it does not spatially tag the cells, and thus cannot provide a mapping of cell types within the tissue. On the other hand, spatial transcriptomics technologies measure gene expression of small (spatially tagged) regions, but most platforms do not have single-cell resolution capabilities, and do not annotate cell types within these regions. This makes it difficult to determine cell-type proportion from the spatial transcriptomics data, which motivates the task of cell-type deconvolution on spatial transcriptomics data. 
The inputs of the task consist of four components: 1) mixed-cell gene expression (to be deconvoluted), 2) reference scRNA-seq, 3) (optional) spot coordinates to indicate the location within the tissue of each spot, and 4) (optional) histology image.

Typically, cell-type deconvolution models are evaluated on datasets with ground truth cell-type proportions using mean squared error (MSE), mean absolute error (MAE), correlation,  cross-entropy and Jensen-Shannon divergence (JSD). In most cases, however, non-simulated datasets do not have ground truth cell-type proportions. In this unsupervised setting, if profiled marker proteins are also provided with the dataset,  one evaluation metric \cite{Danaher2022}  is the correlation between predicted cell-type proportions and the respective marker proteins.

\subsection{Traditional Methods}
Most traditional methods for cell-type deconvolution are based on non-negative linear regression. The most basic method is non-negative least squares (NNLS), where some reference scRNA-seq gene expression is used to create a cell-profile matrix, which is then regressed onto the mixed-cell gene expression. The resulting (non-negative) coefficients are then used as the cell-type proportion predictions. Here, the idea is that the single-cells' expression will aggregate linearly, respective to their proportion in the mixed-cell sample. The cell profile or signature matrix is typically constructed through the median or mean across cells within each cell type of interest. Most other traditional methods build on this key idea. 

NMFreg \cite{Rodriques2019} uses non-negative matrix factorization to construct a basis in the lower dimension gene space, which is then used for the deconvolution of the mixed-cell data via NNLS. Building on NMFReg, SPOTlight \cite{Elosua2021} uses non-negative matrix factorization to produce the cell-topic profile matrix. However, it introduces a spot-topic profile matrix, which is then regressed onto the cell-topic profile matrix, instead of the full mixture/spot data. DWLS \cite{Tsoucas2019} applies a dampened weighting scheme to the standard NNLS framework. Here, each gene's error term is weighted by the squared inverse of the predicted mixed-cell expression level. This is done to reduce bias towards highly expressed genes, or genes that are highly represented across cell types. Extending on DWLS, SpatialDWLS \cite{Yuan2021} first includes the first step of enrichment analysis (using the PAGE method). The fold change for each gene's expression value is computed between each spatial location and mean expression over all spatial locations, which is then standardized and gives an enrichment score. A threshold is then applied to the enrichment scores to determine the subset of cell types to infer to. Altering the classical assumption of an additive error linear model, SpatialDecon \cite{Danaher2022} implements a non-negative linear regression-based method that assumes a log-normal multiplicative error model. 

The first method to incorporate the spatial information of spatial transcriptomics bulk expression regions is Conditional AutoRegressive Model-based Deconvolution (CARD) \cite{Ma2022}. As with most classical methods, CARD assumes a linear model between the mixed-cell expression matrix and a cell-profile matrix, constructed from a single-cell reference. However, spatial information is incorporated through a conditional autoregressive (CAR) assumption on the model, which applies an intrinsic prior on the cell-type proportions by modeling proportions in each location as a weighted combination of proportions in all other locations. The weights in CARD are given by a Gaussian kernel function. Then a maximum-likelihood framework is adopted to optimize the model.

RCTD \cite{cableRobustDecompositionCell2022} and cell2location \cite{kleshchevnikovCell2locationMapsFinegrained2022} are two methods that statistically account for platform effects and other sources of gene expression variations to model cell type compositions in spatial transcriptomic data. As scRNA-seq data are used as references for cell type composition estimation of spatial transcriptomic data, the differences in the library preparation of the two sequencing technologies could lead to systematic bias that confounds cell type deconvolution results, termed as platform effects \cite{cableRobustDecompositionCell2022}. RCTD applies a Poisson distribution to model observed gene counts, and its rate parameter accounts for cell type expression profiles, spot-specific effect, and gene-specific platform effect that follows a normal distribution on the log scale. These parameters are then estimated via MLE. In particular, RCTD assumes that in many cases there are only a few cell types co-present in a spot, and thus supports a 'doublet mode' that restricts the number of cell types present in a given spot to mitigate overfitting. By comparison, cell2location takes a Bayesian approach and models spatial expression counts using a Negative Binomial (NB) distribution. Similar to RCTD, the mean parameter in the NB distribution used by cell2location incorporates gene-specific platform effects, an additive shift, and cell type profile contributions, adjusted by spot-wise sensitivity. All parameters, including cell type proportions, are endowed with informative priors. In particular, cell2location acknowledges that cell type compositions are correlated across locations. Thus, it factorizes cell type proportions into the contribution of multiple latent cell type groups to pool information across spots and thus enhance statistical strength.

\subsection{Deep Learning Methods} Some recent developments in cell-type deconvolution have applied deep learning-based methods. These approaches typically apply a transfer learning scheme wherein they first simulate mixed-cell data from scRNA reference data, and use a common network to predict the cell-type composition of both the simulated and real mixed-cell data. A notable feature of the deep learning-based methods is that they model the cell-type compositions directly, i.e. the model objective is on the predicted cell-type compositions. This contrasts with most classical methods, where the predicted cell-type proportions are the optimal parameters/coefficients from some regression model. 

One of the early deep learning approaches to the cell-type deconvolution problem is the Scaden \cite{Menden2020} method. First, scRNA reference data is randomly sampled from scRNA reference data to generate simulated mixed-cell samples. A fully-connected network is then trained to predict the true cell-type compositions of the simulated mixed-cell data, with cross-entropy loss function. This trained model is then applied to the real mixed-cell data to get cell-type compositions. Building on this approach, DSTG \cite{Song2021} is a Graphical Neural Network (GNN) based method, modeling similarities in expression between different mixed-cell samples. DSTG also generates simulated mixed-cell data from scRNA reference data. Different from Scaden, before training the GNN on this data, DSTG first aligns the simulated and real mixed-cell data in a lower dimensional gene space using Canonical Correlation Analysis (CCA). These embeddings are then used to construct a graph by considering Mutual Nearest Neighbors (MNN) as adjacent in the graph. Here, adjacencies can be between simulated-to-real and real-to-real samples.

\subsection{Tools and Datasets}
We summarize representative tools or baseline methods for the cell-type deconvolution task in Table~\ref{celltype_deconvo_tool_table} and useful benchmarks in Table~\ref{celltype_deconvo_dataset_table}.

\begin{table}[ht]\normalsize
\caption{A summary of tools for cell-type deconvolution.}
\vskip 1em
\resizebox{\textwidth}{!}{%
\begin{tabular}{lllll}
\hline
\textbf{Tool} & \textbf{Algorithm}                                         & \textbf{Description}                                                                                                                                                                                                                                         & \textbf{Language} & \textbf{Availability}                                                                                                                    \\ \hline
NMFReg      & Classical                                                  & \begin{tabular}[c]{@{}l@{}}A non-negative matrix factorization\\ of an annotated scRNA reference matrix \end{tabular}                                                                                                                     & Matlab, Python            & \begin{tabular}[c]{@{}l@{}} \href{https://github.com/broadchenf/Slideseq}{{\textcolor{blue}{NMFReg}}}~\cite{Rodriques2019}
\\ \href{https://github.com/tudaga/NMFreg\_tutorial}{{\textcolor{blue}{NMFReg-Python}}} \end{tabular}          
\\ \hline
SPOTlight     & Classical                                                  & \begin{tabular}[c]{@{}l@{}} Extension of NMFReg, with non-negative matrix \\factorization applied to both the scRNA reference \\ matrix, and the mixed-cell expression matrix  \end{tabular}                                                                                                                                  & R, Python       & \begin{tabular}[c]{@{}l@{}} 
\href{https://github.com/MarcElosua/SPOTlight}{{\textcolor{blue}{SPOTlight}}}~\cite{Elosua2021}
\\ \href{https://github.com/OmicsML/dance}{{\textcolor{blue}{Dance}}}~\cite{link2dance}
\end{tabular}          \\ \hline
DWLS     & Classical                                                  & \begin{tabular}[c]{@{}l@{}}Weighted NNLS; dampened weighting is applied to genes \end{tabular}                                   & R            & \begin{tabular}[c]{@{}l@{}}  
\href{https://github.com/dtsoucas/DWLS}{{\textcolor{blue}{DWLS}}}~\cite{Tsoucas2019}
\end{tabular} \\ \hline
SpatialDWLS   & Classical                                                  & \begin{tabular}[c]{@{}l@{}}A subset of cell-types chosen via PAGE enrichment analysis\end{tabular}                                           & R                 & 
\href{https://github.com/rdong08/spatialDWLS_dataset }{{\textcolor{blue}{SpatialDWLS}}}~\cite{Yuan2021}
\\ \hline
SpatialDecon   & Classical                                                  & \begin{tabular}[c]{@{}l@{}}A multiplicative log-normal error model \end{tabular}                                                                                   & R, Python            &\begin{tabular}[c]{@{}l@{}}  
\href{https://github.com/Nanostring-Biostats/SpatialDecon }{{\textcolor{blue}{SpatialDecon}}}~\cite{Danaher2022}
 \\  \href{https://github.com/OmicsML/dance}{{\textcolor{blue}{Dance}}}~\cite{link2dance}\end{tabular}                                                                                        \\ \hline

cell2location    & Variational Inference                                                  & \begin{tabular}[c]{@{}l@{}}Bayesian hierarchical model of spatial expression counts \\with a spatially informed prior on cell-type compositions  \end{tabular}                                                                                   &  Python            &\begin{tabular}[c]{@{}l@{}} 
\href{https://github.com/BayraktarLab/cell2location}{{\textcolor{blue}{cell2location}}}~\cite{kleshchevnikovCell2locationMapsFinegrained2022}\end{tabular}                                                                                        \\ \hline

CARD     & Classical                                                  & \begin{tabular}[c]{@{}l@{}}Conditional autoregressive based model that incorporates \\ spatial correlation of cell-type compostion \end{tabular}                                                     & R, Python                &\begin{tabular}[c]{@{}l@{}} 
\href{https://github.com/YingMa0107/CARD }{{\textcolor{blue}{CARD}}}~\cite{Ma2022}
 \\\href{https://github.com/OmicsML/dance}{{\textcolor{blue}{Dance}}}~\cite{link2dance}\end{tabular}                                                                                                       \\ \hline
RNA-Sieve      & Classical                                                  & \begin{tabular}[c]{@{}l@{}} A likelihood based inference model that estimates \\ cell-type proportion through a maximum-likelihood method \end{tabular}                                                     & Python                &\begin{tabular}[c]{@{}l@{}}
\href{https://github.com/songlab-cal/rna-sieve}{{\textcolor{blue}{RNA-Sieve}}}~\cite{Pham2021}

\end{tabular}                                                                                                       \\ \hline
Scaden      & \begin{tabular}[c]{@{}l@{}}GNN \end{tabular} & \begin{tabular}[c]{@{}l@{}}A fully-connected network that is trained on simulated \\mixed-cell data, and used to predict cell-type compositions\\ of real mixed-cell data\end{tabular}                                                                   & Python            & \begin{tabular}[c]{@{}l@{}}
\href{https://github.com/KevinMenden/scaden }{{\textcolor{blue}{Scaden}}}~\cite{Menden2020}

\end{tabular}               \\ \hline
DSTG      & \begin{tabular}[c]{@{}l@{}}GNN \end{tabular} & \begin{tabular}[c]{@{}l@{}}A graph convolutional network whose graph is constructed on \\ Mutual Nearest Neighbors of low-dimensional embeddings of \\simulated and real mixed-cell data \end{tabular}                                                                   & R, Python            & \begin{tabular}[c]{@{}l@{}}
\href{https://github.com/Su-informatics-lab/DSTG}{{\textcolor{blue}{DSTG}}}~\cite{Song2021} \\ 

\href{https://github.com/OmicsML/dance}{{\textcolor{blue}{Dance}}}~\cite{link2dance}
\end{tabular}               \\ \hline
\end{tabular}%
}
\label{celltype_deconvo_tool_table}
\end{table}

\begin{table}[ht]\normalsize
\caption{A summary of datasets for cell-type deconvolution.}
\vskip 1em
\resizebox{\textwidth}{!}{%
\begin{tabular}{llllll}
\hline
\textbf{Dataset}                                                                    & \textbf{Species} & \textbf{Tissue}                                                            & \textbf{\begin{tabular}[c]{@{}l@{}}Dataset\\ Dimensions\end{tabular}}                                                              & \textbf{Protocol}       & \textbf{Availability}                                                                                        \\ \hline
\begin{tabular}[c]{@{}l@{}}Mouse Posterior Brain \\ 10x Visium Data\end{tabular}    & Mouse            & Posterior brain                                                            & \begin{tabular}[c]{@{}l@{}}3,353 spots\\ 31,053 genes\end{tabular}                                                                 & 10X Visium              & \begin{tabular}[c]{@{}l@{}}
\href{https://support.10xgenomics.com/spatial-gene-expression/datasets/1.0.0/V1_Mouse_Brain_Sagittal_Posterior}{{\textcolor{blue}{MPB10xV}}}~\cite{link2MPB10xV}

\end{tabular} \\ \hline
\begin{tabular}[c]{@{}l@{}}Mouse Olfactory Bulb\end{tabular} & Mouse            & \begin{tabular}[c]{@{}l@{}} Olfactory bulb\end{tabular}   & \begin{tabular}[c]{@{}l@{}}1,185 spots\\ 11,176 genes\end{tabular} & 10X Visium              & \begin{tabular}[c]{@{}l@{}}
\href{https://www.10xgenomics.com/resources/datasets/adult-mouse-olfactory-bulb-1-standard-1}{{\textcolor{blue}{MOB10xV}}}~\cite{link2MOB10xV}

  \end{tabular}                                                                      \\ \hline
\begin{tabular}[c]{@{}l@{}}HEK293T and CCRF-CEM cell line mixture\end{tabular}   & Human            & \begin{tabular}[c]{@{}l@{}} \end{tabular} & \begin{tabular}[c]{@{}l@{}}56 mixtures\\ 1,414  genes\end{tabular}                                                                   & NanoString GeoMx & \begin{tabular}[c]{@{}l@{}}
\href{https://www.ncbi.nlm.nih.gov/geo/query/acc.cgi?acc=GSE174746 }{{\textcolor{blue}{CelllineGeoMx}}}~\cite{link2CellieGeoMx}
    \end{tabular}                    \\ \hline                        

\begin{tabular}[c]{@{}l@{}}Human PDAC \end{tabular}    & Human            & Pancreas                                                            & \begin{tabular}[c]{@{}l@{}}1,819 spots\\ 19,738 genes\end{tabular}                                                                 & Spatial Transcriptomics              & \begin{tabular}[c]{@{}l@{}}
\href{https://www.ncbi.nlm.nih.gov/geo/query/acc.cgi?acc=GSE111672}{{\textcolor{blue}{HPdacST}}}~\cite{link2HPdacST}
\end{tabular} \\ \hline
\end{tabular}%
}
\label{celltype_deconvo_dataset_table}
\end{table}

\section{Cell Segmentation}\label{segmentation}
Single-cell segmentation aims to achieve cell separation by generating pixel-wise annotations for individual cells in image-based single-cell profiling. Unlike scRNA-seq, tissue imaging obtained from microscopy and stained by various nuclear or membrane markers captures intact specimen imaging of cells and preserves valuable spatial information for transcriptionally distinct cell types and cell states from tissue images at single-cell resolution.

Accurate segmentation of single cells from these microscopy images helps to identify cell morphology and distribution in their spatial context, which benefits cancer diagnosis and clinical treatment. More important, single-cell segmentation not only defines whether a pixel is inside or outside of a cell but also assigns each pixel to each cell. This allows accurate localization of every single cell in images and offers significant implications for the feasibility of cellular-level analysis for downstream tasks like cell-type identification, and RNA/protein expression quantification \cite{Stringer2021, Greenwald2022}. For RNA hybridization-based approaches without measuring the belongings of RNA to which cells, single-cell segmentation is an essential prerequisite for such mRNA assignment. Thus, the RNA detection can be converted into spatial single-cell data \cite{littman2021}.

Due to the variety of cell types, microscopy instruments, and a large number of cells, manual annotation of the whole cell boundaries in microscopy images is highly time-consuming and labor-intensive. It becomes much more challenging to separate adjacent cells when tissue cells are tightly packed with low signal-to-noise ratio staining. Therefore, different imaging-based methods have been widely proposed to realize accurate automatic or semi-automatic segmentation of single cells. Additional cellular or gene information indicating intracellular variability has also been integrated to improve single-cell segmentation performance. 
In general, the inputs of the task are microscopy images with target cells, and the goal is to obtain an intact whole-cell boundary for each cell. To assess the segmentation performance, a one-to-one matching process will be firstly performed between predicted cells and ground truth cells at different matching thresholds, i.e., standard intersection over union metric (IoU). Several commonly used evaluation metrics such as precision, recall, F1 score, Jaccard index, and standard average precision metric (AP) can be further applied to quantify the segmentation results at the single-cell level.

\subsection{Traditional Methods}

Early works typically adopt ideas from the traditional computer vision area. Unlike other segmentation tasks, densely clustered cells suffer from overlapping cell contours. To overcome this problem, Yan et al. \cite{Yan2008} develop an automatic interaction model to segment cells from high-throughput RNAi fluorescent cellular images. They first modify the watershed algorithm \cite{vincent1991} to segment cell nuclei regions and then adopt a multiphase level set method to predict the boundaries of neighboring cells further. 
To perform cellular analysis on epithelial tissues, a similar two-stage pipeline, MxIF \cite{Gerdes2013}, is completed to generate segmentation masks for fluorescent images. Specifically, a wavelet-based detection algorithm is utilized to segment the nucleus, followed by a variation of the probabilistic method \cite{can2009unified} for whole cell segmentation. 
The above methods, based on pre-selected nuclei/membrane fluorescent markers, fail to use multiple, automatically selected markers. Therefore, built on the watershed method, Schüffler et al. \cite{Peter2015} design an automatic marker selection manner with Spearman’s rank correlation to capture multiplexed imaging information and reveal the most useful channels for cell segmentation.
Some other works incorporate cell shape as prior knowledge to enhance segmentation performance. For example, learning from user annotations of the given training data, SVSS \cite{Pang2015} serves as a tissue statistic shape model to enhance segmentation performance and quantify the quality of cell contours. 
Matisse \cite{Krijgsman2021} is a unified workflow based on existing open-source technologies. It combines multiple marker information from imaging mass cytometry (IMC) data and fluorescence microscopy data to boost segmentation performance. They use Ilastikuse tool \cite{Ilastik2011} to generate probability maps, which are then fed into a CellProfiler software \cite{CellProfiler} to acquire the final segmentation maps.

\subsection{Deep Learning Methods}
 Deep learning-based methods could be of great use in assisting single-cell segmentation, as they have shown great success for natural and other medical imaging applications \cite{long2015fully, redmon2016you, chen2016dcan, he2016deep, kamnitsas2017efficient, yadav2019deep}. 
 Existing deep learning-based methods can be categorized into two groups according to their utilized data information and input modality. 
 One is driven by cell morphological characteristics shown on images, which highly depend on the imaging quality and cell distributions. 
 CNN is a fundamental deep neural network for the computer vision task of image segmentation. 
 One of the most popular CNNs, U-Net \cite{ronneberger2015}, has shown significant success in many medical image segmentation tasks \cite{ li2018h, oktay2017anatomically, oktay2018attention, wang2020double, heller2021state, ma2021toward, wang2021acn, xu2021noisy}. It adopts a U-shape encoder-decoder architecture where the encoder serves as a contraction to capture semantically image contextual features. The decoder is a symmetric expanding path recovering spatial information. The two paths are connected using skip connections to recombine with essential high-resolution features from the encoder. 
 To perform a cell segmentation task, the inputs to the U-Net are multi-channel images with different channels corresponding to nuclei or cytoplasmic. After passing through the encoder and decoder, the model will generate a predicted probability map indicating that a pixel is inside or outside a cell.
 Since U-Net performs a pixel-based semantic segmentation and considers a cell segmentation task as a binary classification of pixels into cell foreground and background, it is prone to fail in overlapping cells, which require to be differentiated as individual object instances. Generally, traditional algorithms such as watershed will be combined to locate boundaries and separate touching cells. However, they show little improvement when cells are tightly clustered and stacked with similar brightness contrast. Although a commonly used CNN architecture for instance segmentation, Mask RCNN \cite{he2017mask}, has been dominant in generating a high-quality segmentation mask for each instance in natural imaging tasks, it is ill-equipped to handle high cell density. Recent works carefully incorporate another boundary detection task \cite{chen2020boundary} or introduce contour-aware modules to the decoder path \cite{chen2016dcan, zhou2019cia, huang20212} for edge refinement. However, they are limited to cell nuclei, and thus giving a clear distinction between highly-packed whole cells is still challenging in single-cell segmentation.
 
 Transforming such a pure classification task into a regression model has been proven to be a feasible solution. Instead of directly categorizing each pixel into either foreground or background classes, a regression model allows us to predict a continuous variable (such as gradient and distance value) for each pixel.
 Cellpose \cite{Stringer2021} designs an auxiliary predictive U-Net head to predict two spatial flows for cells, i.e., horizontal and vertical flows, generated by spatial gradients of an energy function. The vector gradient flows lead all pixels towards the horizontal/vertical center of the cells they belong to and these reversible flow maps are powerful representations to reconstruct cells with complex shapes and densities at test time. This framework is further extended to Cellpose 2.0 \cite{Stringer2022} that allows users to adapt their pre-trained segmentation models to their new datasets. 
 Van Valen et al. \cite{van2016} build a deep convolutional neural network to segment single cells in live-cell imaging, which indicates multiple cell types across the domains of life. Another regression model DeepDistance \cite{Koyuncu2020} is trained to concurrently learn two distance metrics for each pixel in live cell images. This multi-task strategy turns out to be effective in capturing shared feature representations to enhance a more accurate cell localization.

To deal with multiple imaging modalities and cell types, a large cell segmentation dataset, TissueNet \cite{Greenwald2022}, is constructed from nine organs and six imaging platforms. Specifically, based on this comprehensive labeled training dataset, they apply a more specified and more robust structure, Mesmer, with two semantic segmentation heads to predict the boundary of cells and two heads to predict the distance of each pixel within a cell to the cell centroid. Their model is built on a ResNet50 \cite{he2016deep} backbone coupled to a Feature Pyramid Network \cite{lin2017feature} and trained in an end-to-end manner. Each cell's obtained centroid and boundary are utilized to generate a final instance segmentation mask using a watershed algorithm.
However, these algorithms are still insufficient for highly multiplexed imaging platforms. MIRIAM \cite{mckinley2022} is another pipeline that can deal with any multiplexed imaging data with various marker compositions. Initial segmentation masks are firstly generated by a random forest pixel classification method and then a seed-based watershed algorithm is applied. To deal with cells with irregular cell shapes and incomplete membrane markers, they extend the cell membranes by ``connecting the dots'' and regulate the cell shapes using an autoencoder neural networking. Their framework is validated to be robust to marker composition and image modality changes.

The above deep learning-based methods in the first category have achieved significant performance, with more accurate and proper whole-cell boundary definition and adjacent single-cell separation. However, these models can only characterize cell morphology from its exhibited appearance in fluorescent images, which is sensitive to cell intensity and staining quality. 

The other category introduces spatial gene expression patterns as extra supervision since the spatial dependency captured by gene information plays a crucial role in defining pixels at the junction of different cell types. Baysor \cite{Petukhov2021} considers the RNA transcriptional composition and optimizes the joint likelihood of transcriptional composition and cell morphology. Specifically, each cell is modeled as a distribution of both spatial position and gene identity of each molecule. The mixture of all the cell distributions is regarded as a Bayesian mixture model (BMM), which is further optimized with Markov random fields (MRFs) before ensuring spatial clustering.
JSTA \cite{littman2021} enhances image-based cell segmentation with the assistance of cell type-specific gene expression. Motivated by the ability to convert raw spatial transcriptomic data into a single cell-level spatial expression map, the authors first train a DNN classifier to assign cell type probabilities to each pixel. These probabilities are utilized to adjust pixels around the cell borders iteratively. In these strategies, RNA information serves as valuable prior knowledge to constrain cell boundaries between neighboring cells.

\subsection{Tools and Datasets}
We summarize representative tools or baseline methods for single-cell segmentation task in Table \ref{cellseg_tool_table} 
and useful benchmarks in Table \ref{cellseg_dataset_table}
\begin{table}[h]\normalsize
\caption{A summary of cell segmentation tools.}
\vskip 1em
\resizebox{\textwidth}{!}{%
\begin{tabular}{lllll}
\hline
\textbf{Tool} & \textbf{Algorithm}                                         & \textbf{Description}                                                                                                                                                                                                                                         & \textbf{Language} & \textbf{Availability}                                                                                                                    \\  \hline

ImageJ   & Classical                                                  & Watershed-based                                                                                                                   & Python, MATLAB etc.           & \begin{tabular}[c]{@{}l@{}}\\ \href{https://github.com/fiji}{{\textcolor{blue}{ImageJ}}}~\cite{vincent1991} \end{tabular}           \\ \hline

Schüffler et al.     & Classical                                                  & \begin{tabular}[c]{@{}l@{}} 
Watershed-based
\end{tabular}    & MATLAB      &  \href{http://www.comp-path.inf.ethz.ch/}{{\textcolor{blue}{Computational Pathology}}}~\cite{Peter2015}        \\ \hline

Ilastik      & Classical                                                  & \begin{tabular}[c]{@{}l@{}} 
Random Forest classifier to perform a semantic segmentation and\\ return a probability map of each class, which can be further \\transformed into individual objects by other methods
\end{tabular}                                                                                                                                  & Python       & \begin{tabular}[c]{@{}l@{}} \href{https://github.com/ilastik/ilastik}{{\textcolor{blue}{Ilastik}}}~\cite{Ilastik2011}
\end{tabular}          \\ \hline

MorphoLibJ     & Classical                                                  & \begin{tabular}[c]{@{}l@{}} 
Watershed + GUI for 2D/3D images of cell tissues
\end{tabular}                                                                                                                                  & Java       & \href{https://github.com/ijpb/MorphoLibJ}{{\textcolor{blue}{MorphoLibJ}}}~\cite{legland2016}\\ \hline

CellProfiler      & Classical                                                  & \begin{tabular}[c]{@{}l@{}} 
Primary objects identification first (often nuclei) and then secondary\\ object identification (cell edges)
\end{tabular}                                                                                                                                  & Python       & \begin{tabular}[c]{@{}l@{}} \href{https://github.com/ijpb/MorphoLibJ}{{\textcolor{blue}{CellProfile}}}~\cite{CellProfiler}\end{tabular}          \\ \hline

MATISSE    & Classical                                                  & \begin{tabular}[c]{@{}l@{}} 
Combination of fluorescence microscopy and multiplex capability of\\ imaging mass cytometry(IMC) into a single workflow, Nuclei and\\ whole cells identification based on probability maps generated by Ilastik
\end{tabular}                  & R & \href{https://github.com/VercoulenLab/MATISSE-Pipeline}{{\textcolor{blue}{MATISSE}}}~\cite{Krijgsman2021} \\ \hline

Oufti  & Classical                                                   & \begin{tabular}[c]{@{}l@{}}  
Laplacian of Gaussian, valley, and cross detection
\end{tabular}                  & MATLAB & \begin{tabular}[c]{@{}l@{}} \href{ http://www.oufti.org/}{{\textcolor{blue}{Oufti}}}~\cite{paintdakhi2016}
 \end{tabular}  \\ \hline

Cellpose   & CNN                                                   & \begin{tabular}[c]{@{}l@{}}  
U-Net with an auxiliary representation
\end{tabular}                  & Python & \begin{tabular}[c]{@{}l@{}} \href{https://github.com/MouseLand/cellpose}{{\textcolor{blue}{Cellpose}}}~\cite{Stringer2021}
\end{tabular}  \\ \hline

Mesmer   & CNN                                                   & \begin{tabular}[c]{@{}l@{}}  
An CNN architecture that generalizes to the full diversity of tissue types
\end{tabular}                  & Python & \begin{tabular}[c]{@{}l@{}} \href{https://github.com/MouseLand/cellpose}{{\textcolor{blue}{Mesmer}}}~\cite{Greenwald2022}
\end{tabular}  \\ \hline

Baysor  & CNN                                                   & \begin{tabular}[c]{@{}l@{}}  
Segmentation performing using only molecule placement data or \\in combination with evidence from auxiliary stains in images
\end{tabular}                  & Python &  \href{https://github.com/kharchenkolab/Baysor}{{\textcolor{blue}{Baysor}}}~\cite{Petukhov2021} \\ \hline

JSTA    & CNN                                                   & \begin{tabular}[c]{@{}l@{}}  
Integration of spatial transcriotomics and scRNAseq information to \\improve segmentation
\end{tabular}                  & Python & \href{https://github.com/wollmanlab/JSTA}{{\textcolor{blue}{JSTA}}}~\cite{littman2021}\\ \hline

MIRIAM  & CNN                                                   & \begin{tabular}[c]{@{}l@{}}  
Initial Cell identification by random forest and cell shapes characterization\\ via an autoencoder
\end{tabular}                  & Python, MATLAB & \href{https://github.com/Coffey-Lab/MIRIAM}{{\textcolor{blue}{MIRIAM}}}~\cite{mckinley2022} \\ \hline
\end{tabular}%
}
\label{cellseg_tool_table}
\end{table}

\begin{table}[h]\normalsize
\caption{A summary of cell segmentation datasets.}
\vskip 1em
\resizebox{\textwidth}{!}{%
\begin{tabular}{llllll}
\hline
\textbf{Dataset}                                                                    & \textbf{Species} & \textbf{Tissue}                                                            & \textbf{\begin{tabular}[c]{@{}l@{}}Dataset\\ Dimensions\end{tabular}}                                                              & \textbf{Protocol}       & \textbf{Availability}                                                                                        \\ \hline
\begin{tabular}[c]{@{}l@{}} ISBI-14 \\ ISBI-15\end{tabular}     & Human            & Cervix                                                        & \begin{tabular}[c]{@{}l@{}}ISBI-14: 16 EDF real cervical cytology images \\ and 945 sythetic images \\ ISBI-15: 17 multi-layer cervical cell volumes\end{tabular} & Microscope   & \begin{tabular}[c]{@{}l@{}} \href{https://cs.adelaide.edu.au/~carneiro/isbi14_challenge/dataset.html}{{\textcolor{blue}{ISBI-14}}}~\cite{link2isbi-14}\\ \href{https://cs.adelaide.edu.au/~zhi/isbi15_challenge/dataset.html}{{\textcolor{blue}{ISBI-15}}}~\cite{link2isbi-15}\end{tabular} \\ \hline

Cellpose & Various species   & Various tissues    & 608 images & Microscope              & \href{https://www.cellpose.org/dataset}{{\textcolor{blue}{Cellpose}}}~\cite{link2Cellpose}                                                                    \\ \hline
TissueNet    & Human            & 9 organs                                                     & \textgreater 1 million cells with nuclear & \begin{tabular}[c]{@{}l@{}} Immersion objective and \\ dual Photometrics Prime  \end{tabular}            & \href{http://netbio.bgu.ac.il/tissuenet}{{\textcolor{blue}{TissueNet}}}~\cite{link2TissueNet} \\ \hline

EVICAN & Human   & Various tissues    & 4600 images, $\sim$26 000 segmented cells & Microscope              & \href{https://edmond.mpdl.mpg.de/dataset.xhtml?persistentId=doi:10.17617/3.AJBV1S}{{\textcolor{blue}{EVICAN}}}~\cite{link2EVICAN} \\ \hline

LiveCell  & Human, Mouse            & Various tissues & 5,239 images, 1.6 million cells &  Phase-contrast microscopy & \href{https://sartorius-research.github.io/LIVECell/}{{\textcolor{blue}{LIVECell}}}~\cite{link2LIVECell} \\ \hline
\end{tabular}%
}
\label{cellseg_dataset_table}
\end{table}


\section{Cell Type Annotation}\label{annotation}

The information with cellular level granularity sc-RNA-seq data allows the identification of cell identity, which can shed light on sample heterogeneity and biological roles of different cell populations in different organs and biological status. Thus, one of the main tasks in scRNA-seq data analysis is cell type annotation. Traditionally, single cells are first clustered and then manually annotated by experts based on genetic profiles \cite{caoComprehensiveSinglecellTranscriptional2017a,fincherCellTypeTranscriptome2018,hanMappingMouseCell2018,thetabulamurisconsortiumSinglecellTranscriptomics202018}. However, such a method faces several drawbacks. Manual annotation results are susceptible to the choice of clustering method and parameters, experts' subjectivity, and the use non-standardized cell type ontologies, which can make them non-comparable across datasets and experiments. Furthermore, this method is also time-consuming and laborious, which makes it less scalable to large datasets \cite{abdelaalComparisonAutomaticCell2019,huangEvaluationCellType2021,pasquiniAutomatedMethodsCell2021}. In response to these challenges, automatic cell type annotation methods have been proposed in recent years. Currently, most cell type annotation algorithms use either marker genes or reference datasets with well-curated cell type labels. As cell type annotation can be treated as a classification task, where each cell is assigned with a label, the performance of these methods can be evaluated using datasets with known cell type labels.

\subsection{Traditional Methods}

Many methods have been developed to identify single cell identities that rely on traditional statistical or machine learning methods. To date, most methods require clustering first and then assign a cell type identity to each cluster. They use either marker genes and/or reference datasets. scCATCH\cite{Liao2020}, CellAssign\cite{zhangProbabilisticCelltypeAssignment2019}, SCINA\cite{zhangSCINASemiSupervisedSubtyping2019}, SCSA\cite{Cao2020}, scSorter\cite{guoScSorterAssigningCells2021}, cellMeSH\cite{maoCellMeSHProbabilisticCelltype2022}, and scType\cite{ianevskiFullyautomatedUltrafastCelltype2022} are based on prior knowledge about cell type specific marker genes. scCATCH, SCSA, and cellMeSH use well-curated cell marker database. scCATCH and SCSA devise scoring models to assign cell types to cell clusters, while cellMeSH applies a probabilistic model to database weighted by the strength of association. However, these methods assign all clusters to some known cell labels and thus could not identify unknown cell types not present in the database, while CellAssign, SCINA, scSorter, and scType support unassigned clusters. scType also devises a scoring metric that is applied to its cell type marker database. CellAssign uses a Bayesian framework where a negative binomial is fitted to the marker gene expression profiles given a cell type. SCINA assumes each marker gene follows a bimodal distribution, where the high mode represents the predicted cell type. scSorter first clusters query cells into known cell types and then separates cells with unknown cell types using information from both marker genes and non-marker genes.\\

Alternatively, many methods utilize reference datasets instead. Among these methods, scmap-cluster\cite{kiselevScmapProjectionSinglecell2018}, SingleR\cite{aranReferencebasedAnalysisLung2019}, CHETAH\cite{dekanterCHETAHSelectiveHierarchical2019}, scMatch\cite{houScMatchSinglecellGene2019}, and CIPR\cite{ekizCIPRWebbasedShiny2020} assign cell types to query clusters based on measures of correlations with reference datasets, such as Pearson correlation, Spearman correlation, or cosine similarity. scmap\cite{kiselevScmapProjectionSinglecell2018} also supports cell type assignment to individual cells. By comparison, a number of methods use supervised models. For example, Garnett\cite{plinerSupervisedClassificationEnables2019} builds a hierarchical tree of cell types based on marker genes and then uses elastic net regression to assign cell types to clusters. SingleCellNet uses the top-scoring pair algorithm \cite{gemanClassifyingGeneExpression2004} to select informative genes, transforms data into a binary form, and train a multi-class random forest \cite{breimanRandomForests2001} classifier using reference datasets. SciBet\cite{liSciBetPortableFast2020} first selects gene features based on statistic differential entropy, fits a multinomial model for gene expressions in each cell type, and then uses a maximum likelihood approach to assign cell types. CellTypist\cite{dominguezcondeCrosstissueImmuneCell2022} uses a logistic regression model with stochastic gradient descent learning for cell type identification using a curated pan-tissue database for immune cell types. By comparison, scPred\cite{alquicira-hernandezScPredAccurateSupervised2019} performs cell type identification for individual cells by first selecting informative PCs from  singular value decomposition and training a support vector machine (SVM) classifier. A special case is scType\cite{ianevskiFullyautomatedUltrafastCelltype2022}, which is unsupervised since it uses neither marker genes nor reference datasets.\\

\subsection{Deep Learning Methods}

Deep learning methods have been applied to address the cell type annotation task in recent years. ACTINN\cite{maACTINNAutomatedIdentification2019} trains a neural network with three hidden layers using reference datasets to predict cell type for each cell. Similarly, SuperCT\cite{xieSuperCTSupervisedlearningFramework2019} trains an artificial neural network model with two hidden layers each followed by a dropout layer against over-fitting using well-curated database including the Mouse-Cell-Atlas. Its inputs are binarized gene expressions, i.e, an entry in the transformed expression matrix is assigned a value of 1 if the corresponding cell expresses the gene and 0 otherwise, while each neuron in the output layer represents a candidate cell type. SuperCT provides three models: v1m and v2m for mice, and v1h for human. New cell types can be incorporated by superCT through transfer learning, where the weights for the first hidden layer are frozen.\\

EnClaSC\cite{chenEnClaSCNovelEnsemble2020} draws upon ensemble learning. Query cells are first screened by a scoring strategy designed to identify rare cell types. Training samples of rare cell types are paired
one by one, and their gene expressions are concatenated to form a paired sample. If the two training cells in a pair are of the same rare cell type, the paired sample is labeled with 1, or 0 otherwise. In the test phase, each query cell is paired with each training cell in a similar fashion. A tree-based model, LightBGM \cite{Ke2017}, is used to predict whether a query cell is of the same rare cell type of a paired training cell. The final score for a query cell and a rare cell type is normalized prediction results for the given query cell and all training cells belonging to the said rare cell type class weighted by Pearson correlations of their gene expressions.
Cells unassigned by this scoring strategy are examined by an ANN model consisting of four dense layers and two dropout layers. In this phase, training cells are sampled with replacement ten times to train ten neural networks, whose prediction results are voted to obtain a final prediction.\\

scNym\cite{kimmelSemisupervisedAdversarialNeural2021} is another deep learning method that combines semi-supervised learning and adversarial neural network. In particular, it utilizes information from both reference and testing datasets during the training phase by using MixMatch \cite{berthelotMixMatchHolisticApproach2019} semi-supervision and domain adversarial \cite{ganinDomainAdversarialTrainingNeural2015} iteratively to accommodate for the differences in data distribution between training and testing datasets due to variability in experiment platforms or conditions. Under the MixMatch framework, query cells are pseudo labeled \cite{Lee2013,vermaInterpolationConsistencyTraining2019} using the current classifier such that the pseudo label's cross-entropy is minimized.  The query cells are then randomly paired with training cells to generate pair-wise weighted averages to train the scNym model. The use of a convex pair-wise combinations of training and testing samples assumes that the linear interpolation of feature vectors leads to that of associated targets \cite{zhangMixupEmpiricalRisk2017}, which encourages generalization between the training data and testing data. Given the embeddings learned by the scNym model, the domain adversarial framework consisting of a two-layer neural network is then used to distinguish whether the example comes from the training or the testing data to compete with the classifier. The model parameters are optimized based on objectives combining classification, MixMatch interpolation consistency, and the domain adversarial.\\

scIAE\cite{yinScIAEIntegrativeAutoencoderbased2022} also uses an ensemble classification framework where the final prediction is based upon multiple base models, each consisting of two stacked sparse denoising autoencoders. The inputs to each base models are gene-wise randomly subsetted expression matrices of both the training and the testing data. The subsetted training data is used to train a stacked sparse denoising autoencoder to project the subsetted testing data to lower dimensional embeddings. The classifier obtained from the training data is then applied to the testing data's lower dimensional representation for prediction. The prediction results generated by all base models are then integrated for final prediction through voting.\\

In addition, scDeepSort \cite{shao2021scdeepsort} and sigGCN \cite{wangSinglecellClassificationUsing2021} are two methods that draw upon graphs. sigGCN employs a graph convolutional network (GCN) to reflect the non-linear topological relationship among cells. It first constructs a gene-wise weighted adjacency matrix using the STRING database \cite{szklarczykSTRINGV11Protein2019} to create a gene interaction network where node features in this network are defined as corresponding gene expressions. This graph is used as the input to a GCN-based autoencoder consisting of a convolutional layer and a maxpooling layer followed by a flatten layer and a fully connected (FC) layer. Another FC layer is then used as the decoder to reconstruct gene expressions from the obtained embeddings, which is associated with gene expression reconstruction loss measured by mean squared errors. Meanwhile, a two-layer neural network is trained using the training gene expression matrix in addition to the GCN-based autoencoder so that The hidden layers produced by the two models are concatenated. Finally, the concatenated features are then used to train classification layer whose loss is given by the negative log-likelihood corresponding to the true cell type label. Furthermore, in addition to the gene expression reconstruction loss and the classification loss, a regularization loss is incorporated to prevent overfitting. The three types of loss are combined to train the sigGCN model to obtain optimized parameters.\\

By comparison, scDeepSort \cite{shao2021scdeepsort} uses a graph neural network using a weighted bipartite graph where both cells and genes are its nodes. The gene expression value for each cell-gene pair is the edge weights between them. Features for gene nodes are obtained from principal component analysis, while those for cell nodes are the weighted aggregation of gene node features they connect to. The gene-cell graph then goes through a weighted graph aggregator layer adapted from GraphSage  \cite{hamiltonInductiveRepresentationLearning2017}, which transforms a node along with information taken from its local neighborhood. In particular, the aggregation layer in scdeepSort incorporated the weighted adjacency matrix normalization to address gene expression variability and the learnable sharing confidence to deal with batch effect and dropouts in sc-RNA-seq data. Each edge will have a corresponding learnable sharing confidence parameter weighing the connection of the neighbors, in addition to another parameter for self-loops for cells. The learned confidence parameters inform which genes are discriminatory for annotation.  Finally, cell node representations in the latent representations produced by the aggregation layer pass a linear classifier to predict cell labels. When the trained model is applied to testing datasets, query cells are connected to gene nodes of the trained cell-gene graph with the corresponding gene expression as the edge weight between the new cell node and gene node.\\

\subsection{Tools and Datasets}

We summarize representative tools or baseline methods for cell type annotation task in Table~\ref{cell_type_annotation_tool_table}, and five curated cell atlas with annotated cell types to train and benchmark cell type annotation algorithms in Table~\ref{cell_type_annotation_dataset_table}.


\begin{table}[ht]\normalsize
    \caption{A summary of clustering analysis tools}
    \vskip 1em
    \resizebox{\textwidth}{!}{
        \begin{tabular}{lllll}
            
            \hline
            \textbf{Tools} & \textbf{Algorithm} & \textbf{Description} & \textbf{Language} & \textbf{Availability} \\ \hline
          
\\ \hline
scmap    & Classical & \begin{tabular}[c]{@{}l@{}}A correlation-based method\end{tabular} & R
& \begin{tabular}[c]{@{}l@{}} \href{https://github.com/hemberg-lab/scmap}{{\textcolor{blue}{scmap}}}~\cite{kiselevScmapProjectionSinglecell2018}
\end{tabular}           \\ \hline

CellAssign    & Classical                                                  
& \begin{tabular}[c]{@{}l@{}} A probabilistic model leveraging gene marker information \end{tabular}                                                                                                                                  & R           & \begin{tabular}[c]{@{}l@{}}\href{https://github.com/Irrationone/cellassign}{{\textcolor{blue}{CellAssign}}}~\cite{zhangProbabilisticCelltypeAssignment2019} \end{tabular}          \\ \hline

Garnett   & Classical                                                  & \begin{tabular}[c]{@{}l@{}} An elastic-net-regression-based classifier that uses selected \\representative cells for each cell type  \end{tabular}                                   & R            & \begin{tabular}[c]{@{}l@{}} \href{https://github.com/cole-trapnell-lab/Garnett}{{\textcolor{blue}{Garnett}}}~\cite{plinerSupervisedClassificationEnables2019}  \end{tabular} \\ \hline

SingleR  & Classical                                                  & \begin{tabular}[c]{@{}l@{}} A correlation-based method\end{tabular}                                           & R                 &\href{https://github.com/dviraran/SingleR }{{\textcolor{blue}{SingleR}}}~\cite{aranReferencebasedAnalysisLung2019}                                                                         \\ \hline
CHETAH  & Classical                                                  & \begin{tabular}[c]{@{}l@{}} A correlation-based method that hierarchically assign cell types\end{tabular}                                                                                   & R           &   \href{https://github.com/jdekanter/CHETAH}{{\textcolor{blue}{CHETAH}}}~\cite{dekanterCHETAHSelectiveHierarchical2019}                                                                                                    \\ \hline
SingleCellNet      & Classical                                                  & \begin{tabular}[c]{@{}l@{}}  A random forest~\cite{breimanRandomForests2001}  classifier \end{tabular}                                                     & R                 & \href{https://github.com/pcahan1/singleCellNet}{{\textcolor{blue}{SingleCellNet}}}~\cite{tanSingleCellNetComputationalTool2019}  \href{https://github.com/OmicsML/dance}{{\textcolor{blue}{DANCE}}}~\cite{link2dance}   \\                                              \hline

scMatch    & \begin{tabular}[c]{@{}l@{}}Classical \end{tabular} & \begin{tabular}[c]{@{}l@{}} A correlation-based method\end{tabular}                                                                   & Python            & \begin{tabular}[c]{@{}l@{}}\href{https://github.com/asrhou/scMatch}{{\textcolor{blue}{scMatch}}}~\cite{houScMatchSinglecellGene2019} \end{tabular}               \\ \hline

SCINA   & \begin{tabular}[c]{@{}l@{}}Classical\end{tabular} & \begin{tabular}[c]{@{}l@{}} An expectation-maximization algorithm assuming a bimodal \\distribution for each signature genes\end{tabular}                                                                         & Python            & \begin{tabular}[c]{@{}l@{}}\href{https://github.com/jcao89757/SCINA}{{\textcolor{blue}{SCINA}}}~\cite{zhangSCINASemiSupervisedSubtyping2019} \end{tabular}            \\ \hline

scPred  & \begin{tabular}[c]{@{}l@{}} Classical \end{tabular}  & \begin{tabular}[c]{@{}l@{}} An SVM-based model\end{tabular}              & R            &  \href{https://github.com/powellgenomicslab/scPred}{{\textcolor{blue}{scPred}}}~\cite{alquicira-hernandezScPredAccurateSupervised2019}                      \\ \hline

CIPR     & Classical       & \begin{tabular}[c]{@{}l@{}}An algorithm that assigns cell types based on similarity with\\ reference samples using marker genes information.\end{tabular} & R            & \begin{tabular}[c]{@{}l@{}}
\href{https://github.com/atakanekiz/CIPR-Package}{{\textcolor{blue}{CIPR}}}~\cite{ekizCIPRWebbasedShiny2020}    
\href{https://aekiz.shinyapps.io/CIPR/}{{\textcolor{blue}{CIPR(Shiny)}}} \end{tabular} \\ \hline

SciBet     & Classical        &  \begin{tabular}[c]{@{}l@{}}An maximum likelihood estimation method that utilizes marker gene\\ information under assumed multinomial distribution for gene expressions\end{tabular} & R            &
\begin{tabular}[c]{@{}l@{}} \href{http://scibet.cancer-pku.cn}{{\textcolor{blue}{SciBet}}}~\cite{liSciBetPortableFast2020} \end{tabular} \\ \hline

SCSA       & Classical     & \begin{tabular}[c]{@{}l@{}}A scoring method that leverages known marker genes\\ and their confidence levels  \end{tabular} & Python            &     \href{http://scibet.cancer-pku.cn}{{\textcolor{blue}{SCSA}}}~\cite{Cao2020}\\ \hline

CellMeSH      & Classical     & \begin{tabular}[c]{@{}l@{}}A probabilistic model using the constructed CellMeSH database \end{tabular} & Python            & \href{https://github.com/shunfumao/cellmesh\#pub}{{\textcolor{blue}{CellMeSH}}}~\cite{maoCellMeSHProbabilisticCelltype2022}    \\ \hline

scSorter        & Classical     & \begin{tabular}[c]{@{}l@{}}A semi-supervised method that clusters and assigns cell types\\ based on marker genes information \end{tabular} & R         &   \href{https://cran.r-project.org/web/packages/scSorter/vignettes/scSorter.html}{{\textcolor{blue}{scSorter}}}~\cite{guoScSorterAssigningCells2021}     \\ \hline

scType          & Classical     & \begin{tabular}[c]{@{}l@{}}A scoring method for cell type assignment based on marker genes\end{tabular} & R            & \href{https://github.com/IanevskiAleksandr/sc-type}{{\textcolor{blue}{scType}}}~\cite{ianevskiFullyautomatedUltrafastCelltype2022}        \\ \hline

CellTypist     & Classical     & \begin{tabular}[c]{@{}l@{}}A logistic regression model with stochastic gradient descent learning  \end{tabular} & Python            &  \href{https://github.com/Teichlab/CellTypist}{{\textcolor{blue}{CellTypist}}}~\cite{dominguezcondeCrosstissueImmuneCell2022}     \href{https://github.com/OmicsML/dance}{{\textcolor{blue}{DANCE}}}~\cite{link2dance}  \\ \hline

SuperCT      & NN     & \begin{tabular}[c]{@{}l@{}} A fully connected ANN model using binarized expression data from MCA\\ with transfer learning to incorporate new cell types.\end{tabular} & Python; R            & 
\begin{tabular}[c]{@{}l@{}} \href{https://github.com/weilin-genomics/SuperCT}{{\textcolor{blue}{SuperCT(Python)}}}~\cite{xieSuperCTSupervisedlearningFramework2019} \href{https://github.com/weilin-genomics/rSuperCT}{{\textcolor{blue}{SuperCT(R)}}}\end{tabular}\\ \hline

ACTINN     & NN     & \begin{tabular}[c]{@{}l@{}} A neural network consisting of three hidden layers  \end{tabular} & Python            & \href{https://github.com/mafeiyang/ACTINN}{{\textcolor{blue}{ACTINN}}}~\cite{maACTINNAutomatedIdentification2019}  \href{https://github.com/OmicsML/dance}{{\textcolor{blue}{DANCE}}}~\cite{link2dance} \\ \hline

EnClaSC       & NN     & \begin{tabular}[c]{@{}l@{}} Combination of a few-sample learning strategy to assign query cells and\\ an ANN model for cells unassigned to rare cell types \end{tabular} & Python            &   \href{https://github.com/xy-chen16/EnClaSC}{{\textcolor{blue}{EnClaSC}}}~\cite{chenEnClaSCNovelEnsemble2020}   \\ \hline

scNym     & Adversarial  NN    & \begin{tabular}[c]{@{}l@{}}  A semi-supervised method that utilizes adversarial neural network \end{tabular} & Python            &   \href{https://github.com/calico/scnym }{{\textcolor{blue}{scNym}}}~\cite{kimmelSemisupervisedAdversarialNeural2021}   \\ \hline

 scDeepSort      & GNN     & \begin{tabular}[c]{@{}l@{}}  Adaptation of GraphSAGE \cite{hamiltonInductiveRepresentationLearning2017} to address a weighted cell-gene graph\\ to produce node representations to pass a linear classifier  \end{tabular} & Python            &  \href{https://github.com/ZJUFanLab/scDeepSort}{{\textcolor{blue}{ scDeepSort}}}~\cite{shao2021scdeepsort}   \href{https://github.com/OmicsML/dance}{{\textcolor{blue}{DANCE}}}~\cite{link2dance}   \\ \hline

sigGCN      & GCN; NN     & \begin{tabular}[c]{@{}l@{}}  Integration of a GCN and an NN whose concatenated learned features\\ pass through the classification layer. \end{tabular} & Python     &  \href{https://github.com/NabaviLab/sigGCN}{{\textcolor{blue}{ sigGCN}}}~\cite{wangSinglecellClassificationUsing2021}       \\ \hline

scIAE      & Autoencoders     & \begin{tabular}[c]{@{}l@{}}  An ensemble framework integrating stacked denoising\\ and sparse autoencoders. \end{tabular} & Python            &    \href{https://github.com/JGuan-lab/scIAE}{{\textcolor{blue}{ scIAE}}}~\cite{yinScIAEIntegrativeAutoencoderbased2022}     \\ \hline
\end{tabular}
}
\label{cell_type_annotation_tool_table}
\end{table}


\begin{table}[ht]\normalsize
\centering
\caption{A summary of cell type annotation datasets.}
\vskip 1em
\resizebox{12cm}{!}{%
\begin{tabular}{llllll}
\hline
\textbf{Cell Atlas}       & \textbf{Species}   & \textbf{\begin{tabular}[c]{@{}l@{}}Atlas\\ Description\end{tabular}}                                                                  & \textbf{Availability}                                                                                        \\ \hline
\begin{tabular}[c]{@{}l@{}}Human Cell Atlas\\ \end{tabular}    & Human     & \begin{tabular}[c]{@{}l@{}}33 organs\\~30 million cells\end{tabular}                                                                              & \begin{tabular}[c]{@{}l@{}}\href{https://data.humancellatlas.org/}{{\textcolor{blue}{Human Cell Atlas}}}~\cite{regevHumanCellAtlas2017}   \end{tabular} \\ \hline

\begin{tabular}[c]{@{}l@{}}Tabula Sapien \\ \end{tabular}    & Human     & \begin{tabular}[c]{@{}l@{}}24 organs\\~500k cells\end{tabular}                                                                              & \begin{tabular}[c]{@{}l@{}}\href{https://tabula-sapiens-portal.ds.czbiohub.org/}{{\textcolor{blue}{Tabula Sapien}}}~\cite{thetabulasapiensconsortium*TabulaSapiensMultipleorgan2022}   \end{tabular} \\ \hline

\begin{tabular}[c]{@{}l@{}}Human Cell Landscape \\ \end{tabular}    & Human     & \begin{tabular}[c]{@{}l@{}}56 tissues\\~700k cells\end{tabular}                                                                              & \begin{tabular}[c]{@{}l@{}}\href{https://db.cngb.org/HCL/}{{\textcolor{blue}{Human Cell Landscape}}}~\cite{hanConstructionHumanCell2020}  \end{tabular} \\ \hline

\begin{tabular}[c]{@{}l@{}}Tabula Muris \\ \end{tabular}    & Mouse    & \begin{tabular}[c]{@{}l@{}}20 organs\\~100k cells\end{tabular}                                                                              & \begin{tabular}[c]{@{}l@{}}\href{https://tabula-muris.ds.czbiohub.org/}{{\textcolor{blue}{Tabula Muris}}}~\cite{thetabulamurisconsortiumSinglecellTranscriptomics202018} \end{tabular} \\ \hline

\begin{tabular}[c]{@{}l@{}}Mouse Cell Atlas \\ \end{tabular}    & Mouse    & \begin{tabular}[c]{@{}l@{}}51 tissues\\ ~400k cells\end{tabular}                                                                              & \begin{tabular}[c]{@{}l@{}} \href{https://tabula-muris.ds.czbiohub.org/}{{\textcolor{blue}{Mouse Cell Atlas}}}~\cite{hanMappingMouseCell2018} \end{tabular} \\ \hline

\end{tabular}%
}
\label{cell_type_annotation_dataset_table}
\end{table}




\section{Conclusion}\label{future}

DL-based methods with their ability to model complex nonlinear relationships via big data have demonstrated their superior performance over classical machine learning methods in a variety of single-cell tasks. 

With the technological advances, we will soon have access to unprecedented cellular and even sub-cellular level multi-omic data together with their \textit{in situ} geographical coordinates in tissues. These data offer the promise of solutions to many biological and medical goals. However, this promise will remain unfulfilled without the development of new computational methodologies that can efficiently resolve the limitations of new technologies, utilize existing the large pool of existing data, and  make sense of results. 

Specifically, despite providing higher resolution, measurements from new technologies are prone to severe drop-out events and structured biases, due to the scarcity of initial target materials such as RNA and protein. Thus, there is an urgent need to develop computational methods that can accurately model the error generation process.
The new knowledge will improve the interpretation of high-dimensional data sets.
In addition, there are a vast amount of single-cell data from older platform in the public domain. The continued development of computational algorithms that can integrate existing data while removing platform differences and batch effect is essential for future single-cell studies. 

Aside from RNA and protein abundance, new technologies also provide extra layers of information, such as images, cell location, cell shape and size, and locations of RNA molecules. Undoubtedly, the considerably enriched information will enable comprehensive analyses of cell states,  the interaction between cells, and the mechanism of gene regulations. Beyond the exciting prospect, it demands the development of sophistical models to integrate different layers of information.  

Despite vast single-cell data, deep learning tools are still not popular among biomedical research groups. A major barrier is the lack of robustness of many deep learning methods whose performance relies on the noise levels in the data. Thus, how to establish a single-cell pipeline with the ability to automatically select models and parameter sets with optimal performance is critical for integrating deep learning into biological data sets. In addition, most deep learning methods behave like black boxes with limited interpretability. Indeed, the lack of interpretability of models presents a significant bottleneck to their widespread acceptance. Hence, it is imperative for us to improve our understanding of the behavior of these models and enhance pipeline interpretability.
\newpage

\bibliographystyle{unsrt}
\bibliography{survey.bib}

\begin{thebibliography}{100}

\bibitem{mcmanus2015}
Joel McManus, Zhe Cheng, and Christine Vogel.
\newblock Next-generation analysis of gene expression regulation--comparing the
  roles of synthesis and degradation.
\newblock {\em Molecular bioSystems}, 11(10):2680--2689, 2015.

\bibitem{patterson2003}
Scott Patterson and Ruedi Aebersold.
\newblock Proteomics: the first decade and beyond.
\newblock {\em Nature Genetics}, 33, 2003.

\bibitem{Eberwine1992}
James Eberwine, Hermes Yeh, Kevin Miyashiro, Yanxiang Cao, Suresh Nair, Richard
  Finnell, Martha Zettel, and Paul Coleman.
\newblock Analysis of gene expression in single live neurons.
\newblock {\em Proceedings of the National Academy of Sciences of the United
  States of America}, 89:3010--3014, 1992.

\bibitem{Brady1990}
Gerard Brady, Mary Barbara, and Norman Iscove.
\newblock Representative in vitro cdna amplification from individual
  hemopoietic cells and colonies.
\newblock {\em Methods in Molecular and Cellular Biology}, 2:17--–25, 1990.

\bibitem{Tang2009}
Fuchou Tang, Catalin Barbacioru, Yangzhou Wang, Ellen Nordman, Clarence Lee,
  Nanlan Xu, Xiaohui Wang, John Bodeau, Brian~B Tuch, Asim Siddiqui, Kaiqin
  Lao, and M~Azim Surani.
\newblock mrna-seq whole-transcriptome analysis of a single cell.
\newblock {\em Nature Methods}, 6(5):377--382, 2009.

\bibitem{wang2015advances}
Yong Wang and Nicholas~E Navin.
\newblock Advances and applications of single-cell sequencing technologies.
\newblock {\em Molecular cell}, 58(4):598--609, 2015.

\bibitem{liang2014single}
Jialong Liang, Wanshi Cai, and Zhongsheng Sun.
\newblock Single-cell sequencing technologies: current and future.
\newblock {\em Journal of Genetics and Genomics}, 41(10):513--528, 2014.

\bibitem{wen2022recent}
Lu~Wen and Fuchou Tang.
\newblock Recent advances in single-cell sequencing technologies.
\newblock {\em Precision Clinical Medicine}, 5(1):pbac002, 2022.

\bibitem{svensson2018exponential}
Valentine Svensson, Roser Vento-Tormo, and Sarah~A Teichmann.
\newblock Exponential scaling of single-cell rna-seq in the past decade.
\newblock {\em Nature protocols}, 13(4):599--604, 2018.

\bibitem{kolodziejczyk2015technology}
Aleksandra~A Kolodziejczyk, Jong~Kyoung Kim, Valentine Svensson, John~C
  Marioni, and Sarah~A Teichmann.
\newblock The technology and biology of single-cell rna sequencing.
\newblock {\em Molecular cell}, 58(4):610--620, 2015.

\bibitem{li2021bulk}
Xinmin Li and Cun-Yu Wang.
\newblock From bulk, single-cell to spatial rna sequencing.
\newblock {\em International Journal of Oral Science}, 13(1):1--6, 2021.

\bibitem{pouyanfar2018survey}
Samira Pouyanfar, Saad Sadiq, Yilin Yan, Haiman Tian, Yudong Tao, Maria~Presa
  Reyes, Mei-Ling Shyu, Shu-Ching Chen, and Sundaraja~S Iyengar.
\newblock A survey on deep learning: Algorithms, techniques, and applications.
\newblock {\em ACM Computing Surveys (CSUR)}, 51(5):1--36, 2018.

\bibitem{dong2021survey}
Shi Dong, Ping Wang, and Khushnood Abbas.
\newblock A survey on deep learning and its applications.
\newblock {\em Computer Science Review}, 40:100379, 2021.

\bibitem{chen2019revolutionizing}
Haide Chen, Fang Ye, and Guoji Guo.
\newblock Revolutionizing immunology with single-cell rna sequencing.
\newblock {\em Cellular \& molecular immunology}, 16(3):242--249, 2019.

\bibitem{giladi2018single}
Amir Giladi and Ido Amit.
\newblock Single-cell genomics: a stepping stone for future immunology
  discoveries.
\newblock {\em Cell}, 172(1-2):14--21, 2018.

\bibitem{valdes2018single}
Fatima Valdes-Mora, Kristina Handler, Andrew~MK Law, Robert Salomon, Samantha~R
  Oakes, Christopher~J Ormandy, and David Gallego-Ortega.
\newblock Single-cell transcriptomics in cancer immunobiology: the future of
  precision oncology.
\newblock {\em Frontiers in Immunology}, 9:2582, 2018.

\bibitem{lahnemann2020}
David L\"ahnemann and et~al.
\newblock Eleven grand challenges in single-cell data science.
\newblock {\em Genome Biology}, 21(1), 2020.

\bibitem{muzio2021}
Giulia Muzio, Leslie O’Bray, and Karsten Borgwardt.
\newblock Biological network analysis with deep learning.
\newblock {\em Briefings in Bioinformatics}, 22(2):1515--1530, 2021.

\bibitem{bansal2007infer}
Mukesh Bansal, Vincenzo Belcastro, Alberto Ambesi-Impiombato, and Diego
  Di~Bernardo.
\newblock How to infer gene networks from expression profiles.
\newblock {\em Molecular systems biology}, 3(1):78, 2007.

\bibitem{li2022decoding}
Haochen Li, Tianxing Ma, Minsheng Hao, Lei Wei, and Xuegong Zhang.
\newblock Decoding functional cell-cell communication events by multi-view
  graph learning on spatial transcriptomics.
\newblock {\em bioRxiv}, 2022.

\bibitem{Houle2010}
David Houle, Diddahally~R. Govindaraju, and Stig Omholt.
\newblock Phenomics: the next challenge.
\newblock {\em Nature Reviews Genetics}, 11(12):855--866, November 2010.

\bibitem{Goldman2019-cx}
Samantha~L Goldman, Matthew MacKay, Ebrahim Afshinnekoo, Ari~M Melnick, Shuxiu
  Wu, and Christopher~E Mason.
\newblock The impact of heterogeneity on single-cell sequencing.
\newblock {\em Front. Genet.}, 10:8, March 2019.

\bibitem{Kulkarni2019-qh}
Ashwinikumar Kulkarni, Ashley~G Anderson, Devin~P Merullo, and Genevieve
  Konopka.
\newblock Beyond bulk: a review of single cell transcriptomics methodologies
  and applications.
\newblock {\em Curr. Opin. Biotechnol.}, 58:129--136, August 2019.

\bibitem{Stegle2015-hu}
Oliver Stegle, Sarah~A Teichmann, and John~C Marioni.
\newblock Computational and analytical challenges in single-cell
  transcriptomics.
\newblock {\em Nat. Rev. Genet.}, 16(3):133--145, March 2015.

\bibitem{nguyenExperimentalConsiderationsSingleCell2018}
Quy~H. Nguyen, Nicholas Pervolarakis, Kevin Nee, and Kai Kessenbrock.
\newblock Experimental {{Considerations}} for {{Single-Cell RNA Sequencing
  Approaches}}.
\newblock {\em Frontiers in Cell and Developmental Biology}, 6:108, 2018.

\bibitem{Guo2013}
Hongshan Guo, Ping Zhu, Xinglong Wu, Xianlong Li, Lu~Wen, and Fuchou Tang.
\newblock Single-cell methylome landscapes of mouse embryonic stem cells and
  early embryos analyzed using reduced representation bisulfite sequencing.
\newblock {\em Genome Research}, 23(12):2126--2135, October 2013.

\bibitem{bartosovic2021single}
Marek Bartosovic, Mukund Kabbe, and Gon{\c{c}}alo Castelo-Branco.
\newblock Single-cell cut\&tag profiles histone modifications and transcription
  factors in complex tissues.
\newblock {\em Nature biotechnology}, 39(7):825--835, 2021.

\bibitem{Brehm2004}
Byron~F. Brehm-Stecher and Eric~A. Johnson.
\newblock Single-cell microbiology: Tools, technologies, and applications.
\newblock {\em Microbiology and Molecular Biology Reviews}, 68(2):538–--559,
  1990.

\bibitem{Whitesides2006}
George~M Whitesides.
\newblock The origins and the future of microfluidics.
\newblock {\em Nature}, 442(7101):368--373, 2006.

\bibitem{Thorsen2001}
Todd Thorsen, Richard~W Roberts, Frances~H Arnold, and Stephen~R Quake.
\newblock Dynamic pattern formation in a vesicle-generating microfluidic
  device.
\newblock {\em Physical Review Letters}, 86(18):4163--4166, 2001.

\bibitem{Kornberg1974}
Roger~D Kornberg.
\newblock Chromatin structure: a repeating unit of histones and dna.
\newblock {\em Science}, 184(4139):868--871, 1974.

\bibitem{Kornberg1992}
Roger~D Kornberg and Yahli Lorch.
\newblock Chromatin structure and transcription.
\newblock {\em Annual Review of Cell Biology}, 8:563--587, 1974.

\bibitem{buenrostro2013transposition}
Jason~D Buenrostro, Paul~G Giresi, Lisa~C Zaba, Howard~Y Chang, and William~J
  Greenleaf.
\newblock Transposition of native chromatin for multimodal regulatory analysis
  and personal epigenomics.
\newblock {\em Nature methods}, 10(12):1213, 2013.

\bibitem{Hendrickson2018}
David Hendrickson, Ilya Soifer, Bernd Wranik, David Botstein, and Scott
  McIsaac.
\newblock Simultaneous profiling of dna accessibility and gene expression
  dynamics with atac-seq and rna-seq.
\newblock {\em Methods in Molecular Biology}, 1819:317--333, 2018.

\bibitem{Thurman2012}
Robert Thurman et~al.
\newblock The accessible chromatin landscape of the human genome.
\newblock {\em Nature}, 489(7414):75--82, 2012.

\bibitem{li2021chromatin}
Zhijian Li, Christoph Kuppe, Susanne Ziegler, Mingbo Cheng, Nazanin Kabgani,
  Sylvia Menzel, Martin Zenke, Rafael Kramann, and Ivan~G Costa.
\newblock Chromatin-accessibility estimation from single-cell atac-seq data
  with scopen.
\newblock {\em Nature communications}, 12(1):1--14, 2021.

\bibitem{bird2007perceptions}
Adrian Bird.
\newblock Perceptions of epigenetics.
\newblock {\em Nature}, 447(7143):396, 2007.

\bibitem{moore2013dna}
Lisa~D Moore, Thuc Le, and Guoping Fan.
\newblock Dna methylation and its basic function.
\newblock {\em Neuropsychopharmacology}, 38(1):23--38, 2013.

\bibitem{singal1999dna}
Rakesh Singal and Gordon~D Ginder.
\newblock Dna methylation.
\newblock {\em Blood, The Journal of the American Society of Hematology},
  93(12):4059--4070, 1999.

\bibitem{farlik2016dna}
Matthias Farlik, Florian Halbritter, Fabian M{\"u}ller, Fizzah~A Choudry, Peter
  Ebert, Johanna Klughammer, Samantha Farrow, Antonella Santoro, Valerio
  Ciaurro, Anthony Mathur, et~al.
\newblock Dna methylation dynamics of human hematopoietic stem cell
  differentiation.
\newblock {\em Cell stem cell}, 19(6):808--822, 2016.

\bibitem{smallwood2014single}
S{\'e}bastien~A Smallwood, Heather~J Lee, Christof Angermueller, Felix Krueger,
  Heba Saadeh, Julian Peat, Simon~R Andrews, Oliver Stegle, Wolf Reik, and
  Gavin Kelsey.
\newblock Single-cell genome-wide bisulfite sequencing for assessing epigenetic
  heterogeneity.
\newblock {\em Nature methods}, 11(8):817--820, 2014.

\bibitem{farlik2015single}
Matthias Farlik, Nathan~C Sheffield, Angelo Nuzzo, Paul Datlinger, Andreas
  Sch{\"o}negger, Johanna Klughammer, and Christoph Bock.
\newblock Single-cell dna methylome sequencing and bioinformatic inference of
  epigenomic cell-state dynamics.
\newblock {\em Cell reports}, 10(8):1386--1397, 2015.

\bibitem{hou2016single}
Yu~Hou, Huahu Guo, Chen Cao, Xianlong Li, Boqiang Hu, Ping Zhu, Xinglong Wu,
  Lu~Wen, Fuchou Tang, Yanyi Huang, et~al.
\newblock Single-cell triple omics sequencing reveals genetic, epigenetic, and
  transcriptomic heterogeneity in hepatocellular carcinomas.
\newblock {\em Cell research}, 26(3):304--319, 2016.

\bibitem{vistain2021single}
Luke~F Vistain and Sava{\c{s}} Tay.
\newblock Single-cell proteomics.
\newblock {\em Trends in biochemical sciences}, 46(8):661--672, 2021.

\bibitem{stoeckius2017large}
Marlon Stoeckius, Christoph Hafemeister, William Stephenson, Brian
  Houck-Loomis, Pratip~K Chattopadhyay, Harold Swerdlow, Rahul Satija, and
  Peter Smibert.
\newblock Large-scale simultaneous measurement of epitopes and transcriptomes
  in single cells.
\newblock {\em Nature methods}, 14(9):865, 2017.

\bibitem{baron2017new}
Maayan Baron and Itai Yanai.
\newblock New skin for the old rna-seq ceremony: the age of single-cell
  multi-omics.
\newblock {\em Genome Biology}, 18(1):1--3, 2017.

\bibitem{Chen2019}
Song Chen, Blue~B. Lake, and Kun Zhang.
\newblock High-throughput sequencing of the transcriptome and chromatin
  accessibility in the same cell.
\newblock {\em Nature Biotechnology}, 37(12):1452--1457, October 2019.

\bibitem{Zhu2019}
Lingxue Zhu, Jing Lei, Lambertus Klei, Bernie Devlin, and Kathryn Roeder.
\newblock Semisoft clustering of single-cell data.
\newblock {\em Proceedings of the National Academy of Sciences},
  116(2):466--471, 2019.
\newblock \href{https://github.com/lingxuez/SOUPR}{Code Link:
  https://github.com/lingxuez/SOUPR}.

\bibitem{Ma2020}
Sai Ma, Bing Zhang, Lindsay~M. LaFave, Andrew~S. Earl, Zachary Chiang, Yan Hu,
  Jiarui Ding, Alison Brack, Vinay~K. Kartha, Tristan Tay, Travis Law, Caleb
  Lareau, Ya-Chieh Hsu, Aviv Regev, and Jason~D. Buenrostro.
\newblock Chromatin potential identified by shared single-cell profiling of
  {RNA} and chromatin.
\newblock {\em Cell}, 183(4):1103--1116.e20, November 2020.

\bibitem{Rosenberg2018}
Alexander~B. Rosenberg, Charles~M. Roco, Richard~A. Muscat, Anna Kuchina, Paul
  Sample, Zizhen Yao, Lucas~T. Graybuck, David~J. Peeler, Sumit Mukherjee, Wei
  Chen, Suzie~H. Pun, Drew~L. Sellers, Bosiljka Tasic, and Georg Seelig.
\newblock Single-cell profiling of the developing mouse brain and spinal cord
  with split-pool barcoding.
\newblock {\em Science}, 360(6385):176--182, April 2018.

\bibitem{Angermueller2016}
Christof Angermueller, Stephen~J Clark, Heather~J Lee, Iain~C Macaulay, Mabel~J
  Teng, Tim~Xiaoming Hu, Felix Krueger, S{\'{e}}bastien~A Smallwood, Chris~P
  Ponting, Thierry Voet, Gavin Kelsey, Oliver Stegle, and Wolf Reik.
\newblock Parallel single-cell sequencing links transcriptional and epigenetic
  heterogeneity.
\newblock {\em Nature Methods}, 13(3):229--232, January 2016.

\bibitem{Macaulay2015}
Iain~C Macaulay, Wilfried Haerty, Parveen Kumar, Yang~I Li, Tim~Xiaoming Hu,
  Mabel~J Teng, Mubeen Goolam, Nathalie Saurat, Paul Coupland, Lesley~M
  Shirley, Miriam Smith, Niels~Van der Aa, Ruby Banerjee, Peter~D Ellis,
  Michael~A Quail, Harold~P Swerdlow, Magdalena Zernicka-Goetz, Frederick~J
  Livesey, Chris~P Ponting, and Thierry Voet.
\newblock G{\&}amp;t-seq: parallel sequencing of single-cell genomes and
  transcriptomes.
\newblock {\em Nature Methods}, 12(6):519--522, April 2015.

\bibitem{Zhu2021}
Chenxu Zhu, Yanxiao Zhang, Yang~Eric Li, Jacinta Lucero, M.~Margarita Behrens,
  and Bing Ren.
\newblock Joint profiling of histone modifications and transcriptome in single
  cells from mouse brain.
\newblock {\em Nature Methods}, 18(3):283--292, February 2021.

\bibitem{Xiong2021}
Haiqing Xiong, Yingjie Luo, Qianhao Wang, Xianhong Yu, and Aibin He.
\newblock Single-cell joint detection of chromatin occupancy and transcriptome
  enables higher-dimensional epigenomic reconstructions.
\newblock {\em Nature Methods}, 18(6):652--660, May 2021.

\bibitem{crosetto2015spatially}
Nicola Crosetto, Magda Bienko, and Alexander Van~Oudenaarden.
\newblock Spatially resolved transcriptomics and beyond.
\newblock {\em Nature Reviews Genetics}, 16(1):57--66, 2015.

\bibitem{moor2017spatial}
Andreas~E Moor and Shalev Itzkovitz.
\newblock Spatial transcriptomics: paving the way for tissue-level systems
  biology.
\newblock {\em Current opinion in biotechnology}, 46:126--133, 2017.

\bibitem{wang2018multiplexed}
Guiping Wang, Jeffrey~R Moffitt, and Xiaowei Zhuang.
\newblock Multiplexed imaging of high-density libraries of rnas with merfish
  and expansion microscopy.
\newblock {\em Scientific reports}, 8(1):1--13, 2018.

\bibitem{marx2021method}
Vivien Marx.
\newblock Method of the year: spatially resolved transcriptomics.
\newblock {\em Nature methods}, 18(1):9--14, 2021.

\bibitem{asp2020spatially}
Michaela Asp, Joseph Bergenstr{\aa}hle, and Joakim Lundeberg.
\newblock Spatially resolved transcriptomes—next generation tools for tissue
  exploration.
\newblock {\em BioEssays}, 42(10):1900221, 2020.

\bibitem{waylen2020whole}
Lisa~N Waylen, Hieu~T Nim, Luciano~G Martelotto, and Mirana Ramialison.
\newblock From whole-mount to single-cell spatial assessment of gene expression
  in 3d.
\newblock {\em Communications biology}, 3(1):1--11, 2020.

\bibitem{teves2020mapping}
Joji~Marie Teves and Kyoung~Jae Won.
\newblock Mapping cellular coordinates through advances in spatial
  transcriptomics technology.
\newblock {\em Molecules and Cells}, 43(7):591, 2020.

\bibitem{SpatialTrans}
Patrik~L. Ståhl, Fredrik Salmén, Sanja Vickovic, Anna Lundmark,
  José~Fernández Navarro, Jens Magnusson, Stefania Giacomello, Michaela Asp,
  Jakub~O. Westholm, Mikael Huss, Annelie Mollbrink, Sten Linnarsson, Simone
  Codeluppi, Åke Borg, Fredrik Pontén, Paul~Igor Costea, Pelin Sahlén, Jan
  Mulder, Olaf Bergmann, Joakim Lundeberg, and Jonas Frisén.
\newblock Visualization and analysis of gene expression in tissue sections by
  spatial transcriptomics.
\newblock {\em Science}, 353(6294):78--82, 2016.

\bibitem{Merritt2020}
Christopher~R Merritt, Giang~T Ong, Sarah~E Church, Kristi Barker, Patrick
  Danaher, Gary Geiss, Margaret Hoang, Jaemyeong Jung, Yan Liang, Jill
  McKay-Fleisch, et~al.
\newblock Multiplex digital spatial profiling of proteins and rna in fixed
  tissue.
\newblock {\em Nature Biotechnology}, 05 2020.

\bibitem{MOFFITT20161}
J.R. Moffitt and X.~Zhuang.
\newblock Chapter one - {RNA} imaging with multiplexed error-robust
  fluorescence in situ hybridization ({MERFISH}).
\newblock In Grigory~S. Filonov and Samie~R. Jaffrey, editors, {\em Visualizing
  {RNA} dynamics in the cell}, volume 572 of {\em Methods in enzymology}, pages
  1--49. Academic Press, 2016.
\newblock ISSN: 0076-6879.

\bibitem{mosesMuseumSpatialTranscriptomics2022}
Lambda Moses and Lior Pachter.
\newblock Museum of spatial transcriptomics.
\newblock {\em Nature Methods}, 19(5):534--546, May 2022.

\bibitem{Lubeck2014}
Eric Lubeck, Ahmet~F Coskun, Timur Zhiyentayev, Mubhij Ahmad, and Long Cai.
\newblock Single-cell in situ {RNA} profiling by sequential hybridization.
\newblock {\em Nature Methods}, 11(4):360--361, March 2014.

\bibitem{Shah2016}
Sheel Shah, Eric Lubeck, Wen Zhou, and Long Cai.
\newblock In situ transcription profiling of single cells reveals spatial
  organization of cells in the mouse hippocampus.
\newblock {\em Neuron}, 92(2):342--357, October 2016.

\bibitem{eng2019transcriptome}
Chee-Huat~Linus Eng, Michael Lawson, Qian Zhu, Ruben Dries, Noushin Koulena,
  Yodai Takei, Jina Yun, Christopher Cronin, Christoph Karp, Guo-Cheng Yuan,
  et~al.
\newblock Transcriptome-scale super-resolved imaging in tissues by rna
  seqfish+.
\newblock {\em Nature}, 568(7751):235--239, 2019.

\bibitem{Rao2021}
Anjali Rao, Dalia Barkley, Gustavo~S. Fran{\c{c}}a, and Itai Yanai.
\newblock Exploring tissue architecture using spatial transcriptomics.
\newblock {\em Nature}, 596(7871):211--220, August 2021.

\bibitem{Sthl2016}
Patrik~L. St{\aa}hl, Fredrik Salm{\'{e}}n, Sanja Vickovic, Anna Lundmark,
  Jos{\'{e}}~Fern{\'{a}}ndez Navarro, Jens Magnusson, Stefania Giacomello,
  Michaela Asp, Jakub~O. Westholm, Mikael Huss, Annelie Mollbrink, Sten
  Linnarsson, Simone Codeluppi, {\AA}ke Borg, Fredrik Pont{\'{e}}n, Paul~Igor
  Costea, Pelin Sahl{\'{e}}n, Jan Mulder, Olaf Bergmann, Joakim Lundeberg, and
  Jonas Fris{\'{e}}n.
\newblock Visualization and analysis of gene expression in tissue sections by
  spatial transcriptomics.
\newblock {\em Science}, 353(6294):78--82, July 2016.

\bibitem{Rodriques2019}
Samuel~G. Rodriques, Robert~R. Stickels, Aleksandrina Goeva, Carly~A. Martin,
  Evan Murray, Charles~R. Vanderburg, Joshua Welch, Linlin~M. Chen, Fei Chen,
  and Evan~Z. Macosko.
\newblock Slide-seq: A scalable technology for measuring genome-wide expression
  at high spatial resolution.
\newblock {\em Science}, 363(6434):1463--1467, 2019.
\newblock \href{https://github.com/broadchenf/Slideseq}{Code Link:
  https://github.com/broadchenf/Slideseq}.

\bibitem{Stickels2020}
Robert~R. Stickels, Evan Murray, Pawan Kumar, Jilong Li, Jamie~L. Marshall,
  Daniela J.~Di Bella, Paola Arlotta, Evan~Z. Macosko, and Fei Chen.
\newblock Highly sensitive spatial transcriptomics at near-cellular resolution
  with slide-{seqV}2.
\newblock {\em Nature Biotechnology}, 39(3):313--319, December 2020.

\bibitem{Chen2022}
Ao~Chen, Sha Liao, Mengnan Cheng, Kailong Ma, Liang Wu, Yiwei Lai, Xiaojie Qiu,
  Jin Yang, Jiangshan Xu, Shijie Hao, Xin Wang, Huifang Lu, Xi~Chen, Xing Liu,
  Xin Huang, Zhao Li, Yan Hong, Yujia Jiang, Jian Peng, Shuai Liu, Mengzhe
  Shen, Chuanyu Liu, Quanshui Li, Yue Yuan, Xiaoyu Wei, Huiwen Zheng, Weimin
  Feng, Zhifeng Wang, Yang Liu, Zhaohui Wang, Yunzhi Yang, Haitao Xiang, Lei
  Han, Baoming Qin, Pengcheng Guo, Guangyao Lai, Pura
  Mu{\~{n}}oz-C{\'{a}}noves, Patrick~H. Maxwell, Jean~Paul Thiery, Qing-Feng
  Wu, Fuxiang Zhao, Bichao Chen, Mei Li, Xi~Dai, Shuai Wang, Haoyan Kuang,
  Junhou Hui, Liqun Wang, Ji-Feng Fei, Ou~Wang, Xiaofeng Wei, Haorong Lu,
  Bo~Wang, Shiping Liu, Ying Gu, Ming Ni, Wenwei Zhang, Feng Mu, Ye~Yin,
  Huanming Yang, Michael Lisby, Richard~J. Cornall, Jan Mulder, Mathias
  Uhl{\'{e}}n, Miguel~A. Esteban, Yuxiang Li, Longqi Liu, Xun Xu, and Jian
  Wang.
\newblock Spatiotemporal transcriptomic atlas of mouse organogenesis using
  {DNA} nanoball-patterned arrays.
\newblock {\em Cell}, 185(10):1777--1792.e21, May 2022.

\bibitem{lewis2022subcellular}
Zachary~R Lewis, Tien Phan-Everson, Gary Geiss, Mithra Korukonda, Ruchir Bhatt,
  Carl Brown, Dwayne Dunaway, Joseph Phan, Alyssa Rosenbloom, Brian Filanoski,
  et~al.
\newblock Subcellular characterization of over 100 proteins in ffpe tumor
  biopsies with cosmx spatial molecular imager.
\newblock {\em Cancer Research}, 82(12\_Supplement):3878--3878, 2022.

\bibitem{coons1942demonstration}
Albert~H Coons, Hugh~J Creech, R~Norman Jones, and Ernst Berliner.
\newblock The demonstration of pneumococcal antigen in tissues by the use of
  fluorescent antibody.
\newblock {\em The Journal of Immunology}, 45(3):159--170, 1942.

\bibitem{Michael2013}
Michael~J. Gerdes, Christopher~J. Sevinsky, Anup Sood, Sudeshna Adak,
  Musodiq~O. Bello, Alexander Bordwell, Ali Can, Alex Corwin, Sean Dinn,
  Robert~J. Filkins, Denise Hollman, Vidya Kamath, Sireesha Kaanumalle, Kevin
  Kenny, Melinda Larsen, Michael Lazare, Qing Li, Christina Lowes, Colin~C.
  McCulloch, Elizabeth McDonough, Michael~C. Montalto, Zhengyu Pang, Jens
  Rittscher, Alberto Santamaria-Pang, Brion~D. Sarachan, Maximilian~L. Seel,
  Antti Seppo, Kashan Shaikh, Yunxia Sui, Jingyu Zhang, and Fiona Ginty.
\newblock Highly multiplexed single-cell analysis of formalin-fixed,
  paraffin-embedded cancer tissue.
\newblock {\em Proceedings of the National Academy of Sciences},
  110(29):11982--11987, 2013.

\bibitem{lin2015highly}
Jia-Ren Lin, Mohammad Fallahi-Sichani, and Peter~K Sorger.
\newblock Highly multiplexed imaging of single cells using a high-throughput
  cyclic immunofluorescence method.
\newblock {\em Nature communications}, 6(1):1--7, 2015.

\bibitem{lin2018highly}
Jia-Ren Lin, Benjamin Izar, Shu Wang, Clarence Yapp, Shaolin Mei, Parin~M Shah,
  Sandro Santagata, and Peter~K Sorger.
\newblock Highly multiplexed immunofluorescence imaging of human tissues and
  tumors using t-cycif and conventional optical microscopes.
\newblock {\em Elife}, 7, 2018.

\bibitem{goltsev2018deep}
Yury Goltsev, Nikolay Samusik, Julia Kennedy-Darling, Salil Bhate, Matthew
  Hale, Gustavo Vazquez, Sarah Black, and Garry~P Nolan.
\newblock Deep profiling of mouse splenic architecture with codex multiplexed
  imaging.
\newblock {\em Cell}, 174(4):968--981, 2018.

\bibitem{giesen2014highly}
Charlotte Giesen, Hao~AO Wang, Denis Schapiro, Nevena Zivanovic, Andrea Jacobs,
  Bodo Hattendorf, Peter~J Sch{\"u}ffler, Daniel Grolimund, Joachim~M Buhmann,
  Simone Brandt, et~al.
\newblock Highly multiplexed imaging of tumor tissues with subcellular
  resolution by mass cytometry.
\newblock {\em Nature methods}, 11(4):417--422, 2014.

\bibitem{keren2019mibi}
Leeat Keren, Marc Bosse, Steve Thompson, Tyler Risom, Kausalia Vijayaragavan,
  Erin McCaffrey, Diana Marquez, Roshan Angoshtari, Noah~F Greenwald, Harris
  Fienberg, et~al.
\newblock Mibi-tof: A multiplexed imaging platform relates cellular phenotypes
  and tissue structure.
\newblock {\em Science advances}, 5(10):eaax5851, 2019.

\bibitem{lecun2015deep}
Yann LeCun, Yoshua Bengio, and Geoffrey Hinton.
\newblock Deep learning.
\newblock {\em nature}, 521(7553):436--444, 2015.

\bibitem{mcculloch43a}
Warren Mcculloch and Walter Pitts.
\newblock A logical calculus of ideas immanent in nervous activity.
\newblock {\em Bulletin of Mathematical Biophysics}, 5:127--147, 1943.

\bibitem{rumelhart1985learning}
David~E Rumelhart, Geoffrey~E Hinton, and Ronald~J Williams.
\newblock Learning internal representations by error propagation.
\newblock Technical report, California Univ San Diego La Jolla Inst for
  Cognitive Science, 1985.

\bibitem{rumelhart1986learning}
David~E Rumelhart, Geoffrey~E Hinton, and Ronald~J Williams.
\newblock Learning representations by back-propagating errors.
\newblock {\em nature}, 323(6088):533--536, 1986.

\bibitem{scarselli2008graph}
Franco Scarselli, Marco Gori, Ah~Chung Tsoi, Markus Hagenbuchner, and Gabriele
  Monfardini.
\newblock The graph neural network model.
\newblock {\em IEEE transactions on neural networks}, 20(1):61--80, 2008.

\bibitem{popescu2009multilayer}
Marius-Constantin Popescu, Valentina~E Balas, Liliana Perescu-Popescu, and
  Nikos Mastorakis.
\newblock Multilayer perceptron and neural networks.
\newblock {\em WSEAS Transactions on Circuits and Systems}, 8(7):579--588,
  2009.

\bibitem{weng2017representation}
Wei-Hung Weng, Mingwu Gao, Ze~He, Susu Yan, and Peter Szolovits.
\newblock Representation and reinforcement learning for personalized glycemic
  control in septic patients.
\newblock {\em arXiv preprint arXiv:1712.00654}, 2017.

\bibitem{jaiswal2018large}
Ayush Jaiswal, Dong Guo, Cauligi~S Raghavendra, and Paul Thompson.
\newblock Large-scale unsupervised deep representation learning for brain
  structure.
\newblock {\em arXiv preprint arXiv:1805.01049}, 2018.

\bibitem{tran2018learning}
Phi~Vu Tran.
\newblock Learning to make predictions on graphs with autoencoders.
\newblock In {\em 2018 IEEE 5th international conference on data science and
  advanced analytics (DSAA)}, pages 237--245. IEEE, 2018.

\bibitem{zeune2020deep}
Leonie~L Zeune, Yoeri~E Boink, Guus van Dalum, Afroditi Nanou, Sanne de~Wit,
  Kiki~C Andree, Joost~F Swennenhuis, Stephan~A van Gils, Leon~WMM Terstappen,
  and Christoph Brune.
\newblock Deep learning of circulating tumour cells.
\newblock {\em Nature Machine Intelligence}, 2(2):124--133, 2020.

\bibitem{kingma2013auto}
Diederik~P Kingma and Max Welling.
\newblock Auto-encoding variational bayes.
\newblock {\em arXiv preprint arXiv:1312.6114}, 2013.

\bibitem{kingma2019introduction}
Diederik~P Kingma, Max Welling, et~al.
\newblock An introduction to variational autoencoders.
\newblock {\em Foundations and Trends{\textregistered} in Machine Learning},
  12(4):307--392, 2019.

\bibitem{goodfellow2014generative}
Ian Goodfellow, Jean Pouget-Abadie, Mehdi Mirza, Bing Xu, David Warde-Farley,
  Sherjil Ozair, Aaron Courville, and Yoshua Bengio.
\newblock Generative adversarial nets.
\newblock {\em Advances in neural information processing systems}, 27, 2014.

\bibitem{lecun1995convolutional}
Yann LeCun, Yoshua Bengio, et~al.
\newblock Convolutional networks for images, speech, and time series.
\newblock {\em The handbook of brain theory and neural networks},
  3361(10):1995, 1995.

\bibitem{o2015introduction}
Keiron O'Shea and Ryan Nash.
\newblock An introduction to convolutional neural networks.
\newblock {\em arXiv preprint arXiv:1511.08458}, 2015.

\bibitem{medsker2001recurrent}
Larry~R Medsker and LC~Jain.
\newblock Recurrent neural networks.
\newblock {\em Design and Applications}, 5:64--67, 2001.

\bibitem{hochreiter1998vanishing}
Sepp Hochreiter.
\newblock The vanishing gradient problem during learning recurrent neural nets
  and problem solutions.
\newblock {\em International Journal of Uncertainty, Fuzziness and
  Knowledge-Based Systems}, 6(02):107--116, 1998.

\bibitem{pascanu2013difficulty}
Razvan Pascanu, Tomas Mikolov, and Yoshua Bengio.
\newblock On the difficulty of training recurrent neural networks.
\newblock In {\em International conference on machine learning}, pages
  1310--1318. PMLR, 2013.

\bibitem{hochreiter1997long}
Sepp Hochreiter and J{\"u}rgen Schmidhuber.
\newblock Long short-term memory.
\newblock {\em Neural computation}, 9(8):1735--1780, 1997.

\bibitem{chung2014empirical}
Junyoung Chung, Caglar Gulcehre, KyungHyun Cho, and Yoshua Bengio.
\newblock Empirical evaluation of gated recurrent neural networks on sequence
  modeling.
\newblock {\em arXiv preprint arXiv:1412.3555}, 2014.

\bibitem{henaff2015deep}
Mikael Henaff, Joan Bruna, and Yann LeCun.
\newblock Deep convolutional networks on graph-structured data.
\newblock {\em arXiv preprint arXiv:1506.05163}, 2015.

\bibitem{ma2021deep}
Yao Ma and Jiliang Tang.
\newblock {\em Deep learning on graphs}.
\newblock Cambridge University Press, 2021.

\bibitem{zhang2020deep}
Ziwei Zhang, Peng Cui, and Wenwu Zhu.
\newblock Deep learning on graphs: A survey.
\newblock {\em IEEE Transactions on Knowledge and Data Engineering}, 2020.

\bibitem{kipf2016semi}
Thomas~N Kipf and Max Welling.
\newblock Semi-supervised classification with graph convolutional networks.
\newblock {\em arXiv preprint arXiv:1609.02907}, 2016.

\bibitem{paszke2019pytorch}
Adam Paszke, Sam Gross, Francisco Massa, Adam Lerer, James Bradbury, Gregory
  Chanan, Trevor Killeen, Zeming Lin, Natalia Gimelshein, Luca Antiga, et~al.
\newblock Pytorch: An imperative style, high-performance deep learning library.
\newblock {\em Advances in neural information processing systems}, 32, 2019.

\bibitem{abadi2016tensorflow}
Mart{\'\i}n Abadi, Paul Barham, Jianmin Chen, Zhifeng Chen, Andy Davis, Jeffrey
  Dean, Matthieu Devin, Sanjay Ghemawat, Geoffrey Irving, Michael Isard, et~al.
\newblock $\{$TensorFlow$\}$: a system for $\{$Large-Scale$\}$ machine
  learning.
\newblock In {\em 12th USENIX symposium on operating systems design and
  implementation (OSDI 16)}, pages 265--283, 2016.

\bibitem{chollet2018keras}
Fran{\c{c}}ois Chollet et~al.
\newblock Keras: The python deep learning library.
\newblock {\em Astrophysics source code library}, pages ascl--1806, 2018.

\bibitem{chen2015mxnet}
Tianqi Chen, Mu~Li, Yutian Li, Min Lin, Naiyan Wang, Minjie Wang, Tianjun Xiao,
  Bing Xu, Chiyuan Zhang, and Zheng Zhang.
\newblock Mxnet: A flexible and efficient machine learning library for
  heterogeneous distributed systems.
\newblock {\em arXiv preprint arXiv:1512.01274}, 2015.

\bibitem{wang2019deep}
Minjie Wang, Da~Zheng, Zihao Ye, Quan Gan, Mufei Li, Xiang Song, Jinjing Zhou,
  Chao Ma, Lingfan Yu, Yu~Gai, et~al.
\newblock Deep graph library: A graph-centric, highly-performant package for
  graph neural networks.
\newblock {\em arXiv preprint arXiv:1909.01315}, 2019.

\bibitem{fey2019fast}
Matthias Fey and Jan~Eric Lenssen.
\newblock Fast graph representation learning with pytorch geometric.
\newblock {\em arXiv preprint arXiv:1903.02428}, 2019.

\bibitem{hagberg2008exploring}
Aric Hagberg, Pieter Swart, and Daniel S~Chult.
\newblock Exploring network structure, dynamics, and function using networkx.
\newblock Technical report, Los Alamos National Lab.(LANL), Los Alamos, NM
  (United States), 2008.

\bibitem{Subramanian2020}
A.K Subramanian.
\newblock Pytorch-vae.
\newblock \url{https://github.com/AntixK/PyTorch-VAE}, 2020.

\bibitem{culjak2012brief}
Ivan Culjak, David Abram, Tomislav Pribanic, Hrvoje Dzapo, and Mario Cifrek.
\newblock A brief introduction to opencv.
\newblock In {\em 2012 proceedings of the 35th international convention MIPRO},
  pages 1725--1730. IEEE, 2012.

\bibitem{islam2011characterization}
Saiful Islam, Una Kj{\"a}llquist, Annalena Moliner, Pawel Zajac, Jian-Bing Fan,
  Peter L{\"o}nnerberg, and Sten Linnarsson.
\newblock Characterization of the single-cell transcriptional landscape by
  highly multiplex rna-seq.
\newblock {\em Genome research}, 21(7):1160--1167, 2011.

\bibitem{natarajan2019single}
Kedar~Nath Natarajan.
\newblock Single-cell tagged reverse transcription (strt-seq).
\newblock In {\em Single Cell Methods}, pages 133--153. Springer, 2019.

\bibitem{haghverdi2018batch}
Laleh Haghverdi, Aaron~TL Lun, Michael~D Morgan, and John~C Marioni.
\newblock Batch effects in single-cell rna-sequencing data are corrected by
  matching mutual nearest neighbors.
\newblock {\em Nature biotechnology}, 36(5):421--427, 2018.

\bibitem{booeshaghi2022depth}
A~Sina Booeshaghi, Ingileif~B Hallgr{\'\i}msd{\'o}ttir, {\'A}ngel
  G{\'a}lvez-Merch{\'a}n, and Lior Pachter.
\newblock Depth normalization for single-cell genomics count data.
\newblock {\em bioRxiv}, 2022.

\bibitem{cole2019performance}
Michael~B Cole, Davide Risso, Allon Wagner, David DeTomaso, John Ngai,
  Elizabeth Purdom, Sandrine Dudoit, and Nir Yosef.
\newblock Performance assessment and selection of normalization procedures for
  single-cell rna-seq.
\newblock {\em Cell systems}, 8(4):315--328, 2019.

\bibitem{shao2020sccatch}
Xin Shao, Jie Liao, Xiaoyan Lu, Rui Xue, Ni~Ai, and Xiaohui Fan.
\newblock sccatch: automatic annotation on cell types of clusters from
  single-cell rna sequencing data.
\newblock {\em Iscience}, 23(3):100882, 2020.

\bibitem{tritschler2019concepts}
Sophie Tritschler, Maren B{\"u}ttner, David~S Fischer, Marius Lange, Volker
  Bergen, Heiko Lickert, and Fabian~J Theis.
\newblock Concepts and limitations for learning developmental trajectories from
  single cell genomics.
\newblock {\em Development}, 146(12):dev170506, 2019.

\bibitem{fan2020single}
Jean Fan, Kamil Slowikowski, and Fan Zhang.
\newblock Single-cell transcriptomics in cancer: computational challenges and
  opportunities.
\newblock {\em Experimental \& Molecular Medicine}, 52(9):1452--1465, 2020.

\bibitem{almet2021landscape}
Axel~A Almet, Zixuan Cang, Suoqin Jin, and Qing Nie.
\newblock The landscape of cell--cell communication through single-cell
  transcriptomics.
\newblock {\em Current opinion in systems biology}, 26:12--23, 2021.

\bibitem{chung2017single}
Woosung Chung, Hye~Hyeon Eum, Hae-Ock Lee, Kyung-Min Lee, Han-Byoel Lee,
  Kyu-Tae Kim, Han~Suk Ryu, Sangmin Kim, Jeong~Eon Lee, Yeon~Hee Park, et~al.
\newblock Single-cell rna-seq enables comprehensive tumour and immune cell
  profiling in primary breast cancer.
\newblock {\em Nature communications}, 8(1):1--12, 2017.

\bibitem{navin2015first}
Nicholas~E Navin.
\newblock The first five years of single-cell cancer genomics and beyond.
\newblock {\em Genome research}, 25(10):1499--1507, 2015.

\bibitem{chattopadhyay2014single}
Pratip~K Chattopadhyay, Todd~M Gierahn, Mario Roederer, and J~Christopher Love.
\newblock Single-cell technologies for monitoring immune systems.
\newblock {\em Nature immunology}, 15(2):128--135, 2014.

\bibitem{soneson2018bias}
Charlotte Soneson and Mark~D Robinson.
\newblock Bias, robustness and scalability in single-cell differential
  expression analysis.
\newblock {\em Nature methods}, 15(4):255--261, 2018.

\bibitem{van2020trajectory}
Koen Van~den Berge, Hector Roux~de B{\'e}zieux, Kelly Street, Wouter Saelens,
  Robrecht Cannoodt, Yvan Saeys, Sandrine Dudoit, and Lieven Clement.
\newblock Trajectory-based differential expression analysis for single-cell
  sequencing data.
\newblock {\em Nature communications}, 11(1):1--13, 2020.

\bibitem{wagner2019single}
Johanna Wagner, Maria~Anna Rapsomaniki, St{\'e}phane Chevrier, Tobias
  Anzeneder, Claus Langwieder, August Dykgers, Martin Rees, Annette Ramaswamy,
  Simone Muenst, Savas~Deniz Soysal, et~al.
\newblock A single-cell atlas of the tumor and immune ecosystem of human breast
  cancer.
\newblock {\em Cell}, 177(5):1330--1345, 2019.

\bibitem{travaglini2020molecular}
Kyle~J Travaglini, Ahmad~N Nabhan, Lolita Penland, Rahul Sinha, Astrid Gillich,
  Rene~V Sit, Stephen Chang, Stephanie~D Conley, Yasuo Mori, Jun Seita, et~al.
\newblock A molecular cell atlas of the human lung from single-cell rna
  sequencing.
\newblock {\em Nature}, 587(7835):619--625, 2020.

\bibitem{hwang2018single}
Byungjin Hwang, Ji~Hyun Lee, and Duhee Bang.
\newblock Single-cell rna sequencing technologies and bioinformatics pipelines.
\newblock {\em Experimental \& molecular medicine}, 50(8):1--14, 2018.

\bibitem{mallory2020methods}
Xian~F Mallory, Mohammadamin Edrisi, Nicholas Navin, and Luay Nakhleh.
\newblock Methods for copy number aberration detection from single-cell
  dna-sequencing data.
\newblock {\em Genome biology}, 21(1):1--22, 2020.

\bibitem{labib2020single}
Mahmoud Labib and Shana~O Kelley.
\newblock Single-cell analysis targeting the proteome.
\newblock {\em Nature Reviews Chemistry}, 4(3):143--158, 2020.

\bibitem{peterson2017multiplexed}
Vanessa~M Peterson, Kelvin~Xi Zhang, Namit Kumar, Jerelyn Wong, Lixia Li,
  Douglas~C Wilson, Renee Moore, Terrill~K McClanahan, Svetlana Sadekova, and
  Joel~A Klappenbach.
\newblock Multiplexed quantification of proteins and transcripts in single
  cells.
\newblock {\em Nature biotechnology}, 35(10):936--939, 2017.

\bibitem{stoeckius2017simultaneous}
Marlon Stoeckius, Christoph Hafemeister, William Stephenson, Brian
  Houck-Loomis, Pratip~K Chattopadhyay, Harold Swerdlow, Rahul Satija, and
  Peter Smibert.
\newblock Simultaneous epitope and transcriptome measurement in single cells.
\newblock {\em Nature methods}, 14(9):865--868, 2017.

\bibitem{cao2018joint}
Junyue Cao, Darren~A Cusanovich, Vijay Ramani, Delasa Aghamirzaie, Hannah~A
  Pliner, Andrew~J Hill, Riza~M Daza, Jose~L McFaline-Figueroa, Jonathan~S
  Packer, Lena Christiansen, et~al.
\newblock Joint profiling of chromatin accessibility and gene expression in
  thousands of single cells.
\newblock {\em Science}, 361(6409):1380--1385, 2018.
\newblock
  \href{https://www.ncbi.nlm.nih.gov/geo/query/acc.cgi?acc=GSE117089}{Dataset
  Link: https://www.ncbi.nlm.nih.gov/geo/query/acc.cgi?acc=GSE117089}.

\bibitem{zhu2019ultra}
Chenxu Zhu, Miao Yu, Hui Huang, Ivan Juric, Armen Abnousi, Rong Hu, Jacinta
  Lucero, M~Margarita Behrens, Ming Hu, and Bing Ren.
\newblock An ultra high-throughput method for single-cell joint analysis of
  open chromatin and transcriptome.
\newblock {\em Nature structural \& molecular biology}, 26(11):1063--1070,
  2019.

\bibitem{chen2019high}
Song Chen, Blue~B Lake, and Kun Zhang.
\newblock High-throughput sequencing of the transcriptome and chromatin
  accessibility in the same cell.
\newblock {\em Nature biotechnology}, 37(12):1452--1457, 2019.
\newblock
  \href{https://www.ncbi.nlm.nih.gov/geo/query/acc.cgi?acc=GSE126074}{Dataset
  Link: https://www.ncbi.nlm.nih.gov/geo/query/acc.cgi?acc=GSE126074}.

\bibitem{hao2021integrated}
Yuhan Hao, Stephanie Hao, Erica Andersen-Nissen, William~M Mauck~III, Shiwei
  Zheng, Andrew Butler, Maddie~J Lee, Aaron~J Wilk, Charlotte Darby, Michael
  Zager, et~al.
\newblock Integrated analysis of multimodal single-cell data.
\newblock {\em Cell}, 184(13):3573--3587, 2021.
\newblock \href{https://github.com/satijalab/seurat}{Code Link:
  https://github.com/satijalab/seurat}.

\bibitem{ma2020chromatin}
Sai Ma, Bing Zhang, Lindsay~M LaFave, Andrew~S Earl, Zachary Chiang, Yan Hu,
  Jiarui Ding, Alison Brack, Vinay~K Kartha, Tristan Tay, et~al.
\newblock Chromatin potential identified by shared single-cell profiling of rna
  and chromatin.
\newblock {\em Cell}, 183(4):1103--1116, 2020.
\newblock
  \href{https://www.ncbi.nlm.nih.gov/geo/query/acc.cgi?acc=GSE140203}{Dataset
  Link: https://www.ncbi.nlm.nih.gov/geo/query/acc.cgi?acc=GSE140203}.

\bibitem{duren2018integrative}
Zhana Duren, Xi~Chen, Mahdi Zamanighomi, Wanwen Zeng, Ansuman~T Satpathy,
  Howard~Y Chang, Yong Wang, and Wing~Hung Wong.
\newblock Integrative analysis of single-cell genomics data by coupled
  nonnegative matrix factorizations.
\newblock {\em Proceedings of the National Academy of Sciences},
  115(30):7723--7728, 2018.

\bibitem{zeng2019}
Wanwen Zeng, Xi~Chen, Zhana Duren, Yong Wang, Rui Jiang, and Wing~Hung Wong.
\newblock Dc3 is a method for deconvolution and coupled clustering from bulk
  and single-cell genomics data.
\newblock {\em Nature communications}, 10(1):1--11, 2019.

\bibitem{jansen2019building}
Camden Jansen, Ricardo~N Ramirez, Nicole~C El-Ali, David Gomez-Cabrero, Jesper
  Tegner, Matthias Merkenschlager, Ana Conesa, and Ali Mortazavi.
\newblock Building gene regulatory networks from scatac-seq and scrna-seq using
  linked self organizing maps.
\newblock {\em PLoS computational biology}, 15(11):e1006555, 2019.
\newblock \href{https://github.com/csjansen/SOMatic}{Code Link:
  https://github.com/csjansen/SOMatic}.

\bibitem{rautenstrauch2021intricacies}
Pia Rautenstrauch, Anna Hendrika~Cornelia Vlot, Sepideh Saran, and Uwe Ohler.
\newblock Intricacies of single-cell multi-omics data integration.
\newblock {\em Trends in Genetics}, 2021.

\bibitem{luecken2021sandbox}
Malte~D Luecken, Daniel~Bernard Burkhardt, Robrecht Cannoodt, Christopher
  Lance, Aditi Agrawal, Hananeh Aliee, Ann~T Chen, Louise Deconinck, Angela~M
  Detweiler, Alejandro~A Granados, et~al.
\newblock A sandbox for prediction and integration of dna, rna, and proteins in
  single cells.
\newblock In {\em Thirty-fifth Conference on Neural Information Processing
  Systems Datasets and Benchmarks Track (Round 2)}, 2021.

\bibitem{lin2022scjoint}
Yingxin Lin, Tung-Yu Wu, Sheng Wan, Jean~YH Yang, Wing~H Wong, and YX~Wang.
\newblock scjoint integrates atlas-scale single-cell rna-seq and atac-seq data
  with transfer learning.
\newblock {\em Nature Biotechnology}, 40(5):703--710, 2022.
\newblock \href{https://github.com/SydneyBioX/scJoint}{Code Link:
  https://github.com/SydneyBioX/scJoint}.

\bibitem{zhang2022scdart}
Ziqi Zhang, Chengkai Yang, and Xiuwei Zhang.
\newblock scdart: integrating unmatched scrna-seq and scatac-seq data and
  learning cross-modality relationship simultaneously.
\newblock {\em Genome biology}, 23(1):1--28, 2022.
\newblock \href{https://github.com/PeterZZQ/scDART\_test}{Code Link:
  https://github.com/PeterZZQ/scDART\_test}.

\bibitem{jin2020scai}
Suoqin Jin, Lihua Zhang, and Qing Nie.
\newblock scai: an unsupervised approach for the integrative analysis of
  parallel single-cell transcriptomic and epigenomic profiles.
\newblock {\em Genome biology}, 21(1):1--19, 2020.

\bibitem{liu2020}
Jialin Liu, Chao Gao, Joshua Sodicoff, Velina Kozareva, Evan~Z Macosko, and
  Joshua~D Welch.
\newblock Jointly defining cell types from multiple single-cell datasets using
  liger.
\newblock {\em Nature protocols}, 15(11):3632--3662, 2020.

\bibitem{kriebel2022uinmf}
April~R Kriebel and Joshua~D Welch.
\newblock Uinmf performs mosaic integration of single-cell multi-omic datasets
  using nonnegative matrix factorization.
\newblock {\em Nature communications}, 13(1):1--17, 2022.

\bibitem{stanley2020}
Jay~S Stanley~III, Scott Gigante, Guy Wolf, and Smita Krishnaswamy.
\newblock Harmonic alignment.
\newblock In {\em Proceedings of the 2020 SIAM International Conference on Data
  Mining}, pages 316--324. SIAM, 2020.

\bibitem{cao2020unsupervised}
Kai Cao, Xiangqi Bai, Yiguang Hong, and Lin Wan.
\newblock Unsupervised topological alignment for single-cell multi-omics
  integration.
\newblock {\em Bioinformatics}, 36(Supplement\_1):i48--i56, 2020.

\bibitem{jain2021}
Mika~Sarkin Jain, Krzysztof Polanski, Cecilia~Dominguez Conde, Xi~Chen, Jongeun
  Park, Lira Mamanova, Andrew Knights, Rachel~A Botting, Emily Stephenson,
  Muzlifah Haniffa, et~al.
\newblock Multimap: dimensionality reduction and integration of multimodal
  data.
\newblock {\em Genome biology}, 22(1):1--26, 2021.

\bibitem{zeng2021}
Pengcheng Zeng and Zhixiang Lin.
\newblock Couple coc+: an information-theoretic co-clustering-based transfer
  learning framework for the integrative analysis of single-cell genomic data.
\newblock {\em PLoS Computational Biology}, 17(6):e1009064, 2021.

\bibitem{welch2017}
Joshua~D Welch, Alexander~J Hartemink, and Jan~F Prins.
\newblock Matcher: manifold alignment reveals correspondence between single
  cell transcriptome and epigenome dynamics.
\newblock {\em Genome biology}, 18(1):1--19, 2017.

\bibitem{lin2020model}
Zhixiang Lin, Mahdi Zamanighomi, Timothy Daley, Shining Ma, and Wing~Hung Wong.
\newblock Model-based approach to the joint analysis of single-cell data on
  chromatin accessibility and gene expression.
\newblock {\em Statistical Science}, 35(1):2--13, 2020.

\bibitem{wangwu2021scamace}
Jiaxuan Wangwu, Zexuan Sun, and Zhixiang Lin.
\newblock scamace: model-based approach to the joint analysis of single-cell
  data on chromatin accessibility, gene expression and methylation.
\newblock {\em Bioinformatics}, 37(21):3874--3880, 2021.

\bibitem{argelaguet2020mofa+}
Ricard Argelaguet, Damien Arnol, Danila Bredikhin, Yonatan Deloro, Britta
  Velten, John~C Marioni, and Oliver Stegle.
\newblock Mofa+: a statistical framework for comprehensive integration of
  multi-modal single-cell data.
\newblock {\em Genome biology}, 21(1):1--17, 2020.

\bibitem{duren2017modeling}
Zhana Duren, Xi~Chen, Rui Jiang, Yong Wang, and Wing~Hung Wong.
\newblock Modeling gene regulation from paired expression and chromatin
  accessibility data.
\newblock {\em Proceedings of the National Academy of Sciences},
  114(25):E4914--E4923, 2017.

\bibitem{yang2016non}
Zi~Yang and George Michailidis.
\newblock A non-negative matrix factorization method for detecting modules in
  heterogeneous omics multi-modal data.
\newblock {\em Bioinformatics}, 32(1):1--8, 2016.

\bibitem{dou2020}
Jinzhuang Dou, Shaoheng Liang, Vakul Mohanty, Xuesen Cheng, Sangbae Kim, Jongsu
  Choi, Yumei Li, Katayoun Rezvani, Rui Chen, and Ken Chen.
\newblock Unbiased integration of single cell multi-omics data.
\newblock {\em BioRxiv}, 2020.

\bibitem{stuart2019}
Tim Stuart, Andrew Butler, Paul Hoffman, Christoph Hafemeister, Efthymia
  Papalexi, William~M Mauck~III, Yuhan Hao, Marlon Stoeckius, Peter Smibert,
  and Rahul Satija.
\newblock Comprehensive integration of single-cell data.
\newblock {\em Cell}, 177(7):1888--1902, 2019.
\newblock
  \href{https://www.ncbi.nlm.nih.gov/geo/query/acc.cgi?acc=GSE128639}{Dataset
  Link: https://www.ncbi.nlm.nih.gov/geo/query/acc.cgi?acc=GSE128639}.

\bibitem{luo2022single}
Chongyuan Luo, Hanqing Liu, Fangming Xie, Ethan~J Armand, Kimberly Siletti,
  Trygve~E Bakken, Rongxin Fang, Wayne~I Doyle, Tim Stuart, Rebecca~D Hodge,
  et~al.
\newblock Single nucleus multi-omics identifies human cortical cell regulatory
  genome diversity.
\newblock {\em Cell genomics}, 2(3):100107, 2022.

\bibitem{singh2020}
Ritambhara Singh, Pinar Demetci, Giancarlo Bonora, Vijay Ramani, Choli Lee,
  He~Fang, Zhijun Duan, Xinxian Deng, Jay Shendure, Christine Disteche, et~al.
\newblock Unsupervised manifold alignment for single-cell multi-omics data.
\newblock In {\em Proceedings of the 11th ACM International Conference on
  Bioinformatics, Computational Biology and Health Informatics}, pages 1--10,
  2020.

\bibitem{lake2018}
Blue~B Lake, Song Chen, Brandon~C Sos, Jean Fan, Gwendolyn~E Kaeser, Yun~C
  Yung, Thu~E Duong, Derek Gao, Jerold Chun, Peter~V Kharchenko, et~al.
\newblock Integrative single-cell analysis of transcriptional and epigenetic
  states in the human adult brain.
\newblock {\em Nature biotechnology}, 36(1):70--80, 2018.

\bibitem{kohonen1982self}
Teuvo Kohonen.
\newblock Self-organized formation of topologically correct feature maps.
\newblock {\em Biological cybernetics}, 43(1):59--69, 1982.

\bibitem{mclean2010great}
Cory~Y McLean, Dave Bristor, Michael Hiller, Shoa~L Clarke, Bruce~T Schaar,
  Craig~B Lowe, Aaron~M Wenger, and Gill Bejerano.
\newblock Great improves functional interpretation of cis-regulatory regions.
\newblock {\em Nature biotechnology}, 28(5):495--501, 2010.

\bibitem{stark2020scim}
Stefan~G Stark, Joanna Ficek, Francesco Locatello, Ximena Bonilla, St{\'e}phane
  Chevrier, Franziska Singer, Gunnar R{\"a}tsch, and Kjong-Van Lehmann.
\newblock Scim: universal single-cell matching with unpaired feature sets.
\newblock {\em Bioinformatics}, 36(Supplement\_2):i919--i927, 2020.
\newblock \href{https://github.com/ratschlab/scim}{Code Link:
  https://github.com/ratschlab/scim}.

\bibitem{cao2022multi}
Zhi-Jie Cao and Ge~Gao.
\newblock Multi-omics single-cell data integration and regulatory inference
  with graph-linked embedding.
\newblock {\em Nature Biotechnology}, pages 1--9, 2022.
\newblock \href{https://github.com/gao-lab/GLUE}{Code Link:
  https://github.com/gao-lab/GLUE}.

\bibitem{cao2022}
Kai Cao, Yiguang Hong, and Lin Wan.
\newblock Manifold alignment for heterogeneous single-cell multi-omics data
  integration using pamona.
\newblock {\em Bioinformatics}, 38(1):211--219, 2022.

\bibitem{wen2022graph}
Hongzhi Wen, Jiayuan Ding, Wei Jin, Yiqi Wang, Yuying Xie, and Jiliang Tang.
\newblock Graph neural networks for multimodal single-cell data integration.
\newblock In {\em Proceedings of the 28th ACM SIGKDD Conference on Knowledge
  Discovery and Data Mining}, pages 4153--4163, 2022.
\newblock \href{https://github.com/OmicsML/dance}{Code Link:
  https://github.com/OmicsML/dance}.

\bibitem{yang2021}
Karren~Dai Yang, Anastasiya Belyaeva, Saradha Venkatachalapathy, Karthik
  Damodaran, Abigail Katcoff, Adityanarayanan Radhakrishnan, GV~Shivashankar,
  and Caroline Uhler.
\newblock Multi-domain translation between single-cell imaging and sequencing
  data using autoencoders.
\newblock {\em Nature communications}, 12(1):1--10, 2021.
\newblock \href{https://github.com/uhlerlab/cross-modal-autoencoders}{Code
  Link: https://github.com/uhlerlab/cross-modal-autoencoders}.

\bibitem{amodio2018magan}
Matthew Amodio and Smita Krishnaswamy.
\newblock Magan: Aligning biological manifolds.
\newblock In {\em International Conference on Machine Learning}, pages
  215--223. PMLR, 2018.
\newblock \href{https://github.com/KrishnaswamyLab/MAGAN}{Code Link:
  https://github.com/KrishnaswamyLab/MAGAN}.

\bibitem{gong2021cobolt}
Boying Gong, Yun Zhou, and Elizabeth Purdom.
\newblock Cobolt: integrative analysis of multimodal single-cell sequencing
  data.
\newblock {\em Genome biology}, 22(1):1--21, 2021.
\newblock \href{https://github.com/epurdom/cobolt\_manuscript}{Code Link:
  https://github.com/epurdom/cobolt\_manuscript}.

\bibitem{xu2022smile}
Yang Xu, Priyojit Das, and Rachel~Patton McCord.
\newblock Smile: mutual information learning for integration of single-cell
  omics data.
\newblock {\em Bioinformatics}, 38(2):476--486, 2022.
\newblock \href{https://https://github.com/rpmccordlab/SMILE}{Code Link:
  https://github.com/rpmccordlab/SMILE}.

\bibitem{wu2021babel}
Kevin~E Wu, Kathryn~E Yost, Howard~Y Chang, and James Zou.
\newblock Babel enables cross-modality translation between multiomic profiles
  at single-cell resolution.
\newblock {\em Proceedings of the National Academy of Sciences},
  118(15):e2023070118, 2021.
\newblock \href{https://github.com/wukevin/babel}{Code Link:
  https://github.com/wukevin/babel}.

\bibitem{minoura2021mixture}
Kodai Minoura, Ko~Abe, Hyunha Nam, Hiroyoshi Nishikawa, and Teppei Shimamura.
\newblock A mixture-of-experts deep generative model for integrated analysis of
  single-cell multiomics data.
\newblock {\em Cell reports methods}, 1(5):100071, 2021.
\newblock \href{https://github.com/kodaim1115/scMM}{Code Link:
  https://github.com/kodaim1115/scMM}.

\bibitem{zuo2021deep}
Chunman Zuo and Luonan Chen.
\newblock Deep-joint-learning analysis model of single cell transcriptome and
  open chromatin accessibility data.
\newblock {\em Briefings in Bioinformatics}, 22(4):bbaa287, 2021.
\newblock \href{https://github.com/cmzuo11/scMVAE}{Code Link:
  https://github.com/cmzuo11/scMVAE}.

\bibitem{zuo2021dcca}
Chunman Zuo, Hao Dai, and Luonan Chen.
\newblock Deep cross-omics cycle attention model for joint analysis of
  single-cell multi-omics data.
\newblock {\em Bioinformatics}, 37(22):4091--4099, 2021.
\newblock \href{https://github.com/cmzuo11/DCCA}{Code Link:
  https://github.com/cmzuo11/DCCA}.

\bibitem{link2dance}
Jiayuan Ding, Hongzhi Wen, Wenzhuo Tang, Renming Liu, Zhaoheng Li, Julian
  Venegas, Runze Su, Dylan Molho, Wei Jin, Wangyang Zuo, et~al.
\newblock Dance: A deep learning library and benchmark for single-cell
  analysis.
\newblock {\em bioRxiv}, 2022.
\newblock \href{https://github.com/OmicsML/dance}{Code Link:
  https://github.com/OmicsML/dance}.

\bibitem{10x_Genomics_CITE-seq}
10x genomics cite-seq.
\newblock
  \url{https://support.10xgenomics.com/single-cell-gene-expression/datasets/3.1.0/5k\_pbmc\_protein\_v3\_nextgem}.

\bibitem{Multiome_Chromium_X}
10x multiome chromium x.
\newblock
  \url{https://support.10xgenomics.com/single-cell-multiome-atac-gex/datasets/2.0.0/10k\_PBMC\_Multiome\_nextgem\_Chromium\_X}.

\bibitem{Multiome_unsorted}
10x multiome unsorted dataset.
\newblock
  \url{https://www.10xgenomics.com/resources/datasets/pbmc-from-a-healthy-donor-no-cell-sorting-10-k-1-standard-2-0-0}.

\bibitem{clark2018scnmt}
Stephen~J Clark, Ricard Argelaguet, Chantriolnt-Andreas Kapourani, Thomas~M
  Stubbs, Heather~J Lee, Celia Alda-Catalinas, Felix Krueger, Guido
  Sanguinetti, Gavin Kelsey, John~C Marioni, et~al.
\newblock scnmt-seq enables joint profiling of chromatin accessibility dna
  methylation and transcription in single cells.
\newblock {\em Nature communications}, 9(1):1--9, 2018.
\newblock
  \href{https://www.ncbi.nlm.nih.gov/geo/query/acc.cgi?acc=GSE109262}{Dataset
  Link: https://www.ncbi.nlm.nih.gov/geo/query/acc.cgi?acc=GSE109262}.

\bibitem{mimitou2021scalable}
Eleni~P Mimitou, Caleb~A Lareau, Kelvin~Y Chen, Andre~L Zorzetto-Fernandes,
  Yuhan Hao, Yusuke Takeshima, Wendy Luo, Tse-Shun Huang, Bertrand~Z Yeung,
  Efthymia Papalexi, et~al.
\newblock Scalable, multimodal profiling of chromatin accessibility, gene
  expression and protein levels in single cells.
\newblock {\em Nature biotechnology}, 39(10):1246--1258, 2021.
\newblock
  \href{https://www.ncbi.nlm.nih.gov/geo/query/acc.cgi?acc=GSE109262}{Dataset
  Link: https://www.ncbi.nlm.nih.gov/geo/query/acc.cgi?acc=GSE156478}.

\bibitem{svensson2017}
Valentine Svensson, Kedar~Nath Natarajan, Lam-Ha Ly, Ricardo~J Miragaia,
  Charlotte Labalette, Iain~C Macaulay, Ana Cvejic, and Sarah Teichmann.
\newblock Power analysis of single-cell rna-sequencing experiments.
\newblock {\em Nature Methods}, 14(4):381--387, 2017.

\bibitem{kharchenko2014}
Peter~V. Kharchenko, Lev Silberstein, and David~T. Scadden.
\newblock Bayesian approach to single-cell differential expression analysis.
\newblock {\em Nature Methods}, 11(7):740–--742, 2014.

\bibitem{qiu2020embracing}
Peng Qiu.
\newblock Embracing the dropouts in single-cell rna-seq analysis.
\newblock {\em Nature communications}, 11(1):1--9, 2020.

\bibitem{van2017magic}
David van Dijk, Juozas Nainys, Roshan Sharma, Pooja Kaithail, Ambrose~J Carr,
  Kevin~R Moon, Linas Mazutis, Guy Wolf, Smita Krishnaswamy, and Dana Pe'er.
\newblock Magic: A diffusion-based imputation method reveals gene-gene
  interactions in single-cell rna-sequencing data.
\newblock {\em BioRxiv}, page 111591, 2017.

\bibitem{ronen2018netsmooth}
Jonathan Ronen and Altuna Akalin.
\newblock netsmooth: Network-smoothing based imputation for single cell
  rna-seq.
\newblock {\em F1000Research}, 7, 2018.

\bibitem{gong2018drimpute}
Wuming Gong, Il-Youp Kwak, Pruthvi Pota, Naoko Koyano-Nakagawa, and Daniel~J
  Garry.
\newblock Drimpute: imputing dropout events in single cell rna sequencing data.
\newblock {\em BMC bioinformatics}, 19(1):1--10, 2018.

\bibitem{chen2018viper}
Mengjie Chen and Xiang Zhou.
\newblock Viper: variability-preserving imputation for accurate gene expression
  recovery in single-cell rna sequencing studies.
\newblock {\em Genome biology}, 19(1):1--15, 2018.

\bibitem{chen2020scrmd}
Chong Chen, Changjing Wu, Linjie Wu, Xiaochen Wang, Minghua Deng, and Ruibin
  Xi.
\newblock scrmd: imputation for single cell rna-seq data via robust matrix
  decomposition.
\newblock {\em Bioinformatics}, 36(10):3156--3161, 2020.

\bibitem{elyanow2020netnmf}
Rebecca Elyanow, Bianca Dumitrascu, Barbara~E Engelhardt, and Benjamin~J
  Raphael.
\newblock netnmf-sc: leveraging gene--gene interactions for imputation and
  dimensionality reduction in single-cell expression analysis.
\newblock {\em Genome research}, 30(2):195--204, 2020.

\bibitem{mongia2019mcimpute}
Aanchal Mongia, Debarka Sengupta, and Angshul Majumdar.
\newblock Mcimpute: Matrix completion based imputation for single cell rna-seq
  data.
\newblock {\em Frontiers in genetics}, 10:9, 2019.

\bibitem{xu2020cmf}
Junlin Xu, Lijun Cai, Bo~Liao, Wen Zhu, and JiaLiang Yang.
\newblock Cmf-impute: an accurate imputation tool for single-cell rna-seq data.
\newblock {\em Bioinformatics}, 36(10):3139--3147, 2020.

\bibitem{linderman2022zero}
George~C Linderman, Jun Zhao, Manolis Roulis, Piotr Bielecki, Richard~A
  Flavell, Boaz Nadler, and Yuval Kluger.
\newblock Zero-preserving imputation of single-cell rna-seq data.
\newblock {\em Nature communications}, 13(1):1--11, 2022.

\bibitem{huang2018saver}
Mo~Huang, Jingshu Wang, Eduardo Torre, Hannah Dueck, Sydney Shaffer, Roberto
  Bonasio, John~I Murray, Arjun Raj, Mingyao Li, and Nancy~R Zhang.
\newblock Saver: gene expression recovery for single-cell rna sequencing.
\newblock {\em Nature methods}, 15(7):539--542, 2018.

\bibitem{tang2020baynorm}
Wenhao Tang, Fran{\c{c}}ois Bertaux, Philipp Thomas, Claire Stefanelli, Malika
  Saint, Samuel Marguerat, and Vahid Shahrezaei.
\newblock baynorm: Bayesian gene expression recovery, imputation and
  normalization for single-cell rna-sequencing data.
\newblock {\em Bioinformatics}, 36(4):1174--1181, 2020.

\bibitem{li2018accurate}
Wei~Vivian Li and Jingyi~Jessica Li.
\newblock An accurate and robust imputation method scimpute for single-cell
  rna-seq data.
\newblock {\em Nature communications}, 9(1):1--9, 2018.

\bibitem{Lin2017}
Peijie Lin, Michael Troup, and Joshua~WK Ho.
\newblock Cidr: Ultrafast and accurate clustering through imputation for
  single-cell rna-seq data.
\newblock {\em Genome biology}, 18(1):1--11, 2017.
\newblock \href{https://github.com/VCCRI/CIDR}{Code Link:
  https://github.com/VCCRI/CIDR}.

\bibitem{kapourani2019melissa}
Chantriolnt-Andreas Kapourani and Guido Sanguinetti.
\newblock Melissa: Bayesian clustering and imputation of single-cell
  methylomes.
\newblock {\em Genome biology}, 20(1):1--15, 2019.

\bibitem{tang2021camelia}
Jianxiong Tang, Jianxiao Zou, Mei Fan, Qi~Tian, Jiyang Zhang, and Shicai Fan.
\newblock Camelia: imputation in single-cell methylomes based on local
  similarities between cells.
\newblock {\em Bioinformatics}, 37(13):1814--1820, 2021.

\bibitem{pereira2020reviewing}
Ricardo~Cardoso Pereira, Miriam~Seoane Santos, Pedro~Pereira Rodrigues, and
  Pedro~Henriques Abreu.
\newblock Reviewing autoencoders for missing data imputation: Technical trends,
  applications and outcomes.
\newblock {\em Journal of Artificial Intelligence Research}, 69:1255--1285,
  2020.

\bibitem{gondara2017multiple}
Lovedeep Gondara and Ke~Wang.
\newblock Multiple imputation using deep denoising autoencoders.
\newblock {\em arXiv preprint arXiv:1705.02737}, 280, 2017.

\bibitem{gondara2018mida}
Lovedeep Gondara and Ke~Wang.
\newblock Mida: Multiple imputation using denoising autoencoders.
\newblock In {\em Pacific-Asia conference on knowledge discovery and data
  mining}, pages 260--272. Springer, 2018.

\bibitem{beaulieu2017missing}
Brett~K Beaulieu-Jones, Jason~H Moore, and POOLED RESOURCE OPEN-ACCESS ALS
  CLINICAL~TRIALS CONSORTIUM.
\newblock Missing data imputation in the electronic health record using deeply
  learned autoencoders.
\newblock In {\em Pacific symposium on biocomputing 2017}, pages 207--218.
  World Scientific, 2017.

\bibitem{boquet2019missing}
Guillem Boquet, Jose~Lopez Vicario, Antoni Morell, and Javier Serrano.
\newblock Missing data in traffic estimation: A variational autoencoder
  imputation method.
\newblock In {\em ICASSP 2019-2019 IEEE International Conference on Acoustics,
  Speech and Signal Processing (ICASSP)}, pages 2882--2886. IEEE, 2019.

\bibitem{talwar2018autoimpute}
Divyanshu Talwar, Aanchal Mongia, Debarka Sengupta, and Angshul Majumdar.
\newblock Autoimpute: Autoencoder based imputation of single-cell rna-seq data.
\newblock {\em Scientific reports}, 8(1):1--11, 2018.

\bibitem{eraslan2019single}
G{\"o}kcen Eraslan, Lukas~M Simon, Maria Mircea, Nikola~S Mueller, and Fabian~J
  Theis.
\newblock Single-cell rna-seq denoising using a deep count autoencoder.
\newblock {\em Nature communications}, 10(1):1--14, 2019.

\bibitem{deng2018massive}
Yue Deng, Feng Bao, Qionghai Dai, Lani~F Wu, and Steven~J Altschuler.
\newblock Massive single-cell rna-seq analysis and imputation via deep
  learning.
\newblock {\em BioRxiv}, page 315556, 2018.

\bibitem{wang2021scgnn}
Juexin Wang, Anjun Ma, Yuzhou Chang, Jianting Gong, Yuexu Jiang, Ren Qi, Cankun
  Wang, Hongjun Fu, Qin Ma, and Dong Xu.
\newblock scgnn is a novel graph neural network framework for single-cell
  rna-seq analyses.
\newblock {\em Nature communications}, 12(1):1--11, 2021.
\newblock \href{https://github.com/juexinwang/scGNN}{Code Link:
  https://github.com/juexinwang/scGNN}.

\bibitem{rao2021imputing}
Jiahua Rao, Xiang Zhou, Yutong Lu, Huiying Zhao, and Yuedong Yang.
\newblock Imputing single-cell rna-seq data by combining graph convolution and
  autoencoder neural networks.
\newblock {\em Iscience}, 24(5):102393, 2021.

\bibitem{angermueller2017deepcpg}
Christof Angermueller, Heather~J Lee, Wolf Reik, and Oliver Stegle.
\newblock Deepcpg: accurate prediction of single-cell dna methylation states
  using deep learning.
\newblock {\em Genome biology}, 18(1):1--13, 2017.

\bibitem{lin2017cidr}
Peijie Lin, Michael Troup, and Joshua~WK Ho.
\newblock Cidr: Ultrafast and accurate clustering through imputation for
  single-cell rna-seq data.
\newblock {\em Genome biology}, 18(1):1--11, 2017.

\bibitem{badsha2020imputation}
Md~Badsha, Rui Li, Boxiang Liu, Yang~I Li, Min Xian, Nicholas~E Banovich,
  Audrey~Qiuyan Fu, et~al.
\newblock Imputation of single-cell gene expression with an autoencoder neural
  network.
\newblock {\em Quantitative Biology}, 8(1):78--94, 2020.

\bibitem{deng2019scalable}
Yue Deng, Feng Bao, Qionghai Dai, Lani~F Wu, and Steven~J Altschuler.
\newblock Scalable analysis of cell-type composition from single-cell
  transcriptomics using deep recurrent learning.
\newblock {\em Nature methods}, 16(4):311--314, 2019.

\bibitem{xu2020scigans}
Yungang Xu, Zhigang Zhang, Lei You, Jiajia Liu, Zhiwei Fan, and Xiaobo Zhou.
\newblock scigans: single-cell rna-seq imputation using generative adversarial
  networks.
\newblock {\em Nucleic acids research}, 48(15):e85--e85, 2020.

\bibitem{tabula2018single}
Tabula~Muris Consortium et~al.
\newblock Single-cell transcriptomics of 20 mouse organs creates a tabula
  muris.
\newblock {\em Nature}, 562(7727):367--372, 2018.

\bibitem{Tabula2018}
Tabula~Muris Consortium et~al.
\newblock Single-cell transcriptomics of 20 mouse organs creates a tabula
  muris.
\newblock {\em Nature}, 562(7727):367--372, 2018.
\newblock \href{https://tabula-muris.ds.czbiohub.org/}{Data Link:
  https://tabula-muris.ds.czbiohub.org/}.

\bibitem{Zheng2017}
Grace~XY Zheng, Jessica~M Terry, Phillip Belgrader, Paul Ryvkin, Zachary~W
  Bent, Ryan Wilson, Solongo~B Ziraldo, Tobias~D Wheeler, Geoff~P McDermott,
  Junjie Zhu, et~al.
\newblock Massively parallel digital transcriptional profiling of single cells.
\newblock {\em Nature communications}, 8(1):1--12, 2017.
\newblock
  \href{https://support.10xgenomics.com/single-cell-gene-expression/datasets/2.1.0/pbmc4k}{Data
  Link:
  https://support.10xgenomics.com/single-cell-gene-expression/datasets/2.1.0/pbmc4k}.

\bibitem{regev2017science}
Aviv Regev, Sarah~A Teichmann, Eric~S Lander, Ido Amit, Christophe Benoist,
  Ewan Birney, Bernd Bodenmiller, Peter Campbell, Piero Carninci, Menna
  Clatworthy, et~al.
\newblock Science forum: the human cell atlas.
\newblock {\em elife}, 6:e27041, 2017.

\bibitem{Gan2022}
Yanglan Gan, Xingyu Huang, Guobing Zou, Shuigeng Zhou, and Jihong Guan.
\newblock Deep structural clustering for single-cell rna-seq data jointly
  through autoencoder and graph neural network.
\newblock {\em Briefings in Bioinformatics}, 23(2):bbac018, 2022.
\newblock \href{https://github.com/DHUDBlab/scDSC}{Code Link:
  https://github.com/DHUDBlab/scDSC}.

\bibitem{Yu2022}
Zhuohan Yu, Yifu Lu, Yunhe Wang, Fan Tang, Ka-Chun Wong, and Xiangtao Li.
\newblock Zinb-based graph embedding autoencoder for single-cell rna-seq
  interpretations.
\newblock {\em Proceedings of the AAAI Conference on Artificial Intelligence},
  36(4):4671--4679, 2022.
\newblock \href{https://github.com/Philyzh8/scTAG}{Code Link:
  https://github.com/Philyzh8/scTAG}.

\bibitem{Tian2019}
Tian Tian, Ji~Wan, Qi~Song, and Zhi Wei.
\newblock Clustering single-cell rna-seq data with a model-based deep learning
  approach.
\newblock {\em Nature Machine Intelligence}, 1(4):191--198, 2019.
\newblock \href{https://github.com/ttgump/scDeepCluster}{Code Link:
  https://github.com/ttgump/scDeepCluster}.

\bibitem{Han2018}
Xiaoping Han, Renying Wang, Yincong Zhou, Lijiang Fei, Huiyu Sun, Shujing Lai,
  Assieh Saadatpour, Ziming Zhou, Haide Chen, Fang Ye, et~al.
\newblock Mapping the mouse cell atlas by microwell-seq.
\newblock {\em Cell}, 172(5):1091--1107, 2018.
\newblock \href{https://figshare.com/s/865e694ad06d5857db4b}{Data Link:
  https://figshare.com/s/865e694ad06d5857db4b}.

\bibitem{Cao2017}
Junyue Cao, Jonathan~S Packer, Vijay Ramani, Darren~A Cusanovich, Chau Huynh,
  Riza Daza, Xiaojie Qiu, Choli Lee, Scott~N Furlan, Frank~J Steemers, et~al.
\newblock Comprehensive single-cell transcriptional profiling of a
  multicellular organism.
\newblock {\em Science}, 357(6352):661--667, 2017.
\newblock \href{http://atlas.gs.washington.edu/worm-rna/docs/}{Data Link:
  http://atlas.gs.washington.edu/worm-rna/docs/}.

\bibitem{Johnson1967}
Stephen~C Johnson.
\newblock Hierarchical clustering schemes.
\newblock {\em Psychometrika}, 32(3):241--254, 1967.

\bibitem{Prabhakaran2016}
Sandhya Prabhakaran, Elham Azizi, Ambrose Carr, and Dana Pe’er.
\newblock Dirichlet process mixture model for correcting technical variation in
  single-cell gene expression data.
\newblock In {\em International conference on machine learning}, pages
  1070--1079. PMLR, 2016.

\bibitem{Guo2015}
Minzhe Guo, Hui Wang, S~Steven Potter, Jeffrey~A Whitsett, and Yan Xu.
\newblock Sincera: a pipeline for single-cell rna-seq profiling analysis.
\newblock {\em PLoS computational biology}, 11(11):e1004575, 2015.

\bibitem{Jiang2018}
Hao Jiang, Lydia~L Sohn, Haiyan Huang, and Luonan Chen.
\newblock Single cell clustering based on cell-pair differentiability
  correlation and variance analysis.
\newblock {\em Bioinformatics}, 34(21):3684--3694, 2018.

\bibitem{MacQueen1967}
J~MacQueen.
\newblock Classification and analysis of multivariate observations.
\newblock In {\em 5th Berkeley Symp. Math. Statist. Probability}, pages
  281--297, 1967.

\bibitem{Jarvis1973}
Raymond~Austin Jarvis and Edward~A Patrick.
\newblock Clustering using a similarity measure based on shared near neighbors.
\newblock {\em IEEE Transactions on computers}, 100(11):1025--1034, 1973.

\bibitem{Chlis2017}
Nikolaos~K Chlis, F~Alexander Wolf, and Fabian~J Theis.
\newblock Model-based branching point detection in single-cell data by
  k-branches clustering.
\newblock {\em Bioinformatics}, 33(20):3211--3219, 2017.
\newblock \href{https://github.com/theislab/kbranches}{Code Link:
  https://github.com/theislab/kbranches}.

\bibitem{Xu2015}
Chen Xu and Zhengchang Su.
\newblock Identification of cell types from single-cell transcriptomes using a
  novel clustering method.
\newblock {\em Bioinformatics}, 31(12):1974--1980, 2015.

\bibitem{Mieth2019}
Bettina Mieth, James~RF Hockley, Nico G{\"o}rnitz, Marina M-C Vidovic,
  Klaus-Robert M{\"u}ller, Alex Gutteridge, and Daniel Ziemek.
\newblock Using transfer learning from prior reference knowledge to improve the
  clustering of single-cell rna-seq data.
\newblock {\em Scientific reports}, 9(1):1--14, 2019.
\newblock \href{https://github.com/nicococo/scRNA}{Code Link:
  https://github.com/nicococo/scRNA}.

\bibitem{Xie2016}
Junyuan Xie, Ross Girshick, and Ali Farhadi.
\newblock Unsupervised deep embedding for clustering analysis.
\newblock In {\em International conference on machine learning}, pages
  478--487. PMLR, 2016.

\bibitem{Eraslan2019}
G{\"o}kcen Eraslan, Lukas~M Simon, Maria Mircea, Nikola~S Mueller, and Fabian~J
  Theis.
\newblock Single-cell rna-seq denoising using a deep count autoencoder.
\newblock {\em Nature communications}, 10(1):1--14, 2019.

\bibitem{Tian2021}
Tian Tian, Jie Zhang, Xiang Lin, Zhi Wei, and Hakon Hakonarson.
\newblock Model-based deep embedding for constrained clustering analysis of
  single cell rna-seq data.
\newblock {\em Nature communications}, 12(1):1--12, 2021.
\newblock \href{https://github.com/ttgump/scDCC}{Code Link:
  https://github.com/ttgump/scDCC}.

\bibitem{Chen2020}
Liang Chen, Weinan Wang, Yuyao Zhai, and Minghua Deng.
\newblock Deep soft k-means clustering with self-training for single-cell rna
  sequence data.
\newblock {\em NAR genomics and bioinformatics}, 2(2):lqaa039, 2020.
\newblock \href{https://github.com/xuebaliang/scziDesk}{Code and Data Link:
  https://github.com/xuebaliang/scziDesk}.

\bibitem{Lopez2018}
Romain Lopez, Jeffrey Regier, Michael~B Cole, Michael~I Jordan, and Nir Yosef.
\newblock Deep generative modeling for single-cell transcriptomics.
\newblock {\em Nature methods}, 15(12):1053--1058, 2018.
\newblock \href{https://github.com/scverse/scvi-tools}{Code Link:
  https://github.com/scverse/scvi-tools}.

\bibitem{Li2020}
Xiangjie Li, Kui Wang, Yafei Lyu, Huize Pan, Jingxiao Zhang, Dwight Stambolian,
  Katalin Susztak, Muredach~P Reilly, Gang Hu, and Mingyao Li.
\newblock Deep learning enables accurate clustering with batch effect removal
  in single-cell rna-seq analysis.
\newblock {\em Nature communications}, 11(1):1--14, 2020.
\newblock \href{https://eleozzr.github.io/desc/}{Code Link:
  https://eleozzr.github.io/desc/}.

\bibitem{Tran2021}
Duc Tran, Hung Nguyen, Bang Tran, Carlo La~Vecchia, Hung~N Luu, and Tin Nguyen.
\newblock Fast and precise single-cell data analysis using a hierarchical
  autoencoder.
\newblock {\em Nature communications}, 12(1):1--10, 2021.
\newblock \href{https://github.com/duct317/scDHA}{Code Link:
  https://github.com/duct317/scDHA}.

\bibitem{Zeng2020}
Yuansong Zeng, Xiang Zhou, Jiahua Rao, Yutong Lu, and Yuedong Yang.
\newblock Accurately clustering single-cell rna-seq data by capturing
  structural relations between cells through graph convolutional network.
\newblock In {\em 2020 IEEE International Conference on Bioinformatics and
  Biomedicine (BIBM)}, pages 519--522. IEEE, 2020.
\newblock \href{https://github.com/biomed-AI/GraphSCC}{Code Link:
  https://github.com/biomed-AI/GraphSCC}.

\bibitem{shao2021scdeepsort}
Xin Shao, Haihong Yang, Xiang Zhuang, Jie Liao, Penghui Yang, Junyun Cheng,
  Xiaoyan Lu, Huajun Chen, and Xiaohui Fan.
\newblock scdeepsort: a pre-trained cell-type annotation method for single-cell
  transcriptomics using deep learning with a weighted graph neural network.
\newblock {\em Nucleic acids research}, 49(21):e122--e122, 2021.

\bibitem{Ciortan2022}
Madalina Ciortan and Matthieu Defrance.
\newblock Gnn-based embedding for clustering scrna-seq data.
\newblock {\em Bioinformatics}, 38(4):1037--1044, 2022.
\newblock \href{https://github.com/ciortanmadalina/graph-sc}{Code Link:
  https://github.com/ciortanmadalina/graph-sc}.

\bibitem{Kiselev2017}
Vladimir~Yu Kiselev, Kristina Kirschner, Michael~T Schaub, Tallulah Andrews,
  Andrew Yiu, Tamir Chandra, Kedar~N Natarajan, Wolf Reik, Mauricio Barahona,
  Anthony~R Green, et~al.
\newblock Sc3: consensus clustering of single-cell rna-seq data.
\newblock {\em Nature methods}, 14(5):483--486, 2017.
\newblock \href{https://github.com/hemberg-lab/sc3}{Code Link:
  https://github.com/hemberg-lab/sc3}.

\bibitem{Zhang2020}
Yuanchao Zhang, Man~S Kim, Erin~R Reichenberger, Ben Stear, and Deanne~M
  Taylor.
\newblock Scedar: A scalable python package for single-cell rna-seq exploratory
  data analysis.
\newblock {\em PLoS computational biology}, 16(4):e1007794, 2020.
\newblock \href{https://pypi.org/project/scedar}{Code Link:
  https://pypi.org/project/scedar}.

\bibitem{Wang2018}
Bo~Wang, Daniele Ramazzotti, Luca De~Sano, Junjie Zhu, Emma Pierson, and
  Serafim Batzoglou.
\newblock Simlr: A tool for large-scale genomic analyses by multi-kernel
  learning.
\newblock {\em Proteomics}, 18(2):1700232, 2018.
\newblock \href{https://github.com/BatzoglouLabSU/SIMLR}{Code Link:
  https://github.com/BatzoglouLabSU/SIMLR}.

\bibitem{Weinreb2018}
Caleb Weinreb, Samuel Wolock, and Allon~M Klein.
\newblock Spring: a kinetic interface for visualizing high dimensional
  single-cell expression data.
\newblock {\em Bioinformatics}, 34(7):1246--1248, 2018.
\newblock \href{https://github.com/AllonKleinLab/SPRING}{Code Link:
  https://github.com/AllonKleinLab/SPRING}.

\bibitem{Gardeux2017}
Vincent Gardeux, Fabrice~PA David, Adrian Shajkofci, Petra~C Schwalie, and Bart
  Deplancke.
\newblock Asap: a web-based platform for the analysis and interactive
  visualization of single-cell rna-seq data.
\newblock {\em Bioinformatics}, 33(19):3123--3125, 2017.
\newblock \href{https://github.com/DeplanckeLab/ASAP}{Code Link:
  https://github.com/DeplanckeLab/ASAP}.

\bibitem{Klein2015}
Allon~M Klein, Linas Mazutis, Ilke Akartuna, Naren Tallapragada, Adrian Veres,
  Victor Li, Leonid Peshkin, David~A Weitz, and Marc~W Kirschner.
\newblock Droplet barcoding for single-cell transcriptomics applied to
  embryonic stem cells.
\newblock {\em Cell}, 161(5):1187--1201, 2015.
\newblock
  \href{https://www.ncbi.nlm.nih.gov/geo/query/acc.cgi?acc=GSE65525}{Data Link:
  https://www.ncbi.nlm.nih.gov/geo/query/acc.cgi?acc=GSE65525}.

\bibitem{Adam2017}
Mike Adam, Andrew~S Potter, and S~Steven Potter.
\newblock Psychrophilic proteases dramatically reduce single-cell rna-seq
  artifacts: a molecular atlas of kidney development.
\newblock {\em Development}, 144(19):3625--3632, 2017.
\newblock
  \href{https://www.ncbi.nlm.nih.gov/geo/query/acc.cgi?acc=GSE94333}{Data Link:
  https://www.ncbi.nlm.nih.gov/geo/query/acc.cgi?acc=GSE94333}.

\bibitem{Muraro2016}
Mauro~J Muraro, Gitanjali Dharmadhikari, Dominic Gr{\"u}n, Nathalie Groen, Tim
  Dielen, Erik Jansen, Leon Van~Gurp, Marten~A Engelse, Francoise Carlotti,
  Eelco~Jp De~Koning, et~al.
\newblock A single-cell transcriptome atlas of the human pancreas.
\newblock {\em Cell systems}, 3(4):385--394, 2016.
\newblock
  \href{https://www.ncbi.nlm.nih.gov/geo/query/acc.cgi?acc=GSE85241}{Data Link:
  https://www.ncbi.nlm.nih.gov/geo/query/acc.cgi?acc=GSE85241}.

\bibitem{Romanov2017}
Roman~A Romanov, Amit Zeisel, Joanne Bakker, Fatima Girach, Arash Hellysaz,
  Raju Tomer, Alan Alpar, Jan Mulder, Frederic Clotman, Erik Keimpema, et~al.
\newblock Molecular interrogation of hypothalamic organization reveals distinct
  dopamine neuronal subtypes.
\newblock {\em Nature neuroscience}, 20(2):176--188, 2017.
\newblock
  \href{https://www.ncbi.nlm.nih.gov/geo/query/acc.cgi?acc=GSE74672}{Data Link:
  https://www.ncbi.nlm.nih.gov/geo/query/acc.cgi?acc=GSE74672}.

\bibitem{hu2021spagcn}
Jian Hu, Xiangjie Li, Kyle Coleman, Amelia Schroeder, Nan Ma, David~J Irwin,
  Edward~B Lee, Russell~T Shinohara, and Mingyao Li.
\newblock Spagcn: Integrating gene expression, spatial location and histology
  to identify spatial domains and spatially variable genes by graph
  convolutional network.
\newblock {\em Nature methods}, 18(11):1342--1351, 2021.
\newblock \href{https://github.com/jianhuupenn/SpaGCN}{Code Link:
  https://github.com/jianhuupenn/SpaGCN}.

\bibitem{kather2019predicting}
Jakob~Nikolas Kather, Johannes Krisam, Pornpimol Charoentong, Tom Luedde,
  Esther Herpel, Cleo-Aron Weis, Timo Gaiser, Alexander Marx, Nektarios~A
  Valous, Dyke Ferber, et~al.
\newblock Predicting survival from colorectal cancer histology slides using
  deep learning: A retrospective multicenter study.
\newblock {\em PLoS medicine}, 16(1):e1002730, 2019.

\bibitem{majeed2019quantitative}
Hassaan Majeed, Adib Keikhosravi, Mikhail~E Kandel, Tan~H Nguyen, Yuming Liu,
  Andre Kajdacsy-Balla, Krishnarao Tangella, Kevin~W Eliceiri, and Gabriel
  Popescu.
\newblock Quantitative histopathology of stained tissues using color spatial
  light interference microscopy (cslim).
\newblock {\em Scientific reports}, 9(1):1--14, 2019.

\bibitem{yeung2001details}
Ka~Yee Yeung and Walter~L Ruzzo.
\newblock Details of the adjusted rand index and clustering algorithms,
  supplement to the paper an empirical study on principal component analysis
  for clustering gene expression data.
\newblock {\em Bioinformatics}, 17(9):763--774, 2001.

\bibitem{blondelvd2008fast}
VD~Blondel, JL~Guillaume, R~Lambiotte, and E~Lefebvre.
\newblock Fast unfolding of community hierarchies in large networks, 2008.
\newblock \href{https://github.com/taynaud/python-louvain}{Code Link:
  https://github.com/taynaud/python-louvain}.

\bibitem{kanungo2002efficient}
Tapas Kanungo, David~M Mount, Nathan~S Netanyahu, Christine~D Piatko, Ruth
  Silverman, and Angela~Y Wu.
\newblock An efficient k-means clustering algorithm: Analysis and
  implementation.
\newblock {\em IEEE transactions on pattern analysis and machine intelligence},
  24(7):881--892, 2002.
\newblock
  \href{https://scikit-learn.org/stable/modules/generated/sklearn.cluster.KMeans.html}{Code
  Link:
  https://scikit-learn.org/stable/modules/generated/sklearn.cluster.KMeans.html}.

\bibitem{pham2020stlearn}
Duy Pham, Xiao Tan, Jun Xu, Laura~F Grice, Pui~Yeng Lam, Arti Raghubar, Jana
  Vukovic, Marc~J Ruitenberg, and Quan Nguyen.
\newblock stlearn: integrating spatial location, tissue morphology and gene
  expression to find cell types, cell-cell interactions and spatial
  trajectories within undissociated tissues.
\newblock {\em BioRxiv}, 2020.
\newblock \href{https://github.com/BiomedicalMachineLearning/stLearn}{Code
  Link: https://github.com/BiomedicalMachineLearning/stLearn}.

\bibitem{zhao2021spatial}
Edward Zhao, Matthew~R Stone, Xing Ren, Jamie Guenthoer, Kimberly~S Smythe,
  Thomas Pulliam, Stephen~R Williams, Cedric~R Uytingco, Sarah~EB Taylor, Paul
  Nghiem, et~al.
\newblock Spatial transcriptomics at subspot resolution with bayesspace.
\newblock {\em Nature Biotechnology}, 39(11):1375--1384, 2021.
\newblock \href{https://github.com/edward130603/BayesSpace}{Code Link:
  https://github.com/edward130603/BayesSpace}.

\bibitem{zhu2018identification}
Qian Zhu, Sheel Shah, Ruben Dries, Long Cai, and Guo-Cheng Yuan.
\newblock Identification of spatially associated subpopulations by combining
  scrnaseq and sequential fluorescence in situ hybridization data.
\newblock {\em Nature biotechnology}, 36(12):1183--1190, 2018.

\bibitem{schurch2020coordinated}
Christian~M Sch{\"u}rch, Salil~S Bhate, Graham~L Barlow, Darci~J Phillips, Luca
  Noti, Inti Zlobec, Pauline Chu, Sarah Black, Janos Demeter, David~R McIlwain,
  et~al.
\newblock Coordinated cellular neighborhoods orchestrate antitumoral immunity
  at the colorectal cancer invasive front.
\newblock {\em Cell}, 182(5):1341--1359, 2020.

\bibitem{singhal2022banksy}
Vipul Singhal, Nigel Chou, Joseph Lee, Jinyue Liu, Wan~Kee Chock, Li~Lin,
  Yun-Ching Chang, Erica Teo, Hwee~Kuan Lee, Kok~Hao Chen, et~al.
\newblock Banksy: A spatial omics algorithm that unifies cell type clustering
  and tissue domain segmentation.
\newblock {\em bioRxiv}, 2022.

\bibitem{dong2022deciphering}
Kangning Dong and Shihua Zhang.
\newblock Deciphering spatial domains from spatially resolved transcriptomics
  with an adaptive graph attention auto-encoder.
\newblock {\em Nature communications}, 13(1):1--12, 2022.
\newblock \href{https://github.com/zhanglabtools/STAGATE}{Code Link:
  https://github.com/zhanglabtools/STAGATE}.

\bibitem{li2022cell}
Jiachen Li, Siheng Chen, Xiaoyong Pan, Ye~Yuan, and Hong-Bin Shen.
\newblock Cell clustering for spatial transcriptomics data with graph neural
  networks.
\newblock {\em Nature Computational Science}, 2(6):399--408, 2022.
\newblock \href{https://github.com/xiaoyeye/CCST}{Code Link:
  https://github.com/xiaoyeye/CCST}.

\bibitem{fu2021unsupervised}
Huazhu Fu, Hang Xu, Kelvin Chong, Mengwei Li, Kok~Siong Ang, Hong~Kai Lee,
  Jingjing Ling, Ao~Chen, Ling Shao, Longqi Liu, et~al.
\newblock Unsupervised spatially embedded deep representation of spatial
  transcriptomics.
\newblock {\em Biorxiv}, 2021.
\newblock \href{https://github.com/JinmiaoChenLab/SEDR}{Code Link:
  https://github.com/JinmiaoChenLab/SEDR}.

\bibitem{svensson2018spatialde}
Valentine Svensson, Sarah~A Teichmann, and Oliver Stegle.
\newblock Spatialde: identification of spatially variable genes.
\newblock {\em Nature methods}, 15(5):343--346, 2018.
\newblock \href{https://github.com/Teichlab/SpatialDE}{Code Link:
  https://github.com/Teichlab/SpatialDE}.

\bibitem{dries2021giotto}
Ruben Dries, Qian Zhu, Rui Dong, Chee-Huat~Linus Eng, Huipeng Li, Kan Liu,
  Yuntian Fu, Tianxiao Zhao, Arpan Sarkar, Feng Bao, et~al.
\newblock Giotto: a toolbox for integrative analysis and visualization of
  spatial expression data.
\newblock {\em Genome biology}, 22(1):1--31, 2021.
\newblock \href{https://github.com/RubD/Giotto}{Code Link:
  https://github.com/RubD/Giotto}.

\bibitem{link2MPB10xV}
Mouse posterior brain 10x visium data.
\newblock
  \url{https://support.10xgenomics.com/spatial-gene-expression/datasets/1.0.0/V1_Mouse_Brain_Sagittal_Posterior}.

\bibitem{link2spatialLIBD}
Libd human dorsolateral prefrontal cortex data.
\newblock \url{http://research.libd.org/spatialLIBD/}.

\bibitem{Danaher2022}
Patrick Danaher, Youngmi Kim, Brenn Nelson, Maddy Griswold, Zhi Yang, Erin
  Piazza, and Joseph~M Beechem.
\newblock Advances in mixed cell deconvolution enable quantification of cell
  types in spatial transcriptomic data.
\newblock {\em Nature Communications}, 13(1), 01 2022.
\newblock \href{https://github.com/Nanostring-Biostats/SpatialDecon}{Code Link:
  https://github.com/Nanostring-Biostats/SpatialDecon}.

\bibitem{Elosua2021}
Marc Elosua-Bayes, Paula Nieto, Elisabetta Mereu, Ivo Gut, and Holger Heyn.
\newblock Spotlight: seeded nmf regression to deconvolute spatial
  transcriptomics spots with single-cell transcriptomes.
\newblock {\em Nucleic Acids Research}, 49(9):e50--e50, 02 2021.
\newblock \href{https://github.com/MarcElosua/SPOTlight}{Code Link:
  https://github.com/MarcElosua/SPOTlight}.

\bibitem{Tsoucas2019}
Daphne Tsoucas, Rui Dong, Haide Chen, Qian Zhu, Guoji Guo, and Guo-Cheng Yuan.
\newblock Accurate estimation of cell-type composition from gene expression
  data.
\newblock {\em Nature Communications}, 10(1), 07 2019.
\newblock \href{https://github.com/dtsoucas/DWLS}{Code Link:
  https://github.com/dtsoucas/DWLS}.

\bibitem{Yuan2021}
R.~Dong and G.~Yuan.
\newblock Spatialdwls: accurate deconvolution of spatial transcriptomic data.
\newblock {\em Genome Biology}, 22(1), 05 2021.
\newblock \href{https://github.com/rdong08/spatialDWLS_dataset}{Code Link:
  https://github.com/rdong08/spatialDWLS\_dataset}.

\bibitem{Ma2022}
Y.~Ma and X.~Zhou.
\newblock Spatially informed cell-type deconvolution for spatial
  transcriptomics.
\newblock {\em Nature Biotechnology}, 05 2022.
\newblock \href{https://github.com/YingMa0107/CARD}{Code Link:
  https://github.com/YingMa0107/CARD}.

\bibitem{cableRobustDecompositionCell2022}
Dylan~M. Cable, Evan Murray, Luli~S. Zou, Aleksandrina Goeva, Evan~Z. Macosko,
  Fei Chen, and Rafael~A. Irizarry.
\newblock Robust decomposition of cell type mixtures in spatial
  transcriptomics.
\newblock {\em Nature Biotechnology}, 40(4):517--526, 2022.

\bibitem{kleshchevnikovCell2locationMapsFinegrained2022}
Vitalii Kleshchevnikov, Artem Shmatko, Emma Dann, Alexander Aivazidis,
  Hamish~W. King, Tong Li, Rasa Elmentaite, Artem Lomakin, Veronika Kedlian,
  Adam Gayoso, Mika~Sarkin Jain, Jun~Sung Park, Lauma Ramona, Elizabeth Tuck,
  Anna Arutyunyan, Roser Vento-Tormo, Moritz Gerstung, Louisa James, Oliver
  Stegle, and Omer~Ali Bayraktar.
\newblock Cell2location maps fine-grained cell types in spatial
  transcriptomics.
\newblock {\em Nature Biotechnology}, 40(5):661--671, 2022.
\newblock \href{https://github.com/BayraktarLab/cell2location}{Code Link:
  https://github.com/BayraktarLab/cell2location}.

\bibitem{Menden2020}
Kevin Menden, Mohamed Marouf, Sergio Oller, Anupriya Dalmia, Daniel~Sumner
  Magruder, Karin Kloiber, Peter Heutink, and Stefan Bonn.
\newblock Deep learning-based cell composition analysis from tissue expression
  profiles.
\newblock {\em Science Advances}, 6(30):eaba2619, 2020.
\newblock \href{https://github.com/KevinMenden/scaden}{Code Link:
  https://github.com/KevinMenden/scaden}.

\bibitem{Song2021}
Qianqian Song and Jing Su.
\newblock Dstg: deconvoluting spatial transcriptomics data through graph-based
  artificial intelligence.
\newblock {\em Briefings in Bioinformatics}, 22(3):1--13, 2021.
\newblock \href{https://github.com/Su-informatics-lab/DSTG}{Code Link:
  https://github.com/Su-informatics-lab/DSTG}.

\bibitem{Pham2021}
Justin~Hong Dan D. Erdmann-Pham, Jonathan~Fischer and Yun~S. Song.
\newblock A likelihood-based deconvolution of bulk gene expression data using
  single-cell references.
\newblock {\em Genome Research}, 07 2021.
\newblock \href{https://github.com/songlab-cal/rna-sieve}{Code Link:
  https://github.com/songlab-cal/rna-sieve}.

\bibitem{link2MOB10xV}
Mouse olfactory bulb data.
\newblock
  \url{https://www.10xgenomics.com/resources/datasets/adult-mouse-olfactory-bulb-1-standard-1}.

\bibitem{link2CellieGeoMx}
Hek293t and ccrf-cem cell line mixture data.
\newblock \url{https://www.ncbi.nlm.nih.gov/geo/query/acc.cgi?acc=GSE174746}.

\bibitem{link2HPdacST}
Human pdac data.
\newblock \url{https://www.ncbi.nlm.nih.gov/geo/query/acc.cgi?acc=GSE111672}.

\bibitem{Stringer2021}
Carsen Stringer, Tim Wang, Michalis Michaelos, and Marius Pachitariu.
\newblock Cellpose: a generalist algorithm for cellular segmentation.
\newblock {\em Nature methods}, 18(1):100—106, January 2021.
\newblock \href{https://github.com/MouseLand/cellpose}{Website Link:
  https://github.com/MouseLand/cellpose}.

\bibitem{Greenwald2022}
Noah Greenwald, Geneva Miller, Erick Moen, Alex Kong, Adam Kagel, Thomas
  Dougherty, Christine Fullaway, Brianna McIntosh, Ke~Leow, Morgan Schwartz,
  Cole Pavelchek, Sunny Cui, Isabella Camplisson, Omer Bar-Tal, Jaiveer Singh,
  Mara Fong, Gautam Chaudhry, Zion Abraham, Jackson Moseley, and David Valen.
\newblock Whole-cell segmentation of tissue images with human-level performance
  using large-scale data annotation and deep learning.
\newblock {\em Nature Biotechnology}, 40, 04 2022.
\newblock \href{https://github.com/vanvalenlab/deepcell-tf}{Code Link:
  https://github.com/vanvalenlab/deepcell-tf}.

\bibitem{littman2021}
Russell Littman, Zachary Hemminger, Robert Foreman, Douglas Arneson, Guanglin
  Zhang, Fernando G{\'o}mez-Pinilla, Xia Yang, and Roy Wollman.
\newblock Joint cell segmentation and cell type annotation for spatial
  transcriptomics.
\newblock {\em Molecular systems biology}, 17(6):e10108, 2021.
\newblock \href{https://github.com/wollmanlab/JSTA}{Code Link:
  https://github.com/wollmanlab/JSTA}.

\bibitem{Yan2008}
Pingkun Yan, Xiaobo Zhou, Mubarak Shah, and Stephen T.~C. Wong.
\newblock Automatic segmentation of high-throughput rnai fluorescent cellular
  images.
\newblock {\em IEEE Transactions on Information Technology in Biomedicine},
  12(1):109--117, 2008.

\bibitem{vincent1991}
Luc Vincent and Pierre Soille.
\newblock Watersheds in digital spaces: an efficient algorithm based on
  immersion simulations.
\newblock {\em IEEE Transactions on Pattern Analysis \& Machine Intelligence},
  13(06):583--598, 1991.
\newblock \href{https://github.com/fiji}{Code Link: https://github.com/fiji}.

\bibitem{Gerdes2013}
Michael~J. Gerdes, Christopher~J. Sevinsky, Anup Sood, Sudeshna Adak,
  Musodiq~O. Bello, Alexander Bordwell, Ali Can, Alex Corwin, Sean Dinn,
  Robert~J. Filkins, Denise Hollman, Vidya Kamath, Sireesha Kaanumalle, Kevin
  Kenny, Melinda Larsen, Michael Lazare, Qing Li, Christina Lowes, Colin~C.
  McCulloch, Elizabeth McDonough, Michael~C. Montalto, Zhengyu Pang, Jens
  Rittscher, Alberto Santamaria-Pang, Brion~D. Sarachan, Maximilian~L. Seel,
  Antti Seppo, Kashan Shaikh, Yunxia Sui, Jingyu Zhang, and Fiona Ginty.
\newblock Highly multiplexed single-cell analysis of formalin-fixed,
  paraffin-embedded cancer tissue.
\newblock {\em Proceedings of the National Academy of Sciences},
  110(29):11982--11987, 2013.

\bibitem{can2009unified}
Ali Can, Musodiq Bello, Harvey~E Cline, Xiaodong Tao, Paulo Mendonca, and
  Michael Gerdes.
\newblock A unified segmentation method for detecting subcellular compartments
  in immunofluroescently labeled tissue images.
\newblock In {\em International Workshop on Microscopic Image Analysis with
  Applications in Biology. Sept}, 2009.

\bibitem{Peter2015}
Peter~J. Schüffler, Denis Schapiro, Charlotte Giesen, Hao A.~O. Wang, Bernd
  Bodenmiller, and Joachim~M. Buhmann.
\newblock Automatic single cell segmentation on highly multiplexed tissue
  images.
\newblock {\em Cytometry Part A}, 87(10):936--942, 2015.
\newblock \href{http://www.comp-path.inf.ethz.ch/}{Website Link:
  http://www.comp-path.inf.ethz.ch/}.

\bibitem{Pang2015}
Alberto Santamaria-Pang, Jens Rittscher, Michael Gerdes, and Dirk Padfield.
\newblock Cell segmentation and classification by hierarchical supervised shape
  ranking.
\newblock In {\em 2015 IEEE 12th International Symposium on Biomedical Imaging
  (ISBI)}, pages 1296--1299, 2015.

\bibitem{Krijgsman2021}
Daniëlle Krijgsman, Neeraj Sinha, Matthijs~J.D. Baars, Stephanie {van Dam},
  Mojtaba Amini, and Yvonne Vercoulen.
\newblock Matisse: An analysis protocol for combining imaging mass cytometry
  with fluorescence microscopy to generate single-cell data.
\newblock {\em STAR Protocols}, 3(1):101034, 2022.

\bibitem{Ilastik2011}
Christoph Sommer, Christoph Straehle, Ullrich Köthe, and Fred~A. Hamprecht.
\newblock Ilastik: Interactive learning and segmentation toolkit.
\newblock In {\em 2011 IEEE International Symposium on Biomedical Imaging: From
  Nano to Macro}, pages 230--233, 2011.
\newblock \href{https://github.com/ilastik/ilastik}{Code Link:
  https://github.com/ilastik/ilastik}.

\bibitem{CellProfiler}
Lee Kamentsky, Thouis~R. Jones, Adam Fraser, Mark-Anthony Bray, David~J. Logan,
  Katherine~L. Madden, Vebjorn Ljosa, Curtis Rueden, Kevin~W. Eliceiri, and
  Anne~E. Carpenter.
\newblock {Improved structure, function and compatibility for CellProfiler:
  modular high-throughput image analysis software}.
\newblock {\em Bioinformatics}, 27(8):1179--1180, 02 2011.
\newblock \href{https://github.com/CellProfiler/CellProfile}{Code Link:
  https://github.com/CellProfiler/CellProfile}.

\bibitem{long2015fully}
Jonathan Long, Evan Shelhamer, and Trevor Darrell.
\newblock Fully convolutional networks for semantic segmentation.
\newblock In {\em Proceedings of the IEEE conference on computer vision and
  pattern recognition}, pages 3431--3440, 2015.

\bibitem{redmon2016you}
Joseph Redmon, Santosh Divvala, Ross Girshick, and Ali Farhadi.
\newblock You only look once: Unified, real-time object detection.
\newblock In {\em Proceedings of the IEEE conference on computer vision and
  pattern recognition}, pages 779--788, 2016.

\bibitem{chen2016dcan}
Hao Chen, Xiaojuan Qi, Lequan Yu, and Pheng-Ann Heng.
\newblock Dcan: deep contour-aware networks for accurate gland segmentation.
\newblock In {\em Proceedings of the IEEE conference on Computer Vision and
  Pattern Recognition}, pages 2487--2496, 2016.

\bibitem{he2016deep}
Kaiming He, Xiangyu Zhang, Shaoqing Ren, and Jian Sun.
\newblock Deep residual learning for image recognition.
\newblock In {\em Proceedings of the IEEE conference on computer vision and
  pattern recognition}, pages 770--778, 2016.

\bibitem{kamnitsas2017efficient}
Konstantinos Kamnitsas, Christian Ledig, Virginia~FJ Newcombe, Joanna~P
  Simpson, Andrew~D Kane, David~K Menon, Daniel Rueckert, and Ben Glocker.
\newblock Efficient multi-scale 3d cnn with fully connected crf for accurate
  brain lesion segmentation.
\newblock {\em Medical image analysis}, 36:61--78, 2017.

\bibitem{yadav2019deep}
Samir~S Yadav and Shivajirao~M Jadhav.
\newblock Deep convolutional neural network based medical image classification
  for disease diagnosis.
\newblock {\em Journal of Big Data}, 6(1):1--18, 2019.

\bibitem{ronneberger2015}
Olaf Ronneberger, Philipp Fischer, and Thomas Brox.
\newblock U-net: Convolutional networks for biomedical image segmentation.
\newblock In {\em International Conference on Medical image computing and
  computer-assisted intervention}, pages 234--241. Springer, 2015.

\bibitem{li2018h}
Xiaomeng Li, Hao Chen, Xiaojuan Qi, Qi~Dou, Chi-Wing Fu, and Pheng-Ann Heng.
\newblock H-denseunet: hybrid densely connected unet for liver and tumor
  segmentation from ct volumes.
\newblock {\em IEEE transactions on medical imaging}, 37(12):2663--2674, 2018.

\bibitem{oktay2017anatomically}
Ozan Oktay, Enzo Ferrante, Konstantinos Kamnitsas, Mattias Heinrich, Wenjia
  Bai, Jose Caballero, Stuart~A Cook, Antonio De~Marvao, Timothy Dawes,
  Declan~P O‘Regan, et~al.
\newblock Anatomically constrained neural networks (acnns): application to
  cardiac image enhancement and segmentation.
\newblock {\em IEEE transactions on medical imaging}, 37(2):384--395, 2017.

\bibitem{oktay2018attention}
Ozan Oktay, Jo~Schlemper, Loic~Le Folgoc, Matthew Lee, Mattias Heinrich,
  Kazunari Misawa, Kensaku Mori, Steven McDonagh, Nils~Y Hammerla, Bernhard
  Kainz, et~al.
\newblock Attention u-net: Learning where to look for the pancreas.
\newblock {\em arXiv preprint arXiv:1804.03999}, 2018.

\bibitem{wang2020double}
Yixin Wang, Yao Zhang, Jiang Tian, Cheng Zhong, Zhongchao Shi, Yang Zhang, and
  Zhiqiang He.
\newblock Double-uncertainty weighted method for semi-supervised learning.
\newblock In {\em International Conference on Medical Image Computing and
  Computer-Assisted Intervention}, pages 542--551. Springer, 2020.

\bibitem{heller2021state}
Nicholas Heller, Fabian Isensee, Klaus~H Maier-Hein, Xiaoshuai Hou, Chunmei
  Xie, Fengyi Li, Yang Nan, Guangrui Mu, Zhiyong Lin, Miofei Han, et~al.
\newblock The state of the art in kidney and kidney tumor segmentation in
  contrast-enhanced ct imaging: Results of the kits19 challenge.
\newblock {\em Medical image analysis}, 67:101821, 2021.

\bibitem{ma2021toward}
Jun Ma, Yixin Wang, Xingle An, Cheng Ge, Ziqi Yu, Jianan Chen, Qiongjie Zhu,
  Guoqiang Dong, Jian He, Zhiqiang He, et~al.
\newblock Toward data-efficient learning: A benchmark for covid-19 ct lung and
  infection segmentation.
\newblock {\em Medical physics}, 48(3):1197--1210, 2021.

\bibitem{wang2021acn}
Yixin Wang, Yang Zhang, Yang Liu, Zihao Lin, Jiang Tian, Cheng Zhong, Zhongchao
  Shi, Jianping Fan, and Zhiqiang He.
\newblock Acn: Adversarial co-training network for brain tumor segmentation
  with missing modalities.
\newblock In {\em International Conference on Medical Image Computing and
  Computer-Assisted Intervention}, pages 410--420. Springer, 2021.

\bibitem{xu2021noisy}
Zhe Xu, Donghuan Lu, Yixin Wang, Jie Luo, Jagadeesan Jayender, Kai Ma, Yefeng
  Zheng, and Xiu Li.
\newblock Noisy labels are treasure: mean-teacher-assisted confident learning
  for hepatic vessel segmentation.
\newblock In {\em International Conference on Medical Image Computing and
  Computer-Assisted Intervention}, pages 3--13. Springer, 2021.

\bibitem{he2017mask}
Kaiming He, Georgia Gkioxari, Piotr Doll{\'a}r, and Ross Girshick.
\newblock Mask r-cnn.
\newblock In {\em Proceedings of the IEEE international conference on computer
  vision}, pages 2961--2969, 2017.

\bibitem{chen2020boundary}
Shengcong Chen, Changxing Ding, and Dacheng Tao.
\newblock Boundary-assisted region proposal networks for nucleus segmentation.
\newblock In {\em International conference on medical image computing and
  computer-assisted intervention}, pages 279--288. Springer, 2020.

\bibitem{zhou2019cia}
Yanning Zhou, Omer~Fahri Onder, Qi~Dou, Efstratios Tsougenis, Hao Chen, and
  Pheng-Ann Heng.
\newblock Cia-net: Robust nuclei instance segmentation with contour-aware
  information aggregation.
\newblock In {\em International conference on information processing in medical
  imaging}, pages 682--693. Springer, 2019.

\bibitem{huang20212}
Jinghan Huang, Yiqing Shen, Dinggang Shen, and Jing Ke.
\newblock Ca 2.5-net nuclei segmentation framework with a microscopy cell
  benchmark collection.
\newblock In {\em International Conference on Medical Image Computing and
  Computer-Assisted Intervention}, pages 445--454. Springer, 2021.

\bibitem{Stringer2022}
Carsen Stringer and Marius Pachitariu.
\newblock Cellpose 2.0: how to train your own model.
\newblock {\em bioRxiv}, 2022.

\bibitem{van2016}
David~A Van~Valen, Takamasa Kudo, Keara~M Lane, Derek~N Macklin, Nicolas~T
  Quach, Mialy~M DeFelice, Inbal Maayan, Yu~Tanouchi, Euan~A Ashley, and
  Markus~W Covert.
\newblock Deep learning automates the quantitative analysis of individual cells
  in live-cell imaging experiments.
\newblock {\em PLoS computational biology}, 12(11):e1005177, 2016.

\bibitem{Koyuncu2020}
Can~Fahrettin Koyuncu, Gozde~Nur Gunesli, Rengul Cetin-Atalay, and Cigdem
  Gunduz-Demir.
\newblock Deepdistance: A multi-task deep regression model for cell detection
  in inverted microscopy images.
\newblock {\em Medical Image Analysis}, 63:101720, 2020.

\bibitem{lin2017feature}
Tsung-Yi Lin, Piotr Doll{\'a}r, Ross Girshick, Kaiming He, Bharath Hariharan,
  and Serge Belongie.
\newblock Feature pyramid networks for object detection.
\newblock In {\em Proceedings of the IEEE conference on computer vision and
  pattern recognition}, pages 2117--2125, 2017.

\bibitem{mckinley2022}
Eliot~T McKinley, Justin Shao, Samuel~T Ellis, Cody~N Heiser, Joseph~T Roland,
  Mary~C Macedonia, Paige~N Vega, Susie Shin, Robert~J Coffey, and Ken~S Lau.
\newblock Miriam: A machine and deep learning single-cell segmentation and
  quantification pipeline for multi-dimensional tissue images.
\newblock {\em Cytometry Part A}, 2022.
\newblock \href{https://github.com/Coffey-Lab/MIRIAM}{Code Link:
  https://github.com/Coffey-Lab/MIRIAM}.

\bibitem{Petukhov2021}
Viktor Petukhov, Rosalind Xu, Ruslan Soldatov, Paolo Cadinu, Konstantin
  Khodosevich, Jeffrey Moffitt, and Peter Kharchenko.
\newblock Cell segmentation in imaging-based spatial transcriptomics.
\newblock {\em Nature Biotechnology}, 40:1--10, 10 2021.
\newblock \href{https://github.com/kharchenkolab/Baysor}{Code Link:
  https://github.com/kharchenkolab/Baysor}.

\bibitem{legland2016}
David Legland, Ignacio Arganda-Carreras, and Philippe Andrey.
\newblock Morpholibj: integrated library and plugins for mathematical
  morphology with imagej.
\newblock {\em Bioinformatics}, 32(22):3532--3534, 2016.
\newblock \href{https://github.com/ijpb/MorphoLibJ}{Code Link:
  https://github.com/ijpb/MorphoLibJ}.

\bibitem{paintdakhi2016}
Ahmad Paintdakhi, Bradley Parry, Manuel Campos, Irnov Irnov, Johan Elf, Ivan
  Surovtsev, and Christine Jacobs-Wagner.
\newblock Oufti: an integrated software package for high-accuracy,
  high-throughput quantitative microscopy analysis.
\newblock {\em Molecular microbiology}, 99(4):767--777, 2016.
\newblock \href{http://www.oufti.org/}{Website Link: http://www.oufti.org/}.

\bibitem{link2isbi-14}
Isbi-14 data.
\newblock
  \url{https://cs.adelaide.edu.au/~carneiro/isbi14_challenge/dataset.html}.

\bibitem{link2isbi-15}
Isbi-15 data.
\newblock \url{https://cs.adelaide.edu.au/~zhi/isbi15_challenge/dataset.html}.

\bibitem{link2Cellpose}
Cellpose data.
\newblock \url{https://www.cellpose.org/dataset}.

\bibitem{link2TissueNet}
Tissuenet data.
\newblock \url{http://netbio.bgu.ac.il/tissuenet}.

\bibitem{link2EVICAN}
Evican data.
\newblock
  \url{https://edmond.mpdl.mpg.de/dataset.xhtml?persistentId=doi:10.17617/3.AJBV1S}.

\bibitem{link2LIVECell}
Livecell data.
\newblock \url{https://sartorius-research.github.io/LIVECell/}.

\bibitem{caoComprehensiveSinglecellTranscriptional2017a}
Junyue Cao, Jonathan~S. Packer, Vijay Ramani, Darren~A. Cusanovich, Chau Huynh,
  Riza Daza, Xiaojie Qiu, Choli Lee, Scott~N. Furlan, Frank~J. Steemers, Andrew
  Adey, Robert~H. Waterston, Cole Trapnell, and Jay Shendure.
\newblock Comprehensive single-cell transcriptional profiling of a
  multicellular organism.
\newblock {\em Science}, 357(6352):661--667, August 2017.

\bibitem{fincherCellTypeTranscriptome2018}
Christopher~T. Fincher, Omri Wurtzel, Thom de~Hoog, Kellie~M. Kravarik, and
  Peter~W. Reddien.
\newblock Cell type transcriptome atlas for the planarian \textit{{Schmidtea}
  mediterranea}.
\newblock {\em Science}, 360(6391):eaaq1736, May 2018.

\bibitem{hanMappingMouseCell2018}
Xiaoping Han, Renying Wang, Yincong Zhou, Lijiang Fei, Huiyu Sun, Shujing Lai,
  Assieh Saadatpour, Ziming Zhou, Haide Chen, Fang Ye, Daosheng Huang, Yang Xu,
  Wentao Huang, Mengmeng Jiang, Xinyi Jiang, Jie Mao, Yao Chen, Chenyu Lu, Jin
  Xie, Qun Fang, Yibin Wang, Rui Yue, Tiefeng Li, He~Huang, Stuart~H. Orkin,
  Guo-Cheng Yuan, Ming Chen, and Guoji Guo.
\newblock Mapping the {Mouse} {Cell} {Atlas} by {Microwell}-{Seq}.
\newblock {\em Cell}, 172(5):1091--1107.e17, February 2018.

\bibitem{thetabulamurisconsortiumSinglecellTranscriptomics202018}
{The Tabula Muris Consortium}, {Overall coordination}, {Logistical
  coordination}, {Organ collection and processing}, {Library preparation and
  sequencing}, {Computational data analysis}, {Cell type annotation}, {Writing
  group}, {Supplemental text writing group}, and {Principal investigators}.
\newblock Single-cell transcriptomics of 20 mouse organs creates a {Tabula}
  {Muris}.
\newblock {\em Nature}, 562(7727):367--372, October 2018.

\bibitem{abdelaalComparisonAutomaticCell2019}
Tamim Abdelaal, Lieke Michielsen, Davy Cats, Dylan Hoogduin, Hailiang Mei,
  Marcel J.~T. Reinders, and Ahmed Mahfouz.
\newblock A comparison of automatic cell identification methods for single-cell
  {RNA} sequencing data.
\newblock {\em Genome Biology}, 20(1):194, December 2019.

\bibitem{huangEvaluationCellType2021}
Qianhui Huang, Yu~Liu, Yuheng Du, and Lana~X. Garmire.
\newblock Evaluation of {Cell} {Type} {Annotation} {R} {Packages} on
  {Single}-cell {RNA}-seq {Data}.
\newblock {\em Genomics, Proteomics \& Bioinformatics}, 19(2):267--281, April
  2021.

\bibitem{pasquiniAutomatedMethodsCell2021}
Giovanni Pasquini, Jesus~Eduardo Rojo~Arias, Patrick Sch{\"a}fer, and Volker
  Busskamp.
\newblock Automated methods for cell type annotation on {scRNA}-seq data.
\newblock {\em Computational and Structural Biotechnology Journal},
  19:961--969, 2021.

\bibitem{Liao2020}
Xin Shao, Jie Liao, Xiaoyan Lu, Rui Xue, Ni~Ai, and Xiaohui Fan.
\newblock sccatch: Automatic annotation on cell types of clusters from
  single-cell rna sequencing data.
\newblock {\em iScience}, 23(3), 2020.

\bibitem{zhangProbabilisticCelltypeAssignment2019}
Allen~W. Zhang, Ciara O'Flanagan, Elizabeth~A. Chavez, Jamie L.~P. Lim,
  Nicholas Ceglia, Andrew McPherson, Matt Wiens, Pascale Walters, Tim Chan,
  Brittany Hewitson, Daniel Lai, Anja Mottok, Clementine Sarkozy, Lauren Chong,
  Tomohiro Aoki, Xuehai Wang, Andrew~P Weng, Jessica~N. McAlpine, Samuel
  Aparicio, Christian Steidl, Kieran~R. Campbell, and Sohrab~P. Shah.
\newblock Probabilistic cell-type assignment of single-cell {{RNA-seq}} for
  tumor microenvironment profiling.
\newblock {\em Nature Methods}, 16(10):1007--1015, 2019.

\bibitem{zhangSCINASemiSupervisedSubtyping2019}
Ze~Zhang, Danni Luo, Xue Zhong, Jin~Huk Choi, Yuanqing Ma, Stacy Wang, Elena
  Mahrt, Wei Guo, Eric~W Stawiski, Zora Modrusan, et~al.
\newblock {{SCINA}}: {{A Semi-Supervised Subtyping Algorithm}} of {{Single
  Cells}} and {{Bulk Samples}}.
\newblock {\em Genes}, 10(7):531, 2019.

\bibitem{Cao2020}
Yinghao Cao, Xiaoyue Wang, and Gongxin Peng.
\newblock Scsa: a cell type annotation tool for single-cell rna-seq data.
\newblock {\em Frontiers in Genetics}, 11, 2020.

\bibitem{guoScSorterAssigningCells2021}
Hongyu Guo and Jun Li.
\newblock {{scSorter}}: Assigning cells to known cell types according to marker
  genes.
\newblock {\em Genome Biology}, 22(1):69, 2021.

\bibitem{maoCellMeSHProbabilisticCelltype2022}
Shunfu Mao, Yue Zhang, Georg Seelig, and Sreeram Kannan.
\newblock {{CellMeSH}}: Probabilistic cell-type identification using indexed
  literature.
\newblock {\em Bioinformatics}, 38(5):1393--1402, 2022.

\bibitem{ianevskiFullyautomatedUltrafastCelltype2022}
Aleksandr Ianevski, Anil~K. Giri, and Tero Aittokallio.
\newblock Fully-automated and ultra-fast cell-type identification using
  specific marker combinations from single-cell transcriptomic data.
\newblock {\em Nature Communications}, 13(1):1246, 2022.

\bibitem{kiselevScmapProjectionSinglecell2018}
Vladimir~Yu Kiselev, Andrew Yiu, and Martin Hemberg.
\newblock Scmap: Projection of single-cell {{RNA-seq}} data across data sets.
\newblock {\em Nature Methods}, 15(5):359--362, 2018.

\bibitem{aranReferencebasedAnalysisLung2019}
Dvir Aran, Agnieszka~P. Looney, Leqian Liu, Esther Wu, Valerie Fong, Austin
  Hsu, Suzanna Chak, Ram~P. Naikawadi, Paul~J. Wolters, Adam~R. Abate, Atul~J.
  Butte, and Mallar Bhattacharya.
\newblock Reference-based analysis of lung single-cell sequencing reveals a
  transitional profibrotic macrophage.
\newblock {\em Nature Immunology}, 2(2):163--172, 2019.

\bibitem{dekanterCHETAHSelectiveHierarchical2019}
Jurrian~K {de~Kanter}, Philip Lijnzaad, Tito Candelli, Thanasis Margaritis, and
  Frank C~P Holstege.
\newblock {{CHETAH}}: A selective, hierarchical cell type identification method
  for single-cell {{RNA}} sequencing.
\newblock {\em Nucleic Acids Research}, 47(16):e95--e95, 2019.

\bibitem{houScMatchSinglecellGene2019}
Rui Hou, Elena Denisenko, and Alistair R~R Forrest.
\newblock {{scMatch}}: A single-cell gene expression profile annotation tool
  using reference datasets.
\newblock {\em Bioinformatics}, 35(22):4688--4695, 2019.

\bibitem{ekizCIPRWebbasedShiny2020}
H.~Atakan Ekiz, Christopher~J. Conley, W.~Zac Stephens, and Ryan~M. O'Connell.
\newblock {{CIPR}}: A web-based {{R}}/shiny app and {{R}} package to annotate
  cell clusters in single cell {{RNA}} sequencing experiments.
\newblock {\em BMC Bioinformatics}, 21(1):191, 2020.

\bibitem{plinerSupervisedClassificationEnables2019}
Hannah~A. Pliner, Jay Shendure, and Cole Trapnell.
\newblock Supervised classification enables rapid annotation of cell atlases.
\newblock {\em Nature Methods}, 16(10):983--986, 2019.

\bibitem{gemanClassifyingGeneExpression2004}
GK~Smyth.
\newblock Classifying {Gene} {Expression} {Profiles} from {Pairwise} {mRNA}
  {Comparisons}.
\newblock {\em Statistical Applications in Genetics and Molecular Biology},
  3(1):1--19, January 2004.

\bibitem{breimanRandomForests2001}
Leo Breiman.
\newblock Random {Forests}.
\newblock {\em Machine Learning}, 45(1):5--32, 2001.

\bibitem{liSciBetPortableFast2020}
Chenwei Li, Baolin Liu, Boxi Kang, Zedao Liu, Yedan Liu, Changya Chen, Xianwen
  Ren, and Zemin Zhang.
\newblock {{SciBet}} as a portable and fast single cell type identifier.
\newblock {\em Nature Communications}, 11(1):1818, 2020.

\bibitem{dominguezcondeCrosstissueImmuneCell2022}
C.~Dom\'inguez~Conde, C.~Xu, L.~B. Jarvis, D.~B. Rainbow, S.~B. Wells,
  T.~Gomes, S.~K. Howlett, O.~Suchanek, K.~Polanski, H.~W. King, L.~Mamanova,
  N.~Huang, P.~A. Szabo, L.~Richardson, L.~Bolt, E.~S. Fasouli, K.~T.
  Mahbubani, M.~Prete, L.~Tuck, N.~Richoz, Z.~K. Tuong, L.~Campos, H.~S. Mousa,
  E.~J. Needham, S.~Pritchard, T.~Li, R.~Elmentaite, J.~Park, E.~Rahmani,
  D.~Chen, D.~K. Menon, O.~A. Bayraktar, L.~K. James, K.~B. Meyer, N.~Yosef,
  M.~R. Clatworthy, P.~A. Sims, D.~L. Farber, K.~Saeb-Parsy, J.~L. Jones, and
  S.~A. Teichmann.
\newblock Cross-tissue immune cell analysis reveals tissue-specific features in
  humans.
\newblock {\em Science}, 376(6594):eabl5197, 2022.

\bibitem{alquicira-hernandezScPredAccurateSupervised2019}
Jose Alquicira-Hernandez, Anuja Sathe, Hanlee~P. Ji, Quan Nguyen, and Joseph~E.
  Powell.
\newblock {{scPred}}: Accurate supervised method for cell-type classification
  from single-cell {{RNA-seq}} data.
\newblock {\em Genome Biology}, 20(1):264, 2019.

\bibitem{maACTINNAutomatedIdentification2019}
Feiyang Ma and Matteo Pellegrini.
\newblock {{ACTINN}}: Automated identification of cell types in single cell
  {{RNA}} sequencing.
\newblock {\em Bioinformatics}, page btz592, 2019.

\bibitem{xieSuperCTSupervisedlearningFramework2019}
Peng Xie, Mingxuan Gao, Chunming Wang, Jianfei Zhang, Pawan Noel, Chaoyong
  Yang, Daniel Von~Hoff, Haiyong Han, Michael~Q Zhang, and Wei Lin.
\newblock {{SuperCT}}: A supervised-learning framework for enhanced
  characterization of single-cell transcriptomic profiles.
\newblock {\em Nucleic Acids Research}, 47(8):e48--e48, 2019.

\bibitem{chenEnClaSCNovelEnsemble2020}
Xiaoyang Chen, Shengquan Chen, and Rui Jiang.
\newblock {{EnClaSC}}: A novel ensemble approach for accurate and robust
  cell-type classification of single-cell transcriptomes.
\newblock {\em BMC Bioinformatics}, 21(S13):392, 2020.

\bibitem{Ke2017}
Guolin Ke, Qi~Meng, Thomas Finley, Taifeng Wang, Wei Chen, Weidong Ma, Qiwei
  Ye, and Tie-Yan Liu.
\newblock Lightgbm: A highly efficient gradient boosting decision tree.
\newblock In {\em Proceedings of the 31st International Conference on Neural
  Information Processing Systems}, NIPS'17, page 3149–3157, Red Hook, NY,
  USA, 2017. Curran Associates Inc.

\bibitem{kimmelSemisupervisedAdversarialNeural2021}
Jacob~C. Kimmel and David~R. Kelley.
\newblock Semisupervised adversarial neural networks for single-cell
  classification.
\newblock {\em Genome Research}, 31(10):1781--1793, 2021.

\bibitem{berthelotMixMatchHolisticApproach2019}
David Berthelot, Nicholas Carlini, Ian Goodfellow, Nicolas Papernot, Avital
  Oliver, and Colin Raffel.
\newblock {{MixMatch}}: {{A Holistic Approach}} to {{Semi-Supervised
  Learning}}.
\newblock {\em Advances in neural information processing systems}, 2019.

\bibitem{ganinDomainAdversarialTrainingNeural2015}
Yaroslav Ganin, Evgeniya Ustinova, Hana Ajakan, Pascal Germain, Hugo
  Larochelle, Fran\c{c}ois Laviolette, Mario Marchand, and Victor Lempitsky.
\newblock Domain-{{Adversarial Training}} of {{Neural Networks}}.
\newblock {\em The journal of machine learning research}, 2015.

\bibitem{Lee2013}
Dong hyun Lee.
\newblock Pseudo-label: The simple and efficient semi-supervised learning
  method for deep neural networks, 2013.

\bibitem{vermaInterpolationConsistencyTraining2019}
Vikas Verma, Kenji Kawaguchi, Alex Lamb, Juho Kannala, Arno Solin, Yoshua
  Bengio, and David Lopez-Paz.
\newblock Interpolation consistency training for semi-supervised learning.
\newblock {\em Neural Networks}, 145:90--106, 2022.

\bibitem{zhangMixupEmpiricalRisk2017}
Hongyi Zhang, Moustapha Cisse, Yann~N Dauphin, and David Lopez-Paz.
\newblock Mixup: {{Beyond Empirical Risk Minimization}}.
\newblock {\em arXiv preprint arXiv:1710.09412}, 2017.

\bibitem{yinScIAEIntegrativeAutoencoderbased2022}
Qingyang Yin, Yang Wang, Jinting Guan, and Guoli Ji.
\newblock {{scIAE}}: An integrative autoencoder-based ensemble classification
  framework for single-cell {{RNA-seq}} data.
\newblock {\em Briefings in Bioinformatics}, 23(1):bbab508, 2022.

\bibitem{wangSinglecellClassificationUsing2021}
Tianyu Wang, Jun Bai, and Sheida Nabavi.
\newblock Single-cell classification using graph convolutional networks.
\newblock {\em BMC Bioinformatics}, 22(1):364, 2021.

\bibitem{szklarczykSTRINGV11Protein2019}
Damian Szklarczyk, Annika~L Gable, David Lyon, Alexander Junge, Stefan Wyder,
  Jaime Huerta-Cepas, Milan Simonovic, Nadezhda~T Doncheva, John~H Morris, Peer
  Bork, Lars~J Jensen, and Christian~von Mering.
\newblock {{STRING}} v11: Protein\textendash protein association networks with
  increased coverage, supporting functional discovery in genome-wide
  experimental datasets.
\newblock {\em Nucleic Acids Research}, 47(D1):D607--D613, 2019.

\bibitem{hamiltonInductiveRepresentationLearning2017}
William~L. Hamilton, Rex Ying, and Jure Leskovec.
\newblock Inductive {{Representation Learning}} on {{Large Graphs}}.
\newblock {\em Advances in neural information processing systems}, 2017.

\bibitem{tanSingleCellNetComputationalTool2019}
Yuqi Tan and Patrick Cahan.
\newblock {{SingleCellNet}}: {{A Computational Tool}} to {{Classify Single Cell
  RNA-Seq Data Across Platforms}} and {{Across Species}}.
\newblock {\em Cell Systems}, 9(2):207--213.e2, 2019.

\bibitem{regevHumanCellAtlas2017}
Aviv Regev, Sarah~A Teichmann, Eric~S Lander, Ido Amit, Christophe Benoist,
  Ewan Birney, Bernd Bodenmiller, Peter Campbell, Piero Carninci, Menna
  Clatworthy, Hans Clevers, Bart Deplancke, Ian Dunham, James Eberwine, Roland
  Eils, Wolfgang Enard, Andrew Farmer, Lars Fugger, Berthold G{\"o}ttgens, Nir
  Hacohen, Muzlifah Haniffa, Martin Hemberg, Seung Kim, Paul Klenerman, Arnold
  Kriegstein, Ed~Lein, Sten Linnarsson, Emma Lundberg, Joakim Lundeberg, Partha
  Majumder, John~C Marioni, Miriam Merad, Musa Mhlanga, Martijn Nawijn, Mihai
  Netea, Garry Nolan, Dana Pe'er, Anthony Phillipakis, Chris~P Ponting, Stephen
  Quake, Wolf Reik, Orit Rozenblatt-Rosen, Joshua Sanes, Rahul Satija, Ton~N
  Schumacher, Alex Shalek, Ehud Shapiro, Padmanee Sharma, Jay~W Shin, Oliver
  Stegle, Michael Stratton, Michael J~T Stubbington, Fabian~J Theis, Matthias
  Uhlen, Alexander van Oudenaarden, Allon Wagner, Fiona Watt, Jonathan
  Weissman, Barbara Wold, Ramnik Xavier, Nir Yosef, and {Human Cell Atlas
  Meeting Participants}.
\newblock The {Human} {Cell} {Atlas}.
\newblock {\em eLife}, 6:e27041, December 2017.

\bibitem{thetabulasapiensconsortium*TabulaSapiensMultipleorgan2022}
{The Tabula Sapiens Consortium*}, Robert~C. Jones, Jim Karkanias, Mark~A.
  Krasnow, Angela~Oliveira Pisco, Stephen~R. Quake, Julia Salzman, Nir Yosef,
  Bryan Bulthaup, Phillip Brown, William Harper, Marisa Hemenez, Ravikumar
  Ponnusamy, Ahmad Salehi, Bhavani~A. Sanagavarapu, Eileen Spallino, Ksenia~A.
  Aaron, Waldo Concepcion, James~M. Gardner, Burnett Kelly, Nikole Neidlinger,
  Zifa Wang, Sheela Crasta, Saroja Kolluru, Maurizio Morri, Angela~Oliveira
  Pisco, Serena~Y. Tan, Kyle~J. Travaglini, Chenling Xu, Marcela
  Alc{\'a}ntara-Hern{\'a}ndez, Nicole Almanzar, Jane Antony, Benjamin
  Beyersdorf, Deviana Burhan, Kruti Calcuttawala, Matthew~M. Carter, Charles
  K.~F. Chan, Charles~A. Chang, Stephen Chang, Alex Colville, Sheela Crasta,
  Rebecca~N. Culver, Ivana Cvijovi{\'c}, Gaetano D'Amato, Camille Ezran,
  Francisco~X. Galdos, Astrid Gillich, William~R. Goodyer, Yan Hang, Alyssa
  Hayashi, Sahar Houshdaran, Xianxi Huang, Juan~C. Irwin, SoRi Jang,
  Julia~Vallve Juanico, Aaron~M. Kershner, Soochi Kim, Bernhard Kiss, Saroja
  Kolluru, William Kong, Maya~E. Kumar, Angera~H. Kuo, Rebecca Leylek, Baoxiang
  Li, Gabriel~B. Loeb, Wan-Jin Lu, Sruthi Mantri, Maxim Markovic, Patrick~L.
  McAlpine, Antoine de~Morree, Maurizio Morri, Karim Mrouj, Shravani Mukherjee,
  Tyler Muser, Patrick Neuh{\"o}fer, Thi~D. Nguyen, Kimberly Perez, Ragini
  Phansalkar, Angela~Oliveira Pisco, Nazan Puluca, Zhen Qi, Poorvi Rao, Hayley
  Raquer-McKay, Nicholas Schaum, Bronwyn Scott, Bobak Seddighzadeh, Joe Segal,
  Sushmita Sen, Shaheen Sikandar, Sean~P. Spencer, Lea~C. Steffes, Varun~R.
  Subramaniam, Aditi Swarup, Michael Swift, Kyle~J. Travaglini, Will
  Van~Treuren, Emily Trimm, Stefan Veizades, Sivakamasundari Vijayakumar,
  Kim~Chi Vo, Sevahn~K. Vorperian, Wanxin Wang, Hannah N.~W. Weinstein, Juliane
  Winkler, Timothy T.~H. Wu, Jamie Xie, Andrea~R. Yung, Yue Zhang, Angela~M.
  Detweiler, Honey Mekonen, Norma~F. Neff, Rene~V. Sit, Michelle Tan, Jia Yan,
  Gregory~R. Bean, Vivek Charu, Erna Forg{\'o}, Brock~A. Martin, Michael~G.
  Ozawa, Oscar Silva, Serena~Y. Tan, Angus Toland, Venkata N.~P. Vemuri, Shaked
  Afik, Kyle Awayan, Olga~Borisovna Botvinnik, Ashley Byrne, Michelle Chen,
  Roozbeh Dehghannasiri, Angela~M. Detweiler, Adam Gayoso, Alejandro~A.
  Granados, Qiqing Li, Gita Mahmoudabadi, Aaron McGeever, Antoine de~Morree,
  Julia~Eve Olivieri, Madeline Park, Angela~Oliveira Pisco, Neha Ravikumar,
  Julia Salzman, Geoff Stanley, Michael Swift, Michelle Tan, Weilun Tan,
  Alexander~J. Tarashansky, Rohan Vanheusden, Sevahn~K. Vorperian, Peter Wang,
  Sheng Wang, Galen Xing, Chenling Xu, Nir Yosef, Marcela
  Alc{\'a}ntara-Hern{\'a}ndez, Jane Antony, Charles K.~F. Chan, Charles~A.
  Chang, Alex Colville, Sheela Crasta, Rebecca Culver, Les Dethlefsen, Camille
  Ezran, Astrid Gillich, Yan Hang, Po-Yi Ho, Juan~C. Irwin, SoRi Jang, Aaron~M.
  Kershner, William Kong, Maya~E. Kumar, Angera~H. Kuo, Rebecca Leylek, Shixuan
  Liu, Gabriel~B. Loeb, Wan-Jin Lu, Jonathan~S. Maltzman, Ross~J. Metzger,
  Antoine de~Morree, Patrick Neuh{\"o}fer, Kimberly Perez, Ragini Phansalkar,
  Zhen Qi, Poorvi Rao, Hayley Raquer-McKay, Koki Sasagawa, Bronwyn Scott, Rahul
  Sinha, Hanbing Song, Sean~P. Spencer, Aditi Swarup, Michael Swift, Kyle~J.
  Travaglini, Emily Trimm, Stefan Veizades, Sivakamasundari Vijayakumar, Bruce
  Wang, Wanxin Wang, Juliane Winkler, Jamie Xie, Andrea~R. Yung, Steven~E.
  Artandi, Philip~A. Beachy, Michael~F. Clarke, Linda~C. Giudice, Franklin~W.
  Huang, Kerwyn~Casey Huang, Juliana Idoyaga, Seung~K. Kim, Mark Krasnow,
  Christin~S. Kuo, Patricia Nguyen, Stephen~R. Quake, Thomas~A. Rando, Kristy
  Red-Horse, Jeremy Reiter, David~A. Relman, Justin~L. Sonnenburg, Bruce Wang,
  Albert Wu, Sean~M. Wu, and Tony Wyss-Coray.
\newblock The {Tabula} {Sapiens}: {A} multiple-organ, single-cell
  transcriptomic atlas of humans.
\newblock {\em Science}, 376(6594):eabl4896, May 2022.

\bibitem{hanConstructionHumanCell2020}
Xiaoping Han, Ziming Zhou, Lijiang Fei, Huiyu Sun, Renying Wang, Yao Chen,
  Haide Chen, Jingjing Wang, Huanna Tang, Wenhao Ge, Yincong Zhou, Fang Ye,
  Mengmeng Jiang, Junqing Wu, Yanyu Xiao, Xiaoning Jia, Tingyue Zhang, Xiaojie
  Ma, Qi~Zhang, Xueli Bai, Shujing Lai, Chengxuan Yu, Lijun Zhu, Rui Lin, Yuchi
  Gao, Min Wang, Yiqing Wu, Jianming Zhang, Renya Zhan, Saiyong Zhu, Hailan Hu,
  Changchun Wang, Ming Chen, He~Huang, Tingbo Liang, Jianghua Chen, Weilin
  Wang, Dan Zhang, and Guoji Guo.
\newblock Construction of a human cell landscape at single-cell level.
\newblock {\em Nature}, 581(7808):303--309, May 2020.

\end{thebibliography}

\end{document}